\pdfoutput=1
\documentclass[a4paper,amsmath,amssymb,prb,preprintnumbers,superscriptaddress,twocolumn,showpacs,longbibliography]{revtex4-2}
\pdfoutput=1
\usepackage{graphicx}
\usepackage{filecontents}
\usepackage{physics}

\usepackage{dcolumn}  
\usepackage{bm}         
\usepackage{amssymb}
\usepackage{amsmath}
\usepackage{epsfig}
\usepackage[dvipsnames]{xcolor}
\usepackage{xr} 
\usepackage{siunitx}
\usepackage{xspace}
\newcommand{\cpl}{\ensuremath{\,\frac{\mathrm{n \Phi_0}}{\mathrm{\mu_B}}}\xspace}

\newcommand{\jn}[1]{\textcolor{blue}{#1}}

\newif\ifcom
\newif\ifdel
\comtrue                    %
\deltrue                      %

\begin{document}

\title{On the coupling of magnetic moments to superconducting quantum interference devices}

\author{J.~Linek}
\affiliation{%
Physikalisches Institut, Center for Quantum Science (CQ) and LISA$^+$,
University of T\"ubingen,
Auf der Morgenstelle 14,
72076 T\"ubingen, Germany}

\author{M.~Wyszynski}
\affiliation{%
Department of Physics \& NANOlab Center of Excellence, 
University of Antwerp, 
Groenenborgerlaan 171, 
B-2020 Antwerp, Belgium}

\author{B.~M\"uller}
\affiliation{%
Physikalisches Institut, Center for Quantum Science (CQ) and LISA$^+$,
University of T\"ubingen,
Auf der Morgenstelle 14,
72076 T\"ubingen, Germany}

\author{D.~Korinski}
\affiliation{%
Physikalisches Institut, Center for Quantum Science (CQ) and LISA$^+$,
University of T\"ubingen,
Auf der Morgenstelle 14,
72076 T\"ubingen, Germany}

\author{M.~V.~Milo\v{s}evi\'c}
\affiliation{%
Department of Physics \& NANOlab Center of Excellence, 
University of Antwerp, 
Groenenborgerlaan 171, 
B-2020 Antwerp, Belgium}

\author{R.~Kleiner}
\affiliation{%
Physikalisches Institut, Center for Quantum Science (CQ) and LISA$^+$,
University of T\"ubingen,
Auf der Morgenstelle 14,
72076 T\"ubingen, Germany}

\author{D.~Koelle}
\email{koelle@uni-tuebingen.de}
\affiliation{%
Physikalisches Institut, Center for Quantum Science (CQ) and LISA$^+$,
University of T\"ubingen,
Auf der Morgenstelle 14,
72076 T\"ubingen, Germany}

\date{\today}


\begin{abstract} 
We investigate the coupling factor $\phi_\mu$ that quantifies the magnetic flux $\Phi$ per magnetic moment $\mu$ of a point-like magnetic dipole that couples to a superconducting quantum interference device (SQUID).
Representing the dipole by a tiny current-carrying (Amperian) loop, the reciprocity of mutual inductances of SQUID and Amperian loop provides an elegant way of calculating $\phi_\mu(\bm r, \hat{\bm e}_\mu)$ vs.~position $\bm r$ and orientation $\hat{\bm e}_\mu$ of the dipole anywhere in space from the magnetic field $B(\bm r)$ produced by a supercurrent circulating in the SQUID loop.
We use numerical simulations based on London and Ginzburg-Landau theory to calculate $\phi_\mu$ from the supercurrent density distributions in various SQUID geometries.
We treat the far-field regime ($r\gtrsim a=$ inner size of the SQUID loop) with the dipole placed on (oriented along) the symmetry axis of circular or square shaped loops.
We compare expressions for $\phi_\mu$ from simple filamentary loop models with simulation results for loops with finite width $w$ (outer size $A>a$), thickness $d$ and London penetration depth $\lambda_{\rm L}$ and show that for thin ($d\ll a$) and narrow ($w < a$) loops the introduction of an effective loop size $a_{\rm eff}$ in the filamentary loop-model expressions results in good agreement with simulations.
For a dipole placed right in the center of the loop, simulations provide an expression $\phi_\mu(a,A,d,\lambda_{\rm L})$ that covers a wide parameter range.
In the near-field regime (dipole centered at small distance $z$ above one SQUID arm) only coupling to a single strip representing the SQUID arm has to be considered.
For this case, we compare simulations with an analytical expression derived for a homogeneous current density distribution, which yields excellent agreement for $\lambda_{\rm L}>w,d$.
Moreover, we analyze the improvement of $\phi_\mu$ provided by the introduction of a narrow constriction in the SQUID arm below the magnetic dipole.
\end{abstract} 



\maketitle

\section{Introduction}
\label{sec:Introduction}

Magnetic properties of micro- and nanoscale objects, are a topic of intensive research\cite{Fernandez-Pacheco17, Ali21}.
Their investigation requires the development of appropriate tools, e.g. for detection of the magnetization reversal of individual magnetic nanoparticles (MNPs) or for imaging magnetic field profiles on the nanoscale \cite{Lipert10, Ganzhorn13, Schaefer-Nolte14, Marchiori22}.
Promising candidates for this task are nanoSQUIDs, i.e., strongly miniaturized superconducting quantum interference devices \cite{Foley09, Granata16, Martinez-Perez17a, Hao17}.
A SQUID consists of a superconducting loop, intersected by one or two weak links (Josephson junctions).
SQUIDs are extremely sensitive detectors for magnetic flux, and their intrinsic thermal flux noise improves with shrinking size (inductance of the SQUID loop) \cite{Tesche77, Chesca-SHB-2}.
As they enable direct detection of magnetization changes in small spin systems, that are placed close to the SQUID loop, nanoSQUIDs are very promising sensors for nanoscale applications \cite{Wernsdorfer01, Wernsdorfer09}.
Moreover, if miniaturized SQUIDs can be brought in close vicinity to sample surfaces, they enable magnetic scanning probe microscopy on the nanoscale \cite{Vasyukov13, Anahory20, Persky22, Wyss22}.

For the application of nanoSQUIDs to the investigation of magnetic nanoparticles, the figure of merit is the spin sensitivity $S_\mu^{1/2} = S_\Phi^{1/2} / \phi_\mu$, which describes the smallest magnetic moment $\bm{\mu}$ of a point-like magnetic nanoparticle (MNP) that can be detected with a certain device at a certain distance.
Here, $S_\Phi^{1/2}$ is the rms flux noise of the SQUID, i.e.~the  square root of the spectral density of equivalent flux noise power $S_\Phi$, which can determined experimentally, and $\phi_\mu =  \Phi / \mu$ is the coupling factor, expressing how much magnetic flux $\Phi$ per magnetic moment $\mu = |\bm{\mu}|$ is coupled to the SQUID loop by a MNP which is placed at position $\bm r$.
$\phi_\mu (\bm r, \hat{\bm e}_\mu)$ depends on the orientation $\hat{\bm e}_\mu=\bm{\mu}/\mu$ and position $\bm r$ (with respect to the nanoSQUID) of the MNP and is usually estimated either via magnetostatic considerations\cite{Ketchen89, Koshnick08, Tilbrook09} or numerical methods \cite{Hasselbach00, Bouchiat09, Granata18}.
However, these methods do not incorporate the finite size of the superconductor and the spatial variation of the supercurrent density across the width of the superconducting structure forming the SQUID loop, which particularly influences the coupling when optimizing $\phi_\mu$ by minimizing the distance between MNP and SQUID loop \cite{Bouchiat09}.
In order to overcome this drawback, one can perform numerical simulations taking into account the precise SQUID geometry and supercurrent distribution within the loop, as done in Refs.~[\onlinecite{Brandt05a,Nagel11, Nagel11a, Woelbing13, Schwarz13, Woelbing14, Schwarz15, Martinez-Perez16, Martinez-Perez17, Martinez-Perez18}].
The downside of this brute-force approach is its complex and time-consuming nature.

In this work, we present a detailed analysis of the dependence of the coupling factor  $\phi_\mu$ on various geometrical parameters of the SQUID loop, on the London penetration depth $\lambda_{\rm L}$ and on the MNP-to-SQUID distance.
This analysis is based on numerical simulations, as mentioned above.
From the results of these simulations we derive then generalized fit functions for the dependence of $\phi_\mu$ on all relevant parameters.
The purpose of this work is to provide such generalized fit functions that can then be used as a straightforward and easy-to-use alternative for estimates of the coupling factor for realistic SQUID geometries.

We will treat two different regimes, which have been already introduced in Ref.~[\onlinecite{Bouchiat09}]: 
(i) In the far-field regime, the distance of an MNP to the superconductor is comparable to or larger than the inner size of the SQUID loop.
For this case, we consider two commonly used simple SQUID geometries, circular and square shaped loops in the $(x,y)$ plane.
We first discuss the dependence $\phi_\mu(z)$ on the position $z$ of a MNP along the $z$ symmetry axis of the SQUID (with the magnetic moment pointing in $z$-direction, i.e., $\hat{\bm e}_\mu=\hat{\bm e}_z$).
Then, we fix that position to $z=0$ (MNP in the center of the loop) and derive expressions for the dependence of $\phi_\mu$ on the inner and outer loop size, and on the finite thickness $d$  and London penetration depth $\lambda_{\rm L}$ of the superconducting film forming the loop.
(ii) In the near-field regime, we consider the situation when an MNP is placed in close vicinity to one arm of the SQUID loop, which can contain a constriction.
This arrangement typically yields larger values for $\phi_\mu$, as compared to placing an MNP in the center of the SQUID loop. 
Here, we derive an expression for the dependence of $\phi_\mu$ on the distance $z$ of a MNP centered above one  arm of the SQUID and on the width $w$, thickness $d$ and the length $l$ of the arm.

\section{Methods}
\label{sec:methods}

Our method of numerically calculating $\phi_\mu$ is described and applied in Refs.~[\onlinecite{Nagel11, Nagel11a, Woelbing13, Schwarz13, Woelbing14, Schwarz15, Martinez-Perez16, Martinez-Perez17, Martinez-Perez18}] (the most detailed description can be found in Ref.~[\onlinecite{Woelbing14}]).
Due to the central importance for this work we give a brief summary and present an alternative derivation of this method.

\begin{figure}[t]
\includegraphics[width=0.9 \columnwidth]{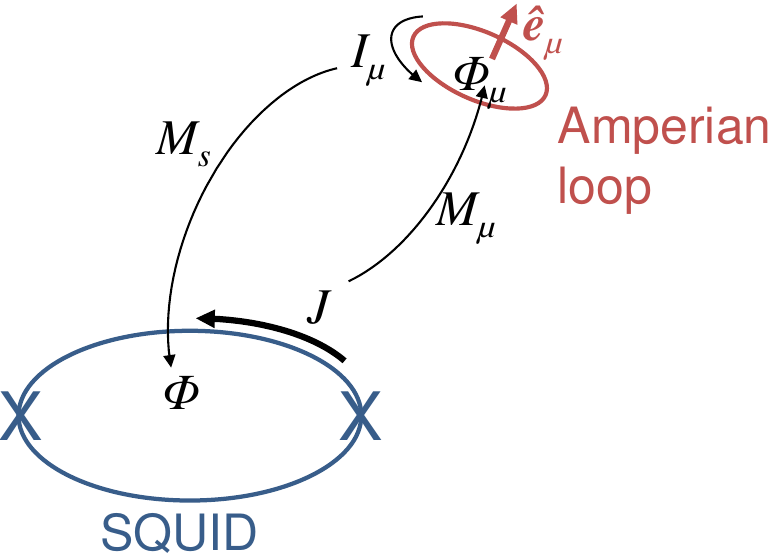} 
\caption{Illustration of the definition of mutual inductances $M_s$ and $M_\mu$ between a SQUID and an Amperian loop (with area $A_\mu$ and normal unit vector $\hat{\bm e}_\mu$), representing a point-like magnetic dipole moment in the limit $A_\mu\to 0$}.
\label{Fig:mutual}
\end{figure}

The coupling factor $\phi_\mu (\bm{r}, \hat{\bm e}_\mu)$ for a point-like MNP with its magnetic moment oriented along $\hat{\bm e}_\mu$ at position $\bm{r}$ can be calculated as
%
\begin{equation}
\phi_\mu (\bm{r}, \hat{\bm e}_\mu) = \hat{\bm e}_\mu \cdot \bm{B}_J (\bm{r}) / J\;,
\label{Eq:phi-mu}
\end{equation}
%
where $\bm B_J(\bm r)$ is the magnetic field that is produced by the current $J$ which is circulating in the SQUID loop \cite{Nagel11}.
We show here, that Eq.~(\ref{Eq:phi-mu}) can also be obtained by employing the Amperian loop model for a magnetic moment placed in the vicinity of the SQUID loop (cf.~Fig.~\ref{Fig:mutual}) and by using the symmetry of the mutual inductances between the two loops.

Within the Amperian loop model the magnetic dipole moment is given as $\bm{\mu}=I_\mu\bm{A_\mu}$, with the current $I_\mu$ flowing through an infinitesimally thin loop, surrounding the oriented flat area $\bm{A_\mu}=A_\mu\hat{\bm e}_\mu$; accordingly, $\mu = I_\mu A_\mu$.
The mutual inductance $M_s=\Phi/I_\mu$ describes the flux $\Phi$ coupled to the SQUID by the current $I_\mu$ flowing through the Amperian loop.
Conversely, the  mutual inductance $M_\mu=\Phi_\mu/J$ describes the flux $\Phi_\mu$ coupled to the Amperian loop by the current $J$ circulating in the SQUID loop.
From the identity $M_s=M_\mu$, and with $I_\mu=\mu/A_\mu$, we obtain
%
\begin{equation}
\frac{\Phi}{\mu} =  \frac{\Phi_\mu}{JA_\mu} = \frac{1}{JA_\mu}\int_{A_\mu} \hat{\bm e}_\mu\cdot\bm{B}_J(\bm{r})\,\mathrm{d}A_\mu \; .
\label{eq:Phi-o-mu}
\end{equation}
%
This equals Eq.~(\ref{Eq:phi-mu}) for an arbitrarily small area $A_\mu$ (i.e., for the Amperian loop describing a point-like elementary magnetic dipole), where a constant magnetic field $\bm{B}_J(\bm{r})$ across the loop area at position $\bm{r}$ can be assumed.

Equation (\ref{Eq:phi-mu}) provides an elegant way of calculating the coupling factor for a magnetic moment placed at any position $\bm r$ with any orientation $\hat{\bm e}_\mu$.
This approach requires the calculation of the supercurrent density distribution $\bm j(\bm r_s)$ in the SQUID loop for a given SQUID geometry.
We denote here the spatial coordinates inside the volume of the SQUID loop as $\bm r_s$, to discriminate this from the position $\bm r$ of the magnetic dipole.
In the following, we present two different methods for calculating $\bm j(\bm r_s)$, which are either based on solving the London equations or the Ginzburg-Landau equations.

We use the finite element software package 3D-MLSI based on London theory \cite{Khapaev01a, Khapaev03, Khapaev15} to calculate $\bm j(\bm r_s)$ for a given current $J$.
In the 3D-MLSI simulations of full loops, we use for the circulating current $J$ the current which flows when one magnetic flux quantum $\Phi_0$ is trapped in the SQUID loop.
In case of single strip conductors, edges can be defined as terminals, serving as current sources or drains for a given current.
In these simulations, terminals introduce a homogeneous current density at the defined edges, which changes to the correct current density over a short length scale.
3D-MLSI calculates discrete sheet current densities $\bm{j}_{2D,n} (x_s,y_s,z_{s,n})$ in 2-dimensional current sheets configured as triangular meshes.
The sheets are parallel to the $(x,y)$-plane and spread over the film thickness $d$.
One can use an arbitrary number $N$ of current sheets distributed throughout the film thickness, where $n$ is the index number and $z_{s,n}$ the position of the current sheet along the $z$-axis.
In our simulations the current sheets are distributed equidistantly.
The resulting $\bm{j}(\bm r_s)$ depends on the thin film geometry and on $\lambda_{\rm L}$.
When simulating a superconducting thin film structure in 3D-MLSI, without an external source of magnetic field (e.g., due to induced currents in a nearby conducting structure), all current sheets carry the same sheet current density distribution, i.e., $\bm{j}(\bm r_s)$ contains no $z$ dependence.
We note that this is a good approximation when $\lambda_{\rm L}>d$, which is often the case in realizations of nanoSQUIDs.
From the simulated supercurrent distribution, we extract the magnetic field $\bm{B}_J (\bm{r})$ at any position $\bm{r}$ by summation of the fields calculated via the Biot-Savart law 
%
\begin{equation}
\bm{B}_J (\bm{r}) = 
\sum_{n=1}^N \frac{\mu_0}{4\pi} \int_{A} \bm{j}_{2D,n}\left(\bm{r}_{s}\right) \times \frac{\bm{r}-\bm{r}_{s}}{\left|\bm{r}-\bm{r}_{s}\right|^3} {\rm d} A_{s}
\label{Eq:Biot_Savart}
\end{equation}
%
from the individual current sheets.
Here $\bm{r} = (x,y,z)$ and $\bm{r}_{s} = (x_{\jn{s}} ,y_{\jn{s}} ,z_n)$ are vectors pointing at the magnetic moment and at the surface elements d$A_{s}$ of the superconducting sheets of area $A_{s}$, respectively.

%
%
%

We also use numerical simulations based on Ginzburg-Landau (GL) theory to calculate $\bm{j}(\bm{r_s})$.
The behavior of the superconducting condensate is described by a complex-valued order parameter $\Psi$, which is allowed to vary in space.
The first GL equation \cite{Milosevic2010} for the order parameter reads
%
\begin{equation}
(-i \grad - \bm{A})^2 \Psi = \Psi (1-\abs{\Psi}^2)\;.
\label{eq:gl1}
\end{equation}
%
Here $\bm{A}$ is the vector potential, for which the second GL equation is solved:
%
\begin{equation}
    -\kappa^2 \Delta \bm{A} = \frac{1}{2i}(\Psi^* \grad \Psi - \Psi \grad \Psi^*) - \abs{\Psi}^2\bm{A}\;,
    \label{eq:gl2}
\end{equation}
%
where $\kappa$ is the Ginzburg-Landau parameter given as the ratio of the London penetration depth $\lambda_{\rm L}$ and the coherence length $\xi$.
The London gauge ($\div \bm{A}=0$) has been used.
The GL equations are written in dimensionless form, i.e., all quantities are measured in their natural units:
distances are measured in $\xi$, the vector potential $\bm{A}$ in $c\hbar / 2e\xi$, the magnetic field $\bm{B}$ in the upper critical field $B_{c2}$ and the order parameter in its bulk value $\Psi_0$ in the absence of field and current.

Equations (\ref{eq:gl1}) and (\ref{eq:gl2}) are solved self-consistently in an iterative manner.
Equation \eqref{eq:gl1} is discretized on a Cartesian grid in real space, with grid spacing below $0.2\,\xi$.
We define $\Psi$ on the nodes and $\bm{A}$ on the links of the grid (link variable approach).
The Neumann boundary condition imposed at the superconductor-vacuum interfaces reads
%
\begin{equation}
\bm{n} \cdot (-i\grad - \bm{A})\Psi \eval_{boundary} = 0\;,
\label{eq:Neumann}
\end{equation}
%
allowing us to simulate arbitrary geometries.
Equation \eqref{eq:gl2} is solved in the Fourier space with periodic boundary conditions (and a sufficient margin of vacuum taken around the superconducting sample so that resultant supercurrent distributions are not affected by boundary effects).
The supercurrent density is the right-hand side of Eq.~\eqref{eq:gl2} and can be written as
\begin{equation}
\bm{j} = \frac{1}{2} \left[\Psi^*\left(-i\grad - \bm{A}\right)\Psi + \Psi\left(-i\grad - \bm{A}\right)^*\Psi^* \right].
\label{eq:supercurrent}
\end{equation}

In the GL calculations, we simulated fully three-dimensional loops (both circular and rectangular), where we used different ways to induce the  circulating current $J$ within the SQUID  loop.
We either induced $J$ by storing a single flux quantum $\Phi_0$ in the SQUID loop (as in 3D-MLSI simulations), or we induced a screening current $J$ by placing a magnetic dipole with magnetic moment $\mu$, typically of the order of $10^7\mu_B$ ($\mu_B$ being the Bohr magneton), at a fixed position $\bm{r_0}$ close to the SQUID loop.
Those different ways to induce $J$ may  result in slightly different supercurrent density distributions $\bm{j}(\bm{r_s})$ and therefore slightly different coupling factors as the physics differs in both cases.
Another reason for slight discrepancies with respect to simulations based on London theory is the finite coherence length $\xi$.

In either case, the Biot-Savart equation was solved for the magnetic field in free space given the supercurrent distribution calculated from Eq.\ \eqref{eq:supercurrent}, as a volume integral:
%
\begin{equation*}
\bm{B}_J(\bm{r}) = \frac{1}{c} \int_V \bm{j}(\bm{r_s}) \times \frac{\bm{r} - \bm{r_s}}{\abs{\bm{r} - \bm{r_s}}^3} \dd^3 \bm{r_s}\,,
\end{equation*}
%
which in turn allowed for obtaining the coupling factors of interest.

\section{Results}
\label{sec:results}

\subsection{Far-field regime: coupling of out-of-plane moments centered on  circular and square loops}
\label{subsec:far-field}

For the far-field regime we consider the coupling for two simple SQUID geometries, circular and square shaped loops in the $(x,y)$-plane, to the magnetic moment of a MNP oriented along $\hat{\bm e}_z$ and placed on the $z$ symmetry axis.
In this case the origin of the coordinate system is placed in the center of the SQUID loop.
As $\hat{\bm e}_\mu = \hat{\bm e}_z$, only the $z$ component $B_{J,z}$ of $\bm{B}_J$ needs to be considered, i.e.~we calculate the coupling factor given as $\phi_{\mu, z}=B_{J,z}/J$.

\begin{figure}[t]
\includegraphics[width = \columnwidth]{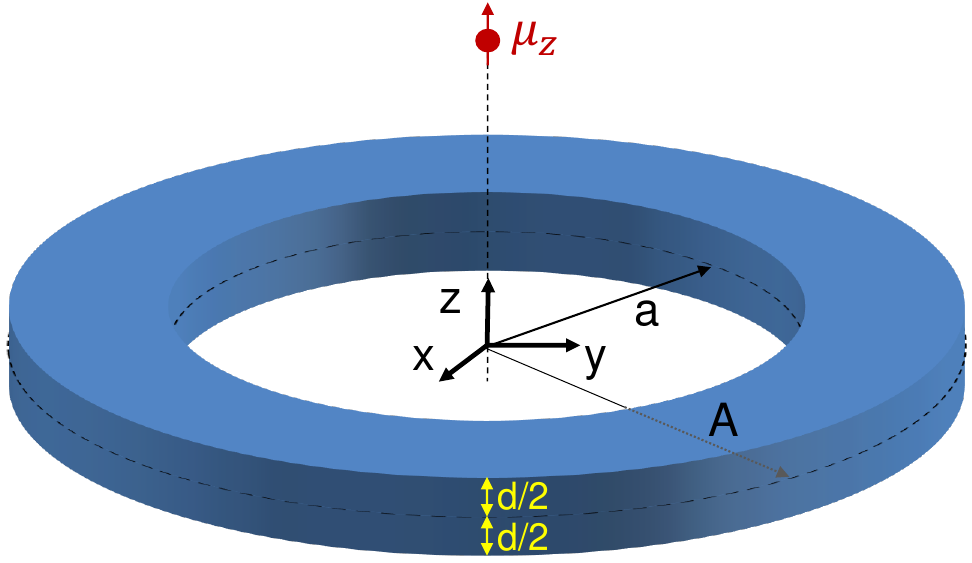}
\caption{
Far-field regime: circular SQUID loop (inner radius $a$, outer radius $A$, film thickness $d$) with the magnetic moment $\bm\mu$ of a MNP on the $z$-axis above the origin pointing along $\hat{\bm e}_z$.
The origin is centered in the loop.}
\label{Fig:circwasher}
\end{figure}

We  first discuss the coupling factor $\phi_{\mu,z,\circ}$ for circular loops with inner radius $a$ and outer radius $A$, as shown scematically in Fig.~\ref{Fig:circwasher}.
For this geometry, Ketchen et al.~[\onlinecite{Ketchen89}] derived a very simple expression for $\phi_\mu(a)$ for the moment of a MNP oriented along $\hat{\bm e}_z$ and placed in the center (at $z=0$) of an infinitely thin ($d = 0$) and narrow ($A = a$), circular SQUID loop:
%
\begin{equation}
\phi_{\mu, {\rm K}, \circ}(a) = \frac{r_e}{a}\,\frac{\mathrm{\Phi_0}}{\mathrm{\mu_B}} = \frac{r_0}{a}\,\frac{\mathrm{n\Phi_0}}{\mathrm{\mu_B}} \approx \frac{2.818\unit{\mu m}}{a}\cpl\;.
\label{Eq:KetchenCirc}
\end{equation}
%
Here, $r_e=\frac{\mu_0}{2}\frac{\mu_{\rm B}}{\Phi_0}\approx 2.818 \times 10^{-15}\unit{m}$ denotes the classical electron radius.
Moreover, we introduce  $r_0=10^9r_e\approx 2.818\,\mu$m, which shows that values for $a$ on the $\mu$m scale produce coupling factors in the range of $1\,\cpl$.
The case treated in Ref.~[\onlinecite{Ketchen89}] has been extended by Tilbrook \cite{Tilbrook09} to arbitrary positions $\bm r$ of MNPs, which yields for MNPs placed along the $z$ symmetry axis
%
\begin{equation}
\phi_{\mu, {\rm T}, \circ}(a,z) = \, \frac{r_0}{a}\frac{1}{(1+\frac{z^2}{a^2})^{3/2}}\,\cpl\;,
\label{Eq:Tilbrook}
\end{equation}
%
which has been also found in \cite{Bouchiat09}.

For an infinitely thin and narrow square shaped SQUID loop with hole size $2a$, the coupling factor for arbitrary position $z$ of an MNP along the $z$ symmetry axis  can be deduced from Eq.~(\ref{Eq:phi-mu}) by calculating the $z$-component of the magnetic field \cite{Giovannetti16} by the Biot-Savart law from a current $J$ flowing through the loop as
%
\begin{equation}
\phi_{\mu, \square}(a,z) = \frac{r_0}{\Delta\cdot a}\frac{1}{(1+\frac{z^2}{a^2})^{3/2}}\sqrt{\frac{2(z^2+a^2)}{z^2+2a^2}}\,\cpl\;.
\label{Eq:Square_Loop}
\end{equation}
%
For $z=0$, Eq.~(\ref{Eq:Square_Loop}) simplifies to an equivalent ``Ketchen expression'' for an infinitely thin and narrow square loop
%
\begin{equation}
\phi_{\mu, {\rm K}, \square}(a) =\frac{r_0}{\Delta\cdot a}\,\frac{\mathrm{n\Phi_0}}{\mathrm{\mu_B}}
\label{Eq:KetchenSquare}
\end{equation}
%
with the geometric scaling factor $\Delta = \pi/\sqrt{8} \approx 1.11 $ \cite{Schmelz12}.

In our simulations, we calculate the coupling factor $\phi_{\mu,z,\circ}$ for the situation described by Eq.~(\ref{Eq:Tilbrook}) , i.e. for variable $a$ and variable $z$ along the symmetry axis of the SQUID loop; in addition, we consider a finite width of the loop with outer radius $A>a$ and finite film thickness $d$ and London penetration depth $\lambda_{\rm L}$.
In case of the far-field regime, fine details on the current density distribution are less important, due to the large distance between the magnetic moment and the superconducting structure.
Therefore all 3D-MLSI simulations for the far-field regime have been carried out with only 11 current sheets distributed equidistantly over the film thickness $d$.

\begin{figure}
\includegraphics[width = \linewidth]{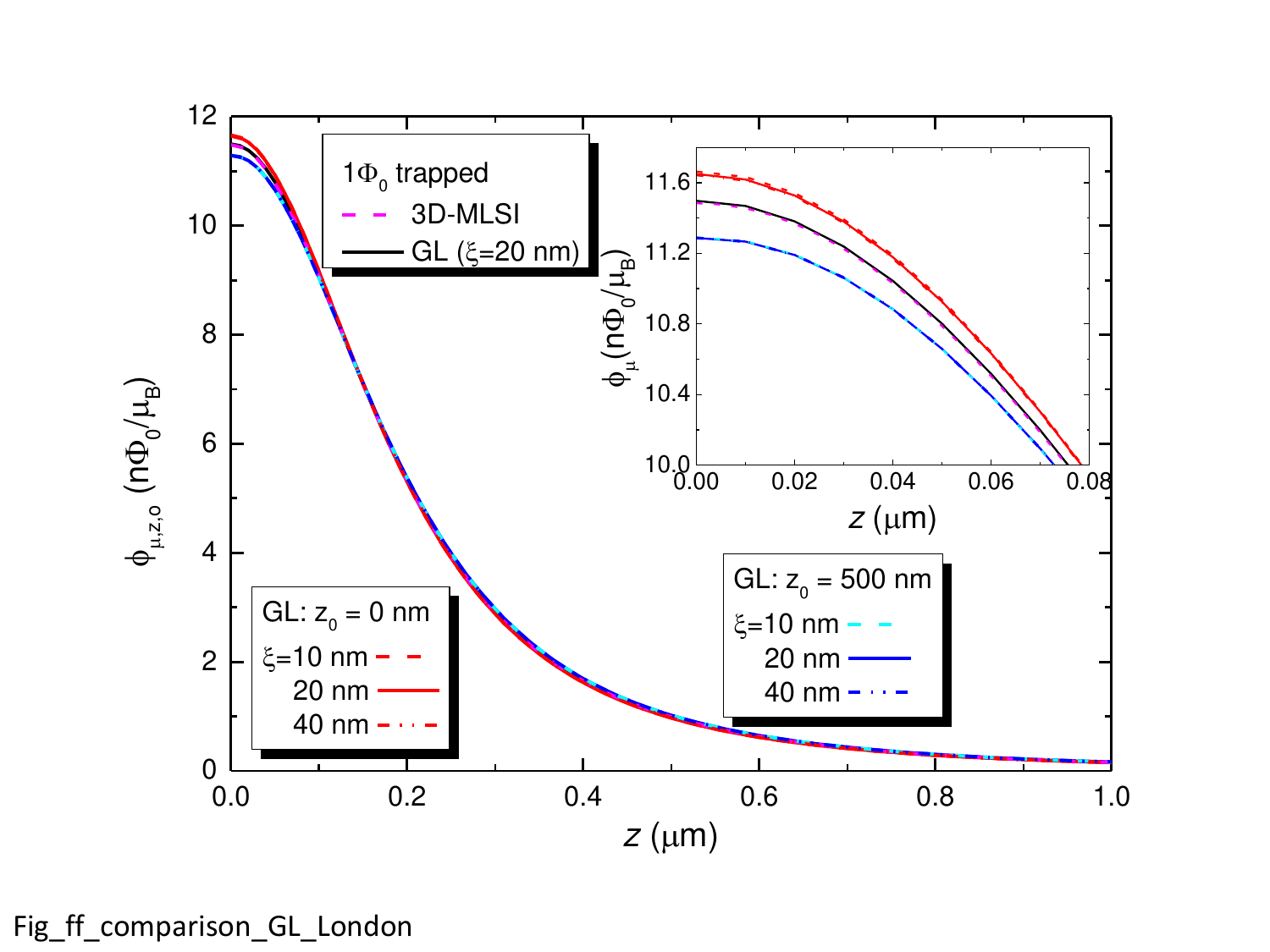}
\caption{Simulated coupling factor $\phi_{\mu,z,\circ}$ vs $z$ position of a MNP placed on the $z$ symmetry axis of a circular SQUID loop with $a = 200\,$nm, $A = 300\,$nm, $d = 50\,$nm and $\lambda_\mathrm{L} = 200\,$nm.
The inset shows a zoom close to $z=0$.
Coupling factors are calculated for three different situations:
(i) Screening currents are induced by trapping one flux quantum in the SQUID loop:
3D-MLSI simulations (dashed pink line) coincide with GL simulations (black solid line; $\xi=20\,$nm).
(ii) Screening currents (from GL simulations for different $\xi$) are induced by placing a magnetic dipole ($\mu\approx35.5 \cdot 10^6 \mu_{\rm B}$) at $z_0 = 0$ (red lines) and $z_0 = 500\,$nm (blue lines).}
\label{Fig_comparison_GL_London}
\end{figure}

To start this analysis we compare coupling factors obtained from both simulation tools and from different scenarios of inducing screening currents in a small loop, with $a=200\,$nm, $A=300\,$nm, $d = 50\,$nm and $\lambda_{\rm L} = 200\,$nm.
The different scenarious produce differences in the induced supercurrent density distributions, and we want to see how strongly those affect the resulting coupling factors.
The results are shown in Fig.~\ref{Fig_comparison_GL_London}.
There, we consider (i) screening currents which are induced by trapping one flux quantum in the loop.
In this case, 3D-MLSI and GL simulations (with $\xi = 20\,$nm) coincide.
(ii) Coupling factors are obtained from GL simulations where the screening currents are induced by a magnetic dipole placed at the origin $z_0=0$ and at $z_0 = 500\,$nm.
These simulations have been carried out for variable coherence length $\xi$, to check the influence on the coupling factor.
The simulation results for fixed $z_0$ (red and blue curves respectively) coincide for all $\xi$, suggesting a negligible effect of $\xi$ on the coupling factor.

One might expect that (ii) provides correct values for the coupling factor only at the position $z=z_0$.
However, Fig.~\ref{Fig_comparison_GL_London} clearly shows that the coupling factors $\phi_{\mu,z,\circ}(z)$ calculated with the two very different values of $z_0$ coincide very well over the entire $z$ range shown here.
Only for $z<100\,$nm slight deviations can be seen (c.f.~inset of Fig.~\ref{Fig_comparison_GL_London}), where $\phi_{\mu,z,\circ}$ calculated with the current density inducded by a dipole placed at $z_0 = 0$ is highest.
Compared to the current density distribution resulting from one flux quantum trapped in the loop, the current is more concentrated at the inner circumference, resulting in a stronger magnetic field along the symmetry axis.
The current density distribution induced by a dipole at $z_0 = 500\,$nm is more homogeneous resulting in smaller coupling factors.
Still, maximum differences in coupling factors (obtained at $z=0$) are only a few percent.
The current density distributions calculated with 3D-MLSI and from GL simulations with $\xi = 20\,$nm for the different cases, can be found in the supplementary material (figures~\ref{suppl-Fig:j_khapaev_lambda_200_d_50_200_300_round}-\ref{suppl-Fig:j_GL_500_xi_20_lambda_200_d_50_200_300_round}).

Altogether, we find that $\phi_{\mu,z,\circ}(z)$ calculated by GL simulations with different values of $\xi$ and $z_0$ has an almost negligible effect on the coupling factor in case of a loop with small line width, and that GL simulations and 3D-MLSI provide consistent results for the coupling factor.
In the supplement a similar comparison of coupling factors (Fig.~\ref{suppl-Fig:khapaev_GL_big_loop}) obtained by 3D-MLSI simulations and GL simulations can be found for a loop with the same inner radius ($a=200\,$nm) but a larger line width ($A = 1.5\,\mu$m).
For the larger line width, differences in the coupling factor for the different methods of inducing screening currents (yielding different current density distributions) become significant (up to about 30\,\% at $z=0$).

\begin{figure}[t]
\includegraphics[width=\columnwidth]{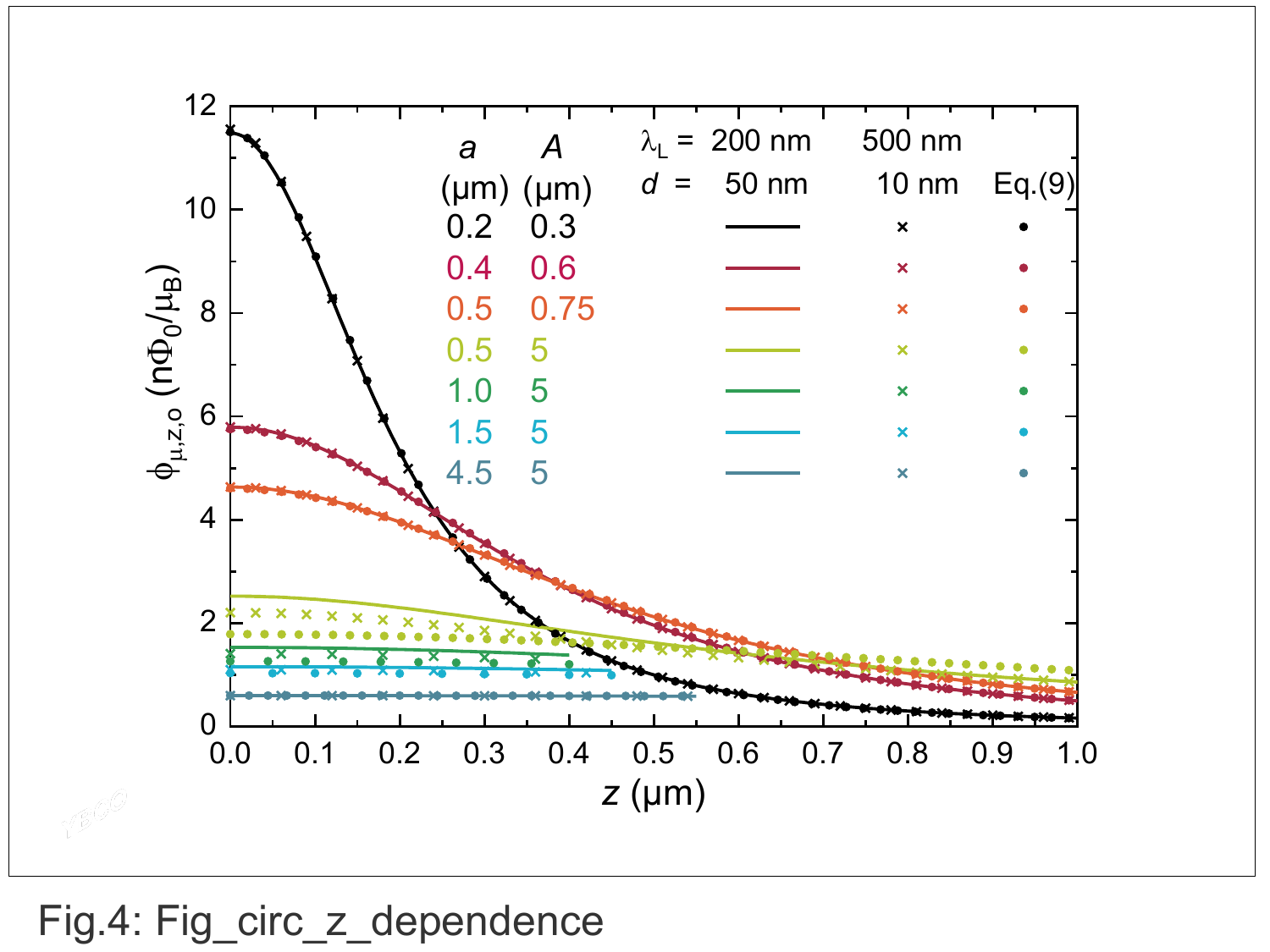}
\caption{With 3D-MLSI simulated coupling factor $\phi_{\mu,z,\circ}$ vs $z$ position of a MNP placed on the $z$ symmetry axis of a circular SQUID loop, for different values of the inner radius $a$ and outer radius $A$.
Results are shown for two sets of $\lambda_{\rm L}$ and $d$ and compared to the expression given in Eq.~(\ref{Eq:Tilbrook}) with $a$ replaced by an effective radius $a_{\rm eff}=\sqrt{a\cdot A}$.}
\label{Fig:washer-z-dep}
\end{figure}

In the following, only data obtained from 3D-MLSI simulations are shown, where the current density originates from trapped flux in the loop.
Figure~\ref{Fig:washer-z-dep} shows simulation results of $\phi_{\mu,z,\circ}(z)$ for different values of $a$ and $A$ and two sets of $\lambda_{\rm L}$ and $d$.
The three curves which yield the largest $\phi_{\mu,z,\circ}$ at $z=0$ are calculated for a fixed ratio $A/a=1.5$, varying $a$ from 0.2 to 0.5\,$\mu$m.
For those, the results for the two different sets of $\lambda_{\rm L}$ and $d$ (solid lines and crosses) coincide, and they also coincide with the modified expression for $\phi_{\mu, {\rm T}, \circ}(a_{\rm eff},z)$ from Eq.~(\ref{Eq:Tilbrook}) (dots), with an effective radius $a_{\rm eff}=\sqrt{a\cdot A}$.
These coincidences get lost when we significantly increase $A$ to 5\,$\mu$m, while keeping $a=0.5\,\mu$m (green curves). 
We note that these coincidences also get lost when $a\gg d $ is not satisfied (not shown).
With the significant increase of $A$, the coupling factor significantly decreases and the $z$ dependence becomes very weak.
Upon further increasing $a$ up to 4.5\,$\mu$m, while keeping $A$ fixed at 5\,$\mu$m, the simulated coupling factor further decreases and becomes almost independent of $z$.
Simulations for an equivalent set of dimensions for square loops exhibit a similar behavior like the data shown in Fig.~\ref{Fig:washer-z-dep}, with slightly reduced coupling factors.
The comparison of simulated coupling factors for square shaped loops with Eq.~(\ref{Eq:Square_Loop}), utilizing the same effective loop size parameter  $a_{\rm eff}$, are shown in the supplementary material, in Sec.~\ref{suppl-subsec:far-field-square}, Fig.~\ref{suppl-fig:square_washer_z_dependence}.

For measurements of the magnetization reversal of individual MNPs, we are mostly interested in the situation which gives strongest coupling.
So in the following we consider $\phi_{\mu,z,\circ}$ for various circular geometries at $z=0$, i.e., for a magnetic dipole placed at the center of the loop, with its moment pointing in $z$ direction.
For this case, we derive from our simulation results a fit formula depending only on the loop geometry $A$, $a$, $d$ and on $\lambda_\mathrm{L}$:
%
\begin{eqnarray}
\label{Eq:washerCirc}
\frac{\phi_{\mu, z, \circ} (A', a', d', \lambda'_\mathrm{L})}{\cpl} &=& \frac{1.57}{a'} + \frac{1.247}{A'} \\
&&- \frac{(A'-a')^{1.5}}{(A' \cdot a')^{1.35}} \cdot\frac{0.43}{(1 + 0.29\, d'/\lambda_\mathrm{L}'^{2})}\nonumber
\end{eqnarray}
%
Here, the primes denote that all parameters are dimensionless and must be given in \unit{\mu m}.
The results given by equation~\eqref{Eq:washerCirc} are within a $\lesssim 10\,\%$ deviation from the results calculated using 3D-MLSI for the parameter range
%
\begin{align*}
a &= 0.2 \ldots 5\,\unit{\mu m} \\
A &= (a + 0.1\unit{\mu m}) \ldots 5\,\unit{\mu m} \\
d &= 10 \ldots 100\,\unit{nm} \\
\lambda_\mathrm{L} &= 10 \ldots 500\,\unit{nm}\;.
\end{align*}
%
Hence, Eq.~\eqref{Eq:washerCirc} covers a wide range of experimentally relevant SQUID loop parameters.

To better facilitate the comparison of our simulation results with the limiting case described by the Ketchen formula for $\phi_{\mu, {\rm K}, \circ}(a)$, given in Eq.~(\ref{Eq:KetchenCirc}), we rewrite Eq.~(\ref{Eq:washerCirc}) by introducing the dimensionless ratio $g\equiv A/a$ and the Pearl length $\Lambda\equiv \lambda_{\rm L}^2/d$ and normalize to $\phi_{\mu, {\rm K}, \circ}(a)$.
This results in
%
\begin{equation}
\label{Eq:washerCirc-norm}
\frac{\phi_{\mu, z, \circ} (g, a, \Lambda)}{\phi_{\mu, {\rm K}, \circ}(a)}  = 0.557 + \frac{0.443}{g} - f_1(g)\cdot f_2(a)\cdot f_3(\Lambda)
\end{equation}
%
with
%
\begin{eqnarray}
\label{Eq:washerCirc-norm2}
f_1(g)&=& \frac{(g-1)^{1.5}}{g^{1.35}} \nonumber\\
f_2(a)&=&\left(\frac{r_0}{a}\right)^{0.2}\\
f_3(\Lambda)&=&\frac{0.124}{1+0.103\,\frac{r_0}{\Lambda}}\;.\nonumber
\end{eqnarray}

\begin{figure}[t]
\includegraphics[width=\columnwidth]{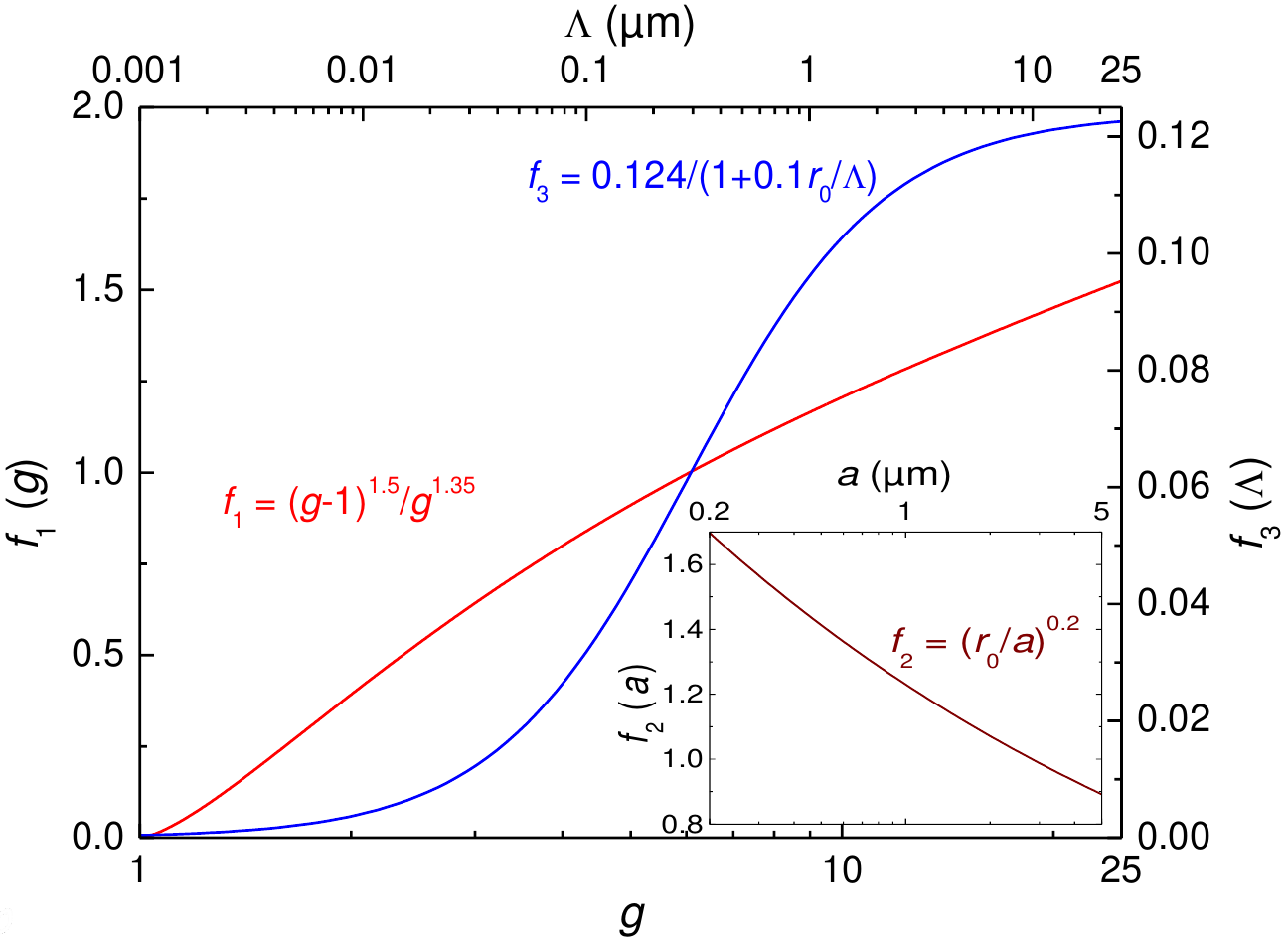}
\caption{Correction functions from Eq.~(\ref{Eq:washerCirc-norm2}) which enter into the expression for the coupling factor of a circuclar SQUID loop (magnetic moment at $z=0$ along $\hat{\bm e}_z$), given in Eq.~(\ref{Eq:washerCirc-norm}).}
\label{Fig:circwasher2}
\end{figure}

When approaching the limit $A \rightarrow a$ (i.e., $g \rightarrow 1$) of a SQUID loop with infinitely narrow linewidth, Eq.~\eqref{Eq:washerCirc-norm} reproduces the Ketchen expression~\eqref{Eq:KetchenCirc}.
Upon increasing $g$ (our simulations cover $1.02\le g \le 25$), the second term on the right-hand-side (r.h.s.) of Eq.~(\ref{Eq:washerCirc-norm}) becomes negligible, i.e., the coupling factor is reduced to $\sim 56\,\%$ of the value obtained from the Ketchen expression, if we neglect the third term in the r.h.s of Eq.~(\ref{Eq:washerCirc-norm}).
This third term, which can lead to a further reduction of the coupling factor, is the product of the functions $f_1(g)$, $f_2(a)$ and $f_3(\Lambda)$, as  defined in Eq.~(\ref{Eq:washerCirc-norm2}) and shown in Fig.~\ref{Fig:circwasher2}.
$f_1(g)$ vanishes for $g\rightarrow 1$; with increasing $g$ it increases monotonically and reaches $\sim 1.5$ for $g=25$.
$f_2(a)$ has a weak dependence on $a$, and monotonically decreases from $\sim 1.7$ for $a=0.2\,\mu$m to $\sim 0.9$ for $a=5\,\mu$m.
$f_3(\Lambda)$ increases monotonically with $\Lambda$, from $\sim 0$ for $\Lambda=1\,$nm to $\sim 0.12$ for $\Lambda=25\,\mu$m (the parameter range for $\Lambda$ covered in the simulations).
Hence, the third term in the r.h.s.~of Eq.~\eqref{Eq:washerCirc-norm} is negligible either for small $g$ approaching one, or for very small $\Lambda$, i.e., thick films.
It has a maximum value of $\sim 0.3$ for largest $g$ and $\Lambda$ and smallest $a$ covered by our simulations.
In this case, the coupling factor reduces to $\sim 25\,\%$ of the value obtained from the Ketchen expression.

\begin{figure}[t]
\includegraphics[width=\columnwidth]{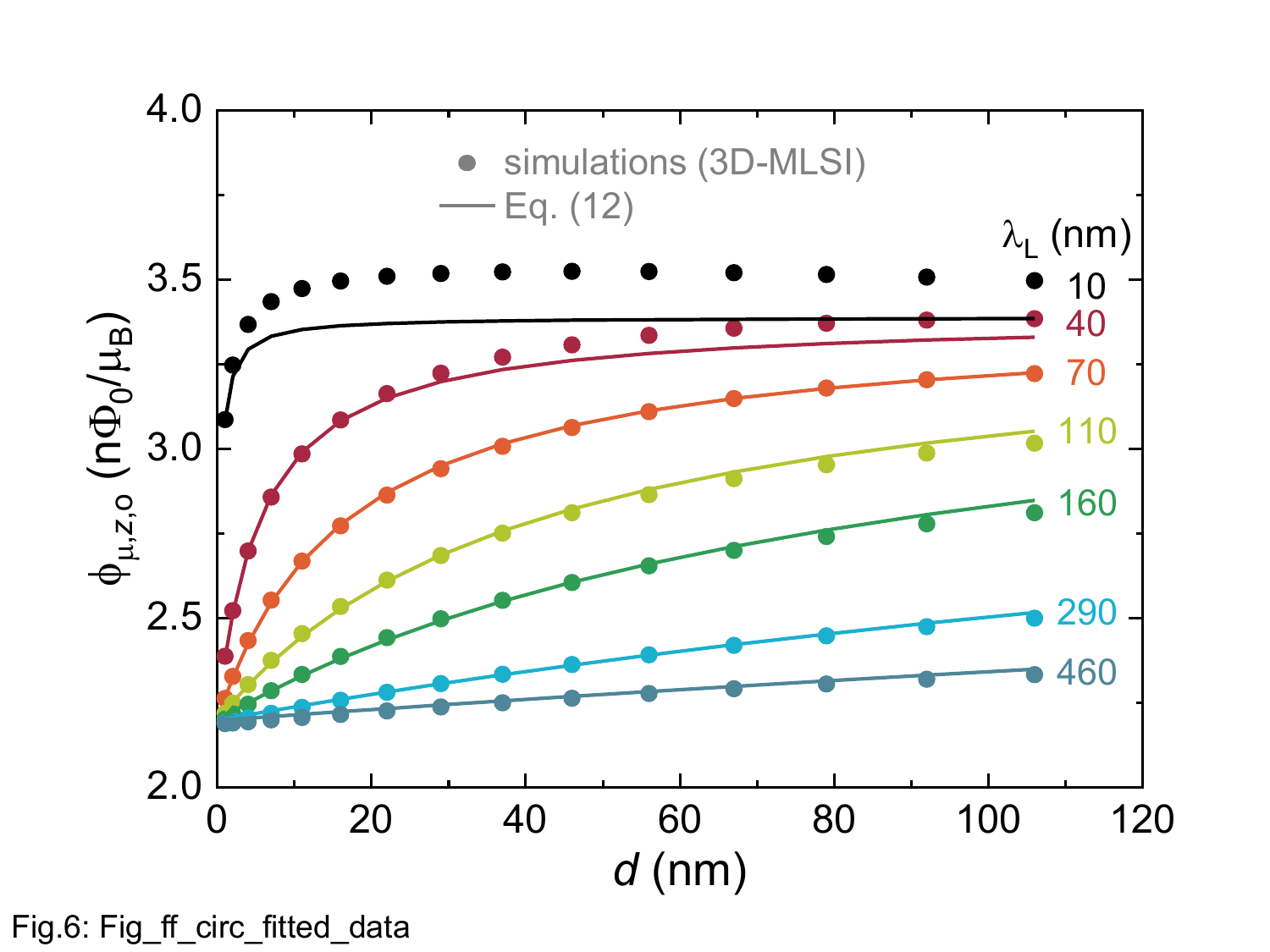}
\caption{3D-MLSI simulations (dots) and results from fit Eq.~(\ref{Eq:washerCirc}) (lines) for the coupling factor $\phi_{\mu,z,\circ}$ vs film thickness $d$ for a circular loop, with $A = 5\,\mu$m and $a = 0.5\,\mu$m for different $\lambda_\mathrm{L}$ (magnetic moment at $z=0$ along $\hat{\bm e}_z$).
Note that the vertical axis covers the range from 2 to 4\,n$\Phi_0/\mu_{\rm B}$.}
\label{Fig:CircleSquare}
\end{figure}

Figure~\ref{Fig:CircleSquare} shows a representative plot of the coupling factor $\Phi_{\mu,z,\circ}$ vs film thickness, obtained from 3D-MLSI  simulation (dots) together with the results obtained from Eq.~\eqref{Eq:washerCirc}  (solid lines) for a circular loop, with $A = 5\unit{\mu m}$ and $a = 0.5\unit{\mu m}$ for different $\lambda_\mathrm{L}$.
For the smallest value of $\lambda_\mathrm{L}=10\,$nm, Eq.~\eqref{Eq:washerCirc} underestimates the coupling factor by up to $\sim 4\,\%$.
However, upon increasing $\lambda_{\rm L}$ this deviation quickly decreases.
Figure \ref{Fig:CircleSquare} clearly shows that upon decreasing $d$ and increasing $\lambda_{\rm L}$ (i.e., increasing $\Lambda$), the coupling factor decreases.
This is a consequence of the scaling of $f_3(\Lambda)$ shown in Fig.~\ref{Fig:circwasher2}.
Additional coupling factors simulated with 3D-MLSI of circular washers with different size and comparison to Eq.~\eqref{Eq:washerCirc} as well as simulations of different sized square washers compared to a slightly adjusted equation can be found in the supplementary material (figures~\ref{suppl-Fig:square_A} and \ref{suppl-Fig:Square_a0_5_A5}).
Those data show a good agreement of the fit over the whole parameter range investigated.

So far, we have shown that by taking into account a finite width of the SQUID loop and a finite Pearl length, we find  that the simple Ketchen expressions [Eqs.~(\ref{Eq:KetchenCirc}) and (\ref{Eq:KetchenSquare})] overestimate the coupling factor, by up to a factor of $\approx 4$ for large $g$ and $\Lambda$ and small $a$.
Like in the comparison of simulations for variable $z$ with Eq.\eqref{Eq:Tilbrook}, a simple correction can be done by replacing in the Ketchen expressions the inner loop size $a$ by an effective loop size $a_{\rm eff}\equiv \sqrt{a\cdot A} = \sqrt{g}\cdot a$, to take into account the finite width of the loop.
This approach corresponds to taking into account the effective area of the SQUID $A_{\rm eff}=b\cdot A_{\rm h}\cdot\sqrt{A_{\rm w}/A_{\rm h}}$, where $A_{\rm w}$ is the outer loop area, $A_{\rm h}$ is the  area of the inner SQUID hole, and $b$ is a geometric factor close to one ($b=1/\Delta^2=8/\pi^2$ for circular and $\approx 1$ for  square loops) \cite{Ketchen85}.
This approach, yields the modified Ketchen expressions
%
\begin{equation}
\tilde{\phi}_{\mu, {\rm K}, i} = \frac{1}{\sqrt{g}}\,\cdot\phi_{\mu, {\rm K}, i}\quad(i=\circ, \square)\;.
\label{Eq:ketchen-mod}
\end{equation}

\begin{figure}[t]
\includegraphics[width=\columnwidth]{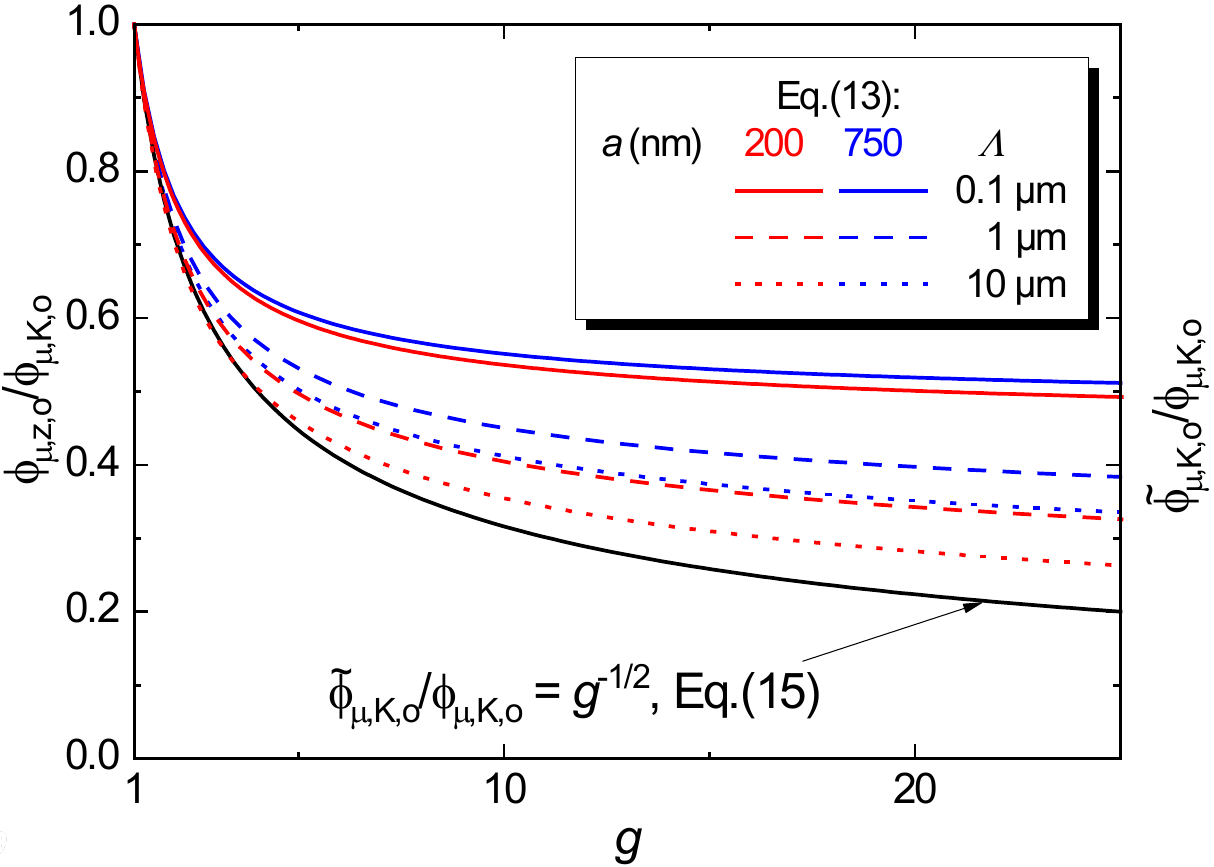}
\caption{Coupling factors vs $g$ for cicular loops (magnetic moment at $z=0$ along $\hat{\bm e}_z$): Comparison of functions calculated from Eq.~(\ref{Eq:washerCirc-norm}) for different $\Lambda$ and $a$ with the modified Ketchen expression Eq.~(\ref{Eq:ketchen-mod}) for a circular loop.}
\label{Fig:modK}
\end{figure}

For a circular loop, Fig.~\ref{Fig:modK} shows for comparison the modfied Ketchen expression $\tilde{\phi}_{\mu, {\rm K}, \circ}(g)$ [Eq.~(\ref{Eq:ketchen-mod})] together with the 3D-MLSI simulation results $\phi_{\mu, z, \circ}(g)$ [Eq.~(\ref{Eq:washerCirc-norm})], both normalized to $\phi_{\mu, {\rm K}, \circ}$.
In the latter case, we show three curves for different $\Lambda$ with fixed $a=750\,$nm (blue) and three curves with $a = 200\,$nm (red).
For $g$ close to one, all curves approach each other; however, for larger $g$, the modified Ketchen expression lies clearly below the $\phi_{\mu, z, \circ}(g)$ curves, and for the latter one clearly sees the dependence on $\Lambda$.
Obviously,  the $\phi_{\mu, z, \circ}(g)$ curves also depend on $a$.
For $a < 80\,$nm, in the limit $\Lambda \rightarrow \infty$, one can find $\phi_{\mu, z, \circ}(g)<\tilde{\phi}_{\mu, {\rm K}, \circ}$ for a wide range of $g$; however, these small values of $a$ are outside the range for which we obtain a $\lesssim 10\,\%$ deviation of the fit functions from our 3D-MLSI simulation results.
Altogether, these results demonstrate, that the simple approximations of the coupling factor by the Ketchen expressions -- even using the expressions with an effective loop size $a_{\rm{eff}}$ -- still  can significantly deviate from numerical simulations with 3D-MLSI.

\subsection{Near-field regime: coupling to SQUID arms}
\label{subsec:near-field}

\begin{figure}[t]
\includegraphics[width= \columnwidth]{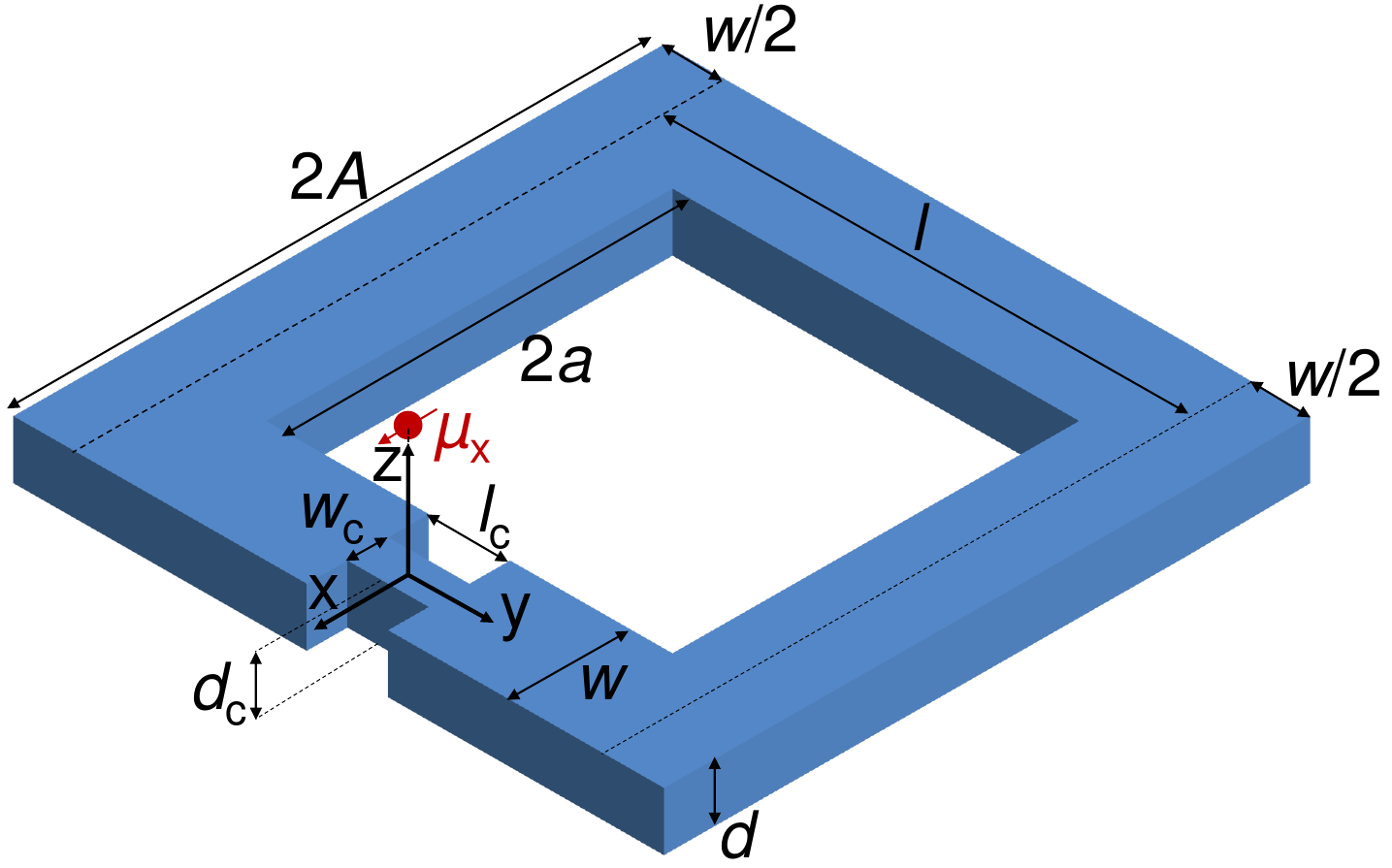}
\caption{Geometry of a square-shaped SQUID loop with inner size $2a$ and outer size $2A$ considered for the calculation of the coupling factor $\phi_\mu(x,y,z)$ in the near-field regime: 
A magnetic moment pointing in $x$-direction is placed at height $z$ above the symmetry point of the top surface of a SQUID arm extending along the $y$-direction.
This arm can contain a constriction (width $w_c$, length $l_c$ and thickness $d_c$).
The film thickness of the loop is $d$, and the width of the arm is $w = A-a$.
Simulations of this geometry are compared to simulations of single strip conductors of length $l = A+a$, which is also indicated.
}

\label{Fig:constriction}
\end{figure}

For the near-field regime, we consider the situation when an MNP is placed in close vicinity to one arm of a square shaped SQUID loop in the $(x,y)$-plane.
We start with showing that this situation can be reduced to only considering the coupling between the SQUID arm, represented by a cuboid shaped conductor (`single strip') and the MNP.
For this coupling, we derive for a homogeneous current density distribution an expression for the dependence of $\phi_\mu$ on the distance $z$ of a magnetic moment centered above a strip and on the width $w$, thickness $d$ and the length $l$ of the strip; 3D-MLSI simulations show that in the near-field regime $\lambda_{\rm L}$ has only a minor effect on the coupling factor.
We note that this approach can also be applied to the situation when a constriction is introduced in the SQUID arm, which leads to increased coupling, when a MNP ist placed close to the conctriction; this situation is indicated in Fig.~\ref{Fig:constriction}.

We consider a square shaped SQUID loop with inner dimension $2a$, outer dimension $2A$, film thickness $d$ and London penetration depth $\lambda_{\rm L}$.
We place the origin of the coordinate system on the upper surface of one SQUID arm, that extends along the $y$-axis.
A magnetic dipole is then placed at position $\bm r=(0,0,z)$, i.e., it is centered above the arm at distance $z$ from the upper surface.
We then assume that the moment of the magnetic dipole is oriented along the $x$-axis, i.e.~$\hat{\bm e}_\mu =\hat{\bm e}_{x}$ (see Fig.~\ref{Fig:constriction});  hence, only the $x$ component $B_{J,x}$ of $\bm{B}_J$ needs to be considered, i.e.~we calculate the coupling factor with Eq.~(\ref{Eq:phi-mu}) as $\phi_{\mu, x}=B_{J,x}/J$.

\begin{figure}[t]
\includegraphics[width=\columnwidth]{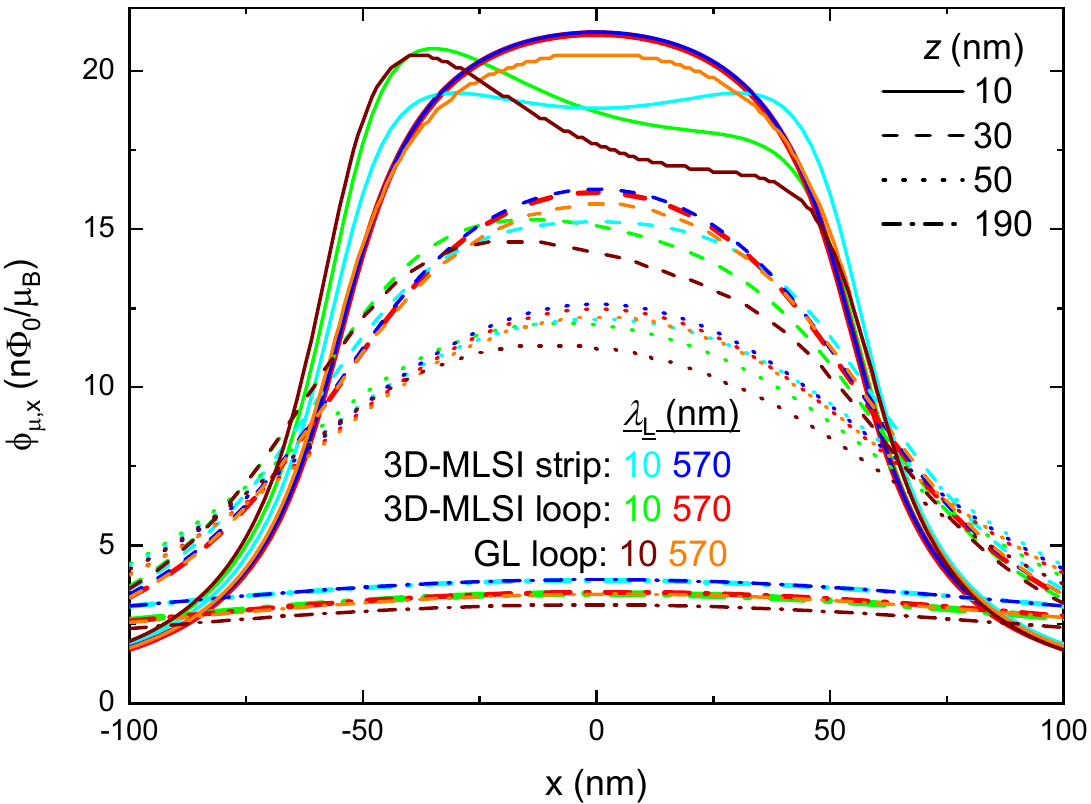}
\caption{Comparison of simulated coupling factors $\phi_{\mu,x}$ vs.~position $x$ of a magnetic moment above one arm of a full square loop ($2a= 480\,$nm, $2A = 700\,$nm) and above a single strip ($w = 110\,$nm, $l = 590\,$nm), both with $d = 10\,$nm. 
Full loops are simulated with 3D-MLSI as well as GL simulations with 1 $\Phi_0$ trapped in the loop.
Simulations of the single strip are done in 3D-MLSI, with current introduced via terminals at the short edges of the strip.
Coupling factors are calculated for $\lambda_{\rm L}= 10\,$nm and 570\,nm with four values of fixed height $z$.
The center of the arm of the full loop and of the single strip is at $x = 0$.}
\label{Fig:loop_vs_constriction}
\end{figure}

For sufficiently large SQUID loops or close proximity of the magnetic moment to the SQUID arm with small cross-section, the dominating part of the coupling results from the interaction of the magnetic moment with the arm only.
This is illustrated in Fig.~\ref{Fig:loop_vs_constriction}, where we compare calculated coupling factors across position $x$ above one arm of a full loop and above a single strip (with $w = A-a$ and $l=a+A$) for two values of $\lambda_{\rm L}$ and four positions $z$ above the upper surface of the structures.
Results on full loops are obtained from 3D-MLSI and GL simulations where 1 $\Phi_0$ is trapped in the loop.
Single strips are only simulated within 3D-MLSI, were the current is introduced via terminals at the short edges of the single strip.
All simulations were done with 3\,nm resolution, again with a triangular mesh and a square mesh for 3D-MLSI and GL simulations respectively.

For $\lambda_{\rm L} = 570\,$nm, which is significantly above the line width $w$, the current density distribution is homogeneous across the cross section of the single strip.
For most of the arm of the full loop this is also true, but towards the corners the current gets inhomogeneous (c.f.~Fig.~\ref{suppl-Fig:j_khapaev_full_loop_lambda_570} in the supplement.)	
Therefore for the smallest distance $z=10\,$nm, both coupling factors calculated with the full loop and the single strip simulated with 3D-MLSI coincide (red and blue solid lines in Fig.~\ref{Fig:loop_vs_constriction}), whereas for increasing $z$ small deviations are visible.
These deviations are due to reduced currents in $y$-direction at the corners of the loop, leading to reduced contributions from these regions, which gets more important with increasing $z$.
Maximum coupling in case of homogeneous current densities is achieved above the center of the arm of the loop/single strip ($x = 0$).
For the largest $z=190\,$nm (dashed-dotted lines in Fig.~\ref{Fig:loop_vs_constriction}) the two simulations (for small and large $\lambda_{\rm L}$) with the full loop (cyan and red) coincide and the simulations with only the single strip (green and blue) coincide.
This shows that for large enough distances of an MNP from the SQUID loop, inhomogeneities in the current distribution do not affect the coupling factor.

For the simulations with $\lambda_{\rm L}  = 10\,$nm, more deviations between the simulation of the full loop with 3D-MLSI and the single strip are visible, especially for the smallest $z$.
This is due to inhomogeneous current densities in these simulations which slightly differ for the two different geometries.
In the case of the single strip, the current density peaks at both edges, symmetrically around the center of the conductor.
This produces a $\phi_{\mu,x}(x)$ dependence (green curves) which is also symmetric about the center of the strip ($x = 0$).
The simulation with the full loop results in a current density with its maximum at the inner edge of the loop (at $x = -55\,$nm), which drops towards the outer edge (at $x = 55\,$nm).
This behavior is also reflected in the asymmetry of the resulting coupling factors (cyan curves), peaking close to the inner edge.
Nevertheless, at the center ($x = 0$) the coupling factors are almost the same.

GL simulations of full loops yield closely the same coupling factor (only slightly reduced values) as for full loop simulations by 3D-MLSI.
The current density distributions with identical $\lambda_{\rm L}$ are similar and can be found in the supplement Figs.~\ref{suppl-Fig:j_khapaev_full_loop_lambda_10} to \ref{suppl-Fig_j_GL_full_loop_lambda_570}.

In the remaining part of this section, we treat only the simpler system of coupling between a single strip (with its length $l$ along the $y$-axis) and a magnetic moment pointing along the $x$-axis, and we consider coupling factors obtained above the center of the conductor.
To start this analysis we investigate the situation $\lambda_{\rm L}>w,d$, where a homogeneous current density in the strip can be assumed.
For this case an analytical expression for the coupling factor for a dipole moment at position $\bm r$ can be found by utilizing Eq.~(\ref{Eq:phi-mu}) and the magnetic field 
$\bm B_{1D}(\bm r)$ generated by a current $I_{1D}$ running in $y$-direction through an infinitesimally thin and narrow conductor (1D-wire) of length $l$
%
\begin{equation}
\bm B_{1D}(\bm r)\!
= \!
\frac{\mu_0I_{1D}}{4\pi\rho}\!
\left(\sin\arctan\frac{\frac{l}{2}+y}{\rho}+\sin\arctan\frac{\frac{l}{2}-y}{\rho}\right)\!
\hat{\bm e}_\varphi .
\label{Eq:magnetic_field_1D_wire}
\end{equation}
%
Here, the origin is set to the midpoint of the 1D-wire along the $y$-axis.
$\rho$ gives the radial distance between the wire and the position of the magnetic moment, and $\hat{\bm e}_\varphi$ is the polar coordinate base vector.
For simplicity, we consider in the following only the case $y = 0$.

To calculate the magnetic field $B_{J,x}$ at $\bm r=(0,0,z)$, i.e., at the height $z$ above the symmetry point of the top surface of an extended strip (with finite $d$ and $w$), we integrate Eq.~(\ref{Eq:magnetic_field_1D_wire}) over the cross-section of the strip.
For this calculation we change the origin to the position of the magnetic moment, reversing the sign of Eq.~(\ref{Eq:magnetic_field_1D_wire}).
In this case $\rho = \sqrt{z'^2+x'^2}$ and the $x'$-component of $\hat{\bm e}_\varphi$  is $\cos \varphi = z'/\rho$ with $\varphi$ being the angle between the $z$-axis and the line connecting the MNP with the midpoint of the 1D wire.
With this, we can integrate Eq.~(\ref{Eq:magnetic_field_1D_wire}) over all possible distances of 1D-wires, filling the strip with cross-sectional area $wd$.
In the integral, we have to replace $I_{1D}$ by the current density $J/(wd)$; this yields

\begin{footnotesize}
\begin{align}
B_{J,x}(z,w,d,l) =&  \frac{\mu_0 J}{\pi w d} \int_{0}^{\frac{w}{2}} \int_{z}^{z+d} \frac{\cos\varphi}{\rho}\sin\arctan\left(\frac{l}{2\rho}\right) \rm d\rm z' \rm d\rm x' 		 \notag \\
=& \frac{\mu_0 J}{\pi w d}\Biggl[ \frac{l}{2}\ln\left(\frac{\left(\rho_z +\frac{w}{2}\right)\sqrt{\left(\frac{l}{2}\right)^2+\left(z+d\right)^2}}{\left(\rho_d+\frac{w}{2}\right)\sqrt{\left(\frac{l}{2}\right)^2+z^2}}\right)  \notag \\
&+\frac{w}{2}{\rm arcsinh}\left(\frac{l/2}{\sqrt{\left(\frac{w}{2}\right)^2+z^2}}\right) - z\arctan\left(\frac{lw}{4z\rho_z}\right) \notag \\
&-\frac{w}{2}{\rm arcsinh}\left(\frac{l/2}{\sqrt{\left(\frac{w}{2}\right)^2+\left(z+d\right)^2}}\right) \notag \\
&+ \left(z+d\right)\arctan\left(\frac{lw}{4\left(z+d\right)\rho_d}\right) \Biggr] \notag \\
=& \frac{\mu_0 J}{\pi w d}\, \Theta(z,w,d,l)   \qquad.
\label{Eq:magnetic_field_extended constriction}
\end{align}
\end{footnotesize}
%
Here, we introduce $\rho_z = \sqrt{\left(\frac{l}{2}\right)^2+\left(\frac{w}{2}\right)^2+z^2}$,  $\rho_d = \sqrt{\left(\frac{l}{2}\right)^2+\left(\frac{w}{2}\right)^2+\left(z+d\right)^2} $ and $\Theta(z,w,d,l)$ represents the expression in square brackets.
From this and Eq.~\eqref{Eq:phi-mu} we obtain the coupling factor
%
\begin{equation}
\phi_{\mu,x}(z,w,d,l)  = \frac{2\Theta(z,w,d,l) r_0}{\pi wd} \cpl \qquad,
\label{Eq:coupling_constriction}
\end{equation}
where again $r_0$ is introduced to provide the coupling factor in units of \cpl.

In the following we compare the analytical result given in Eq.~(\ref{Eq:coupling_constriction}) for a homogeneous current density distribution to simulations obtained with 3D-MLSI, where a broad range of parameters
\begin{align*}
z &= 10 \ldots 190\,\unit{nm} \\
w, d &= 10 \ldots 370\,\unit{nm} \\
l & = 200 \ldots 1000 \,\unit{nm} \\
\lambda_{\rm L} &= 10 \ldots 570 \,\unit{nm}\qquad ,
\end{align*}
was covered.
Here its important to again note that the 3D-MLSI results do not contain a $z$-dependence of the current density distribution, when the current sheets are parallel to the $(x,y)$-plane.
Therefore, simulations with current sheets in the $(x,y)$-plane only capture current density variations across the width $w$ and not across the thickness $d$ of the strip.
This situation is more appropriate for $\lambda_{\rm L}<w$.
To capture the effects of current density variations along the film thickness (in  case $\lambda_{\rm L}<d$), the coupling factor was also simulated with current sheets parallel to the $(y,z)$-plane.
In this second case, variations of the current density along the film thickness are captured, but those along the strip width are neglected.
Due to the smallest distance  $z = 10\,$nm between the upper surface of the strip and magnetic moment, the triangular mesh size and distance between current sheets has to be chosen sufficiently small.
For simulations with the current sheets in the $(x,y)$-plane, the triangular mesh size was chosen as 1\,nm and 2\,nm for simulations with $w < 100\,$nm and $w > 100\,$nm respectively, and 1\,nm distance in $z$-direction between current sheets was used for all simulations.
Simulations with current sheets in the $(y,z)$-plane were done with the same current sheet spacing along $x$, and mesh size was analogously chosen depending on $d$.
We note that the grid size in Fig.~\ref{Fig:loop_vs_constriction} was 3\,nm.
At $x=0$ and $z=10\,$nm, for the larger values of $\lambda_{\rm L}=570\,$nm (homogeneous current density distributions; c.f.~Figs.~\ref{suppl-Fig:j_khapaev_full_loop_lambda_570}, \ref{suppl-Fig_j_GL_full_loop_lambda_570} and \ref{suppl-Fig:j_khapaev_conductor}(b)), the 3D-MLSI simulations (both for single strip and loop) yield the same value as obtained from Eq.~(\ref{Eq:coupling_constriction}), as expected, while for the GL simulation the value lies slightly below.
This deviation might be caused by the different grid geometry used for GL vs 3D-MLSI simulations.

\begin{figure}[t]
\includegraphics[width = \columnwidth]{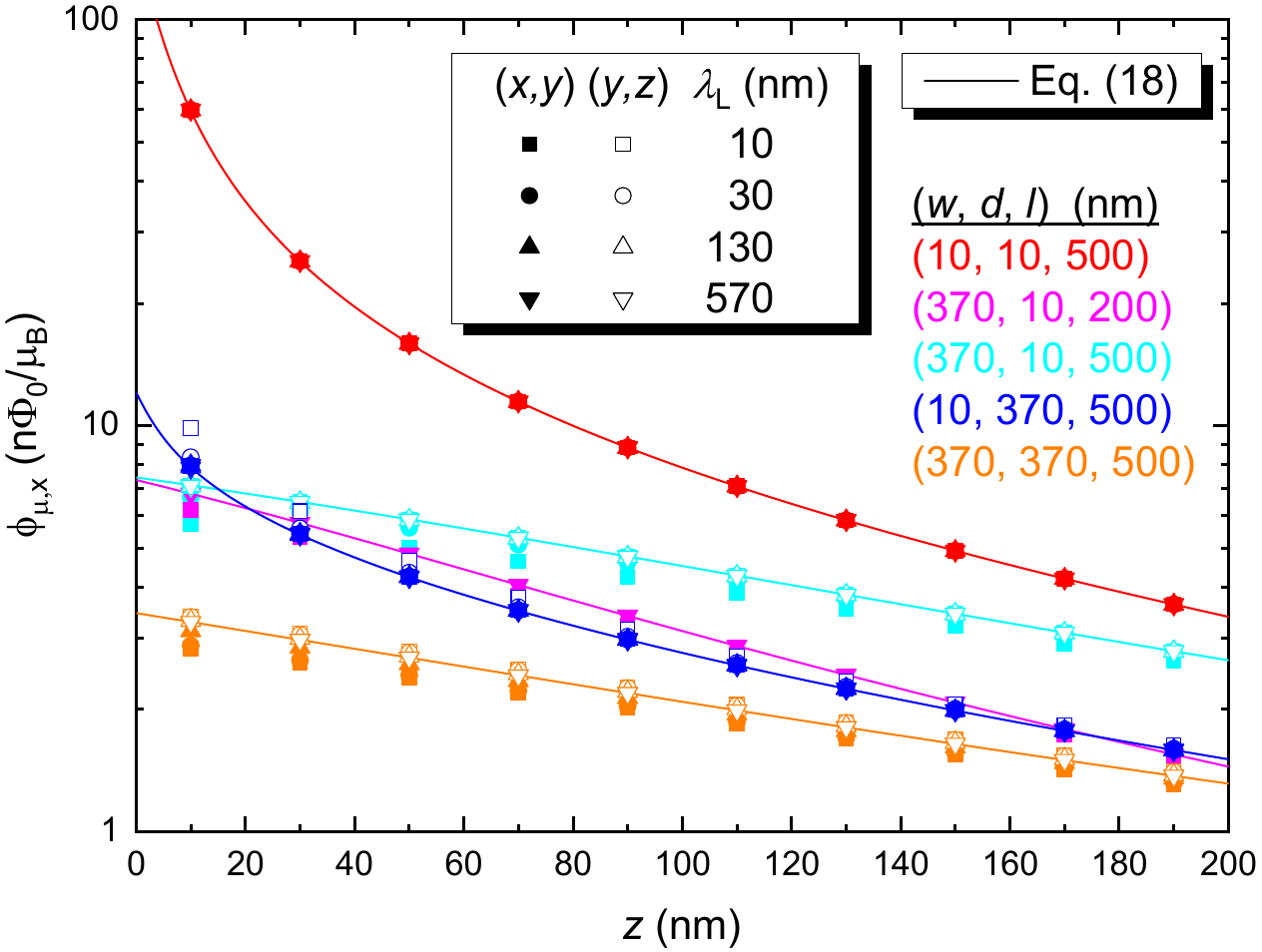}
\caption{Coupling factor vs distance from a single strip calculated with 3D-MLSI (symbols) and with Eq.~(\ref{Eq:coupling_constriction}) (lines).
Different colors correspond to different sets of geometrical parameters $w,d,l$.
Different symbols correspond to different values of $\lambda_{\rm L}$, with current sheets in the $(x,y)$-plane (solid symbols) and $(y,z)$-plane (open symbols).}
\label{Fig:near_field_eval}
\end{figure}

Some exemplary simulation results of the coupling factors $\phi_{\mu, x} (z,w,d,l,\lambda_{\rm L})$ (symbols) are shown in Fig.~\ref{Fig:near_field_eval}, and are compared to Eq.~(\ref{Eq:coupling_constriction}) (solid lines).
More precisely, we plot $\phi_{\mu, x}$ vs $z$ for five different sets of $w,d,l$ (represented by different colors) and -- in case of 3D-MLSI simulations -- each set for four different values of $\lambda_{\rm L}$ (represented by different symbols).
Solid symbols indicate current sheets in the $(x,y)$-plane and open symbols in the $(y,z)$-plane.

Clearly, the coupling factor increases with decreasing distance $z$ to the surface of the strip.
This increase is most pronounced for small $z$ when the cross section is smallest (red data).
Notably, by considering an extended conductor, the divergence at $z = 0$ of the coupling factor, which is observed in filamentary models, is not present.
Moreover, $\phi_{\mu,x}$ increases with decreasing cross-section of the strip (c.f.~red, cyan, blue and orange data for fixed $l$).
For the single strip with $d = w = 10\,$nm (red data) the simulation results are independent of $\lambda_{\rm L}$ (and current sheet orientation) and coincide with Eq.~(\ref{Eq:coupling_constriction}).
For small $\lambda_{\rm L}$, simulations with current sheets along the $(x,y)$-plane and large $w$ (pink, cyan and orange solid symbols) result in lower coupling compared to Eq.~(\ref{Eq:coupling_constriction}).
Simulations with small $\lambda_{\rm L}$ and current sheets along the $(y,z)$-plane and large $d$ (blue and orange open symbols) result in higher coupling compared to Eq.~(\ref{Eq:coupling_constriction}).
It can be seen that an inhomogeneous  current density distribution has most influence for small distances $z$.
Further, the length of the strip gets important for larger distances $z$ (compare pink and cyan).

The strongest deviations between Eq.~(\ref{Eq:coupling_constriction}) and simulation results can be seen for small $z$ and $\lambda_{\rm L}$ and large $w$ and $d$.
Understandably, the limiting case for $\lambda_{\rm L}>w,d$ is represented by Eq.~(\ref{Eq:coupling_constriction}), when the current density in the conductor approaches a homogeneous distribution.
When $\lambda_{\rm L}<w$ for simulations with current sheets along $(x,y)$ or $\lambda_{\rm L}<d$ with current sheets along $(y,z)$, the current is higher at the edges of the current sheet, symmetrically around the center of the strip.
In the first case this leads to the current being more distant to the point where the coupling factor is calculated compared to a homogeneous current density, leading to lower coupling factors in simulations compared to Eq.~(\ref{Eq:coupling_constriction}).
In the second case the symmetrically increased current density at the bottom and top surface of the single strip together with the inverse dependence on distance of the Biot-Savart law leads to higher coupling factors compared to Eq.~(\ref{Eq:coupling_constriction}).

Finally we want to look at the coupling factor above a SQUID arm oriented along the $y$-direction, containing a constriction at the center of the arm.
A magnetic dipole shall be placed above the center of the constriction, with its moment along the $x$-direction.
The parameters $w$, $d$ and $l$ still denote the width, thickness and total length of the SQUID arm.
The smaller dimensions of the constriction are given accordingly by $w_{\rm c}$, $d_{\rm c}$ and $l_{\rm c}$ (see Fig.~\ref{Fig:constriction}).
Eq.~(\ref{Eq:coupling_constriction}) gives the contribution of the constriction $\phi_{\mu, \rm c}(z) = \phi_{\mu, x}(z,w_{\rm c},d_{\rm c},l_{\rm c})$.
The contribution of the SQUID arm can be calculated as $\phi_{\mu, \rm arm} (z) = \phi_{\mu, x}(z,w,d,l) - \phi_{\mu, x}(z,w,d,l_{\rm c}) $.
So in total the coupling factor is $\phi_{\mu, {\rm tot}} = \phi_{\mu, \rm c} + \phi_{\mu, \rm arm} $.
%
\begin{figure}[t]
\includegraphics[width = \columnwidth]{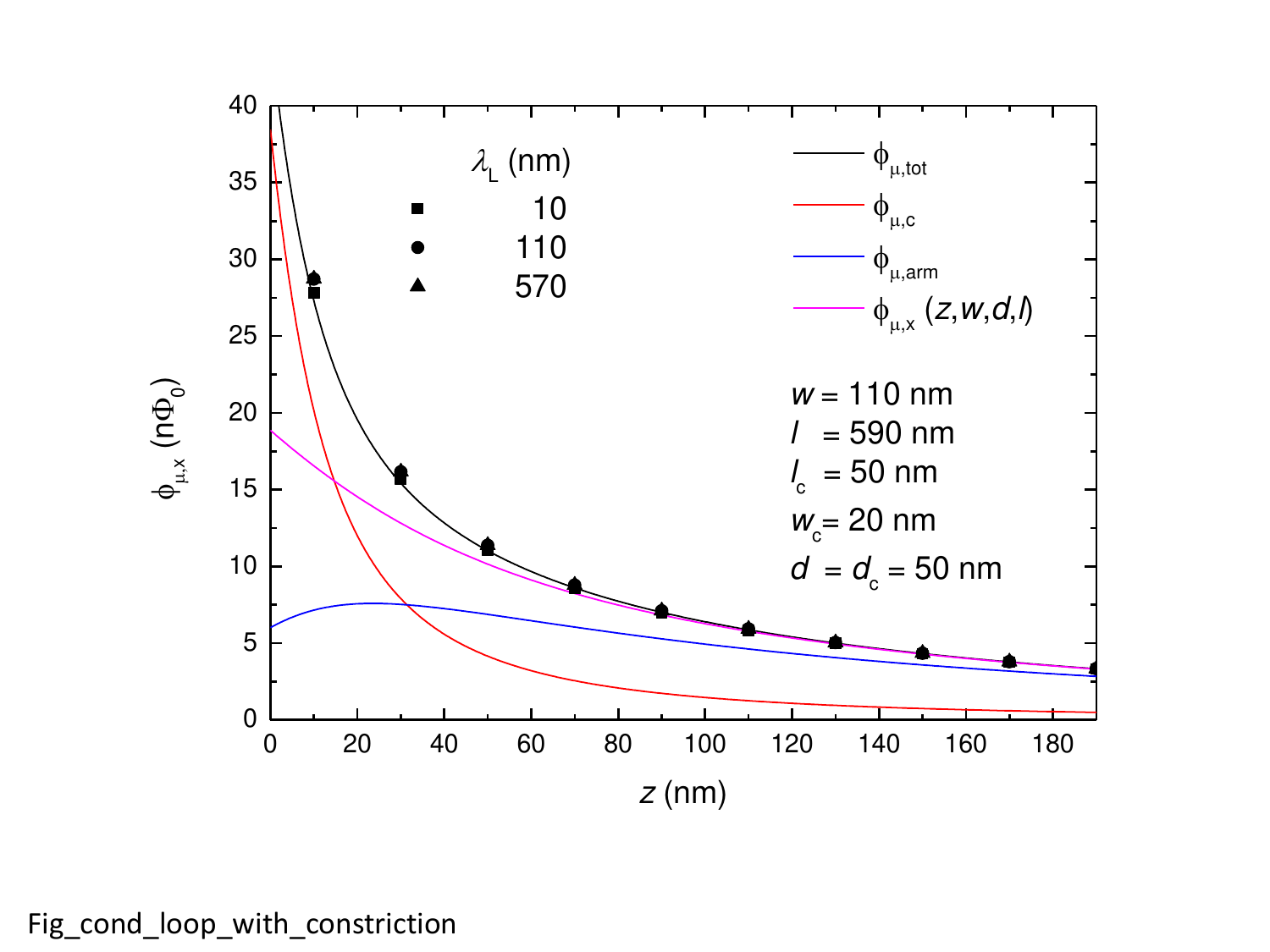}
\caption{Coupling factor of a magnetic moment placed above a SQUID arm (with $l = 590\,$nm, $w = 110\,$nm and $d = 50\,$nm) containing a constriction (with $l_{\rm c} = 50\,$nm, $w_{\rm c} = 20\,$nm and $d_{\rm c} = d$)  vs distance $z$.
Symbols show results from 3D-MLSI simulations with current sheet orientation along the $(x,y)$-plane.
Different symbols represent different values of $\lambda_{\rm L}$ in the simulations.
The simulation data are compared to the coupling factor $\phi_{\mu, {\rm tot}}$ (black line) calcluated from Eq.~(\ref{Eq:coupling_constriction}), consisting of the contribution of the constriction $\phi_{\mu, \rm c}$ (red line) and the contribution of the surrounding arms $\phi_{\mu, \rm arm}$ (blue line).
Also the coupling factor $\phi_{\mu, x} (z,w,d,l)$ of the full arm without a constriction (pink line) is shown.	
}
\label{Fig:strip_with_constriction}
\end{figure}

Figure \ref{Fig:strip_with_constriction} shows 3D-MLSI simulation results (symbols) with current sheet orientation along the $(x,y)$-plane of the coupling factor $\phi_{\mu,x}$ vs $z$ position between a magnetic dipole and a strip with $l = 590\,$nm, $w = 110\,$nm and $d = 50\,$nm, containing a constriction with $l_{\rm c} = 50\,$nm, $w_{\rm c} = 20\,$nm and $d_{\rm c} = d$.
Again, different symbols indicate different values for $\lambda_{\rm L}.$
The calculation with Eq.~(\ref{Eq:coupling_constriction}) of $\phi_{\mu, {\rm tot}}$ (black line) describes the simulation data well, with a slight deviation at smallest $z$.
This deviation is largest for largest $\lambda_{\rm L}$ and originates from the regions where the constriction connects to the wider arms.
There the current density adjusts to the change of width over a small length scale linked to $\lambda_{\rm L}$, not captured by our calculations (Eq.~(\ref{Eq:coupling_constriction})).
For small values of $\lambda_{\rm L}$ the length over which the current density adjusts is smaller.
The simulated current densities can be found in the supplement (Fig.~\ref{suppl-Fig:j_khapaev_conductor_with_constriction}).
For small $z$ the main contribution of the total coupling factor results from the constriction (red line) and for big $z$ from the surrounding arms (blue line).
From the comparison of the arm with and without a constriction (black and pink line) for small $z$, a significant increase in coupling due to the introduction of the constriction is visible.

\section{Conclusion}
In this paper we investigated the coupling factor $\phi_\mu$, that quantifies the coupling of a magnetic dipole to a SQUID, for various SQUID geometries by means of numerical simulations based on London theory (3D-MLSI) and Ginzburg Landau (GL) theory.
We investigated the far field regime, where the distance $r$ between magnetic dipole and SQUID is comparable to or larger than the inner size $a$ of the SQUID loop and the near field regime, where the magnetic dipole is in close proximity to the SQUID.

In the far field regime, we reviewed formulas for the calculation of the coupling factor for a magnetic dipole placed on the symmetry axis of circular and square shaped SQUID loops, obtained from filamentary loop models.
We investigated the applicability of these formulas on spatially extended SQUID loops with outer size $A$, film thickness $d$ and under consideration of the London penetration depth $\lambda_{\rm L}$ and the Ginburg-Landau coherence length $\xi$, by simulating the coupling factor of various circular and square shaped loops.
For loops with small film thickness $d \ll a$ and small line width $A-a < a$, the introduction of an effective loop size $a_{\rm eff}$ (connected to the effective area $A_{\rm eff}$ of the SQUID) in the filamentary loop model formulas results in good agreement with simulations.

Furthermore, the coupling factor for circular and square loops was investigated for magnetic dipoles placed at the center of the loop, where the coupling factor is largest for a given geometry in the far field regime.
From 3D-MLSI simulations we derive a fit formula for the coupling factor $\phi_\mu(a,A,d,\lambda_{\rm L})$ that covers a wide parameter range of the inner and outer size of the loop, film thickness and London penetration depth.

In the near field regime we investigated the coupling factor for a magnetic dipole centered at a distance $z$ above the upper surface of one SQUID arm.
3D-MLSI and GL simulations show that in this case, it is sufficient to consider the coupling between a single strip conductor representing the SQUID arm and the magnetic dipole, to reduce the computational effort.
The coupling between a single strip and a magnetic dipole centered above the strip at distance $z$ was then calculated analytically for a homogeneous current density distribution across the strip with variable film thickness $d$, width $w$ and length $l$.
The obtained expression showed an excellent agreement with coupling factors calculated numerically with 3D-MLSI, for single strips of various geometries and variable London penetration depth for $\lambda_{\rm L}> w,d$.
When $\lambda_{\rm L}<w$ or $\lambda_{\rm L}<d$ the analytical calculation slightly overestimates or underestimates, respectively, the results of the simulations for the investigated parameter range.

Finally, we investigated the coupling factor for a magnetic dipole centered at height $z$ above a constriction introduced to the center of the SQUID arm.
Here, the simulated coupling factor can be well described by calculating the superposition of the coupling factor of the constriction and the surrounding SQUID arm.
For small distances $z$ the coupling factor increases significantly by the introduction of the constriction into the SQUID arm, while for larger distances the coupling factor is determined by the SQUID arm geometry.

Altogether, our studies show that, by taking into account the finite cross-section of a SQUID loop and finite $\lambda_{\rm L}$, results for the coupling factor (and accordingly the spin sensitivity) can significantly deviate from simple approximations based on filamentary loop structures.
Still one can avoid time-consuming numerical simulations of the coupling factor by using approximate results that have been derived here for various situations and for ranges of geometric parameters that are relevant for practical applications.
On the other hand, we demonstrate that numerical simulations based on Ginzburg-Landau theory provide results that are consistent with our 3D-MLSI simulations that have been frequently used in the past for calculating the coupling factor. 
This is important if one needs to calculate more complex three-dimensional structures (e.g., vector nanoSQUIDs \cite{Martinez-Perez16}), that cannot be treated anymore by 3D-MLSI.
The results presented here may be useful for providing good estimates of signals that are produced upon magnetization reversal of magnetic nanoparticles that are coupled to nanoSQUIDs.
Moreover, these results could also be helpful for the estimation of the spatial resolution of scanning SQUID microscopy, in the case of imaging dipole-like sources.

\acknowledgments
B.~ M\"{u}ller acknowledges funding by the German Academic Scholarship Foundation.
This work was supported by the Deutsche Forschungsgemeinschaft (DFG) (KO 1303/13-2),  by the European Commission under H2020 FET Open grant ‘FIBsuperProbes’ (number 892427) and by the COST actions NANOCOHYBRI (CA16218) and SUPERQUMAP (CA21144).

\bibliography{CouplingSQUIDs_v15.bib}
\makeatletter\@input{yy.tex}\makeatother
\end{document}


\title{Supplementary material: On the coupling of magnetic moments to superconducting quantum interference devices}

\author{J.~Linek}
\affiliation{%
	Physikalisches Institut, Center for Quantum Science (CQ) and LISA$^+$,
	University of T\"ubingen,
	Auf der Morgenstelle 14,
	72076 T\"ubingen, Germany}

\author{M.~Wyszynski}
\affiliation{%
	Department of Physics \& NANOlab Center of Excellence, 
	University of Antwerp, 
	Groenenborgerlaan 171, 
	B-2020 Antwerp, Belgium}

\author{B.~M\"uller}
\affiliation{%
	Physikalisches Institut, Center for Quantum Science (CQ) and LISA$^+$,
	University of T\"ubingen,
	Auf der Morgenstelle 14,
	72076 T\"ubingen, Germany}

\author{D.~Korinski}
\affiliation{%
	Physikalisches Institut, Center for Quantum Science (CQ) and LISA$^+$,
	University of T\"ubingen,
	Auf der Morgenstelle 14,
	72076 T\"ubingen, Germany}


\author{M.~V.~Milo\v{s}evi\'c}
\affiliation{%
	Department of Physics \& NANOlab Center of Excellence, 
	University of Antwerp, 
	Groenenborgerlaan 171, 
	B-2020 Antwerp, Belgium}

\author{R.~Kleiner}
\affiliation{%
	Physikalisches Institut, Center for Quantum Science (CQ) and LISA$^+$,
	University of T\"ubingen,
	Auf der Morgenstelle 14,
	72076 T\"ubingen, Germany}

\author{D.~Koelle}
\email{koelle@uni-tuebingen.de}
\affiliation{%
	Physikalisches Institut, Center for Quantum Science (CQ) and LISA$^+$,
	University of T\"ubingen,
	Auf der Morgenstelle 14,
	72076 T\"ubingen, Germany}

\date{\today}




\maketitle

In this supplementary material, we present in section \ref{sec:far-field} additional simulation results on the coupling factor in the far-field regime for circular and square shaped loops.
Section \ref{sec:near-field} includes additional simulation results of coupling factors for the near-field regime.
Section \ref{sec:current-density} provides a comparison of current density distributions simulated with 3D-MLSI, based on London theory, and Ginzburg-Landau (GL) simulations.

\section{Far field regime}
\label{sec:far-field}

In the far-field regime we consider the coupling for circular and square shaped SQUID loops in the $(x, y)$-plane, to a magnetic dipole moment oriented along $\hat{\bm e}_z$ and placed on the $z$ symmetry axis.
In this case the origin of the coordinate system is placed in the center of the SQUID loop. 

\subsection{Additional simulation results on circular loops}
\label{subsec:far-field-circular}

\begin{figure}
\includegraphics[width = \linewidth]{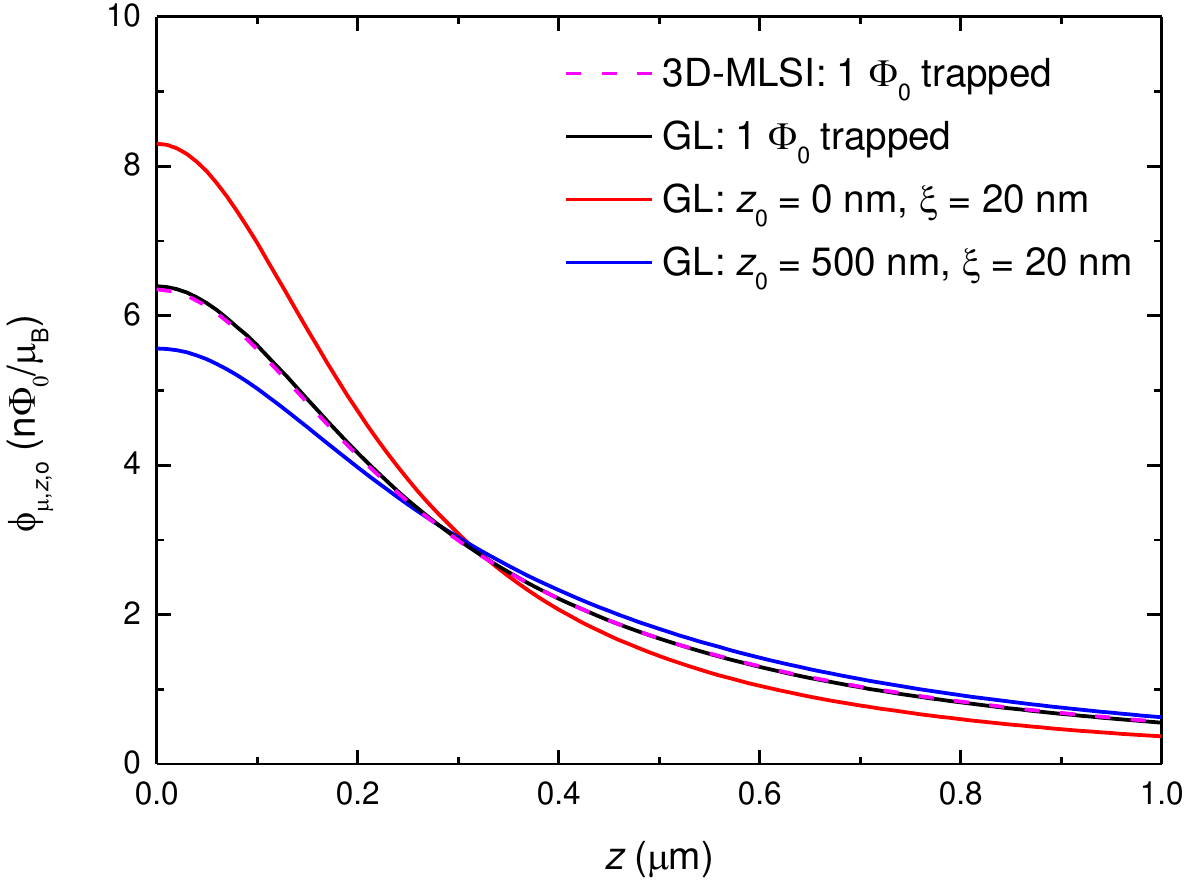}
\caption{Comparison of simulated coupling factors by 3D-MLSI and GL simulations of a circular loop with $a = \qty{200}{\nano\meter}$, $A = \qty{1.5}{\micro\meter}$, $d = \qty{50}{\nano\meter}$ and $\lambda_{\rm L} = \qty{200}{\nano\meter}$ ($\xi = \qty{20}{\nano\meter}$ in GL simulations).
The current in the 3D-MLSI simulation (pink) corresponds to $1\Phi_0$ in the loop.
In case of GL different current densities were used for the determination of the coupling factor.
Also the situation corresponding to $1\Phi_0$ in the loop was simulated (black).
Further the current was induced by screening the magnetic field of a magnetic dipole ($\mu\approx35.5 \cdot 10^6 \mu_{\rm B}$) at $z_0 = \qty{0}{\nano\meter}$ (red) and $z_0 = \qty{500}{\nano\meter}$ (blue).
}
\label{Fig:khapaev_GL_big_loop}
\end{figure}

Figure \ref{Fig:khapaev_GL_big_loop} shows a comparison of GL and 3D-MLSI simulation results of the coupling factor $\phi_{\mu,z}$ vs position $z$ along the symmetriy axis, similar to Fig.~\ref{main-Fig_comparison_GL_London} of the main manuscript, but here for a circular loop with much larger outer radius $A = \qty{1.5}{\micro\meter}$.
The other parameters are kept the same with $a = \qty{200}{\nano\meter}$, $d = \qty{50}{\nano\meter}$ and $\lambda_{\rm L} = \qty{200}{\nano\meter}$.
A sketch of the geometry can be found in the main manuscript (Fig.~\ref{main-Fig:circwasher}).
Here, coupling factors are calculated for different methods of inducing currents.
The resulting current density distributions can be seen in Figs.~\ref{Fig:j_khapaev_lambda_200_d_50_200_1500_round} to \ref{Fig:j_GL_500_xi_20_lambda_200_d_50_200_1500_round}.
The coupling factor calculated by 3D-MLSI, with a flux quantum trapped in the loop (pink dashed line), coincides with the coupling obtained from GL simulations with a trapped flux quantum (black solid line).
In contrast, the coupling factor obtained from GL simulations with screening currents induced by a magnetic moment placed at $z_0 = 0$ (red line) shows deviations.
For small $z$ and especially for $z =0$, where this simulation strictly is only valid, the coupling factor is higher.
This is due the current being more concentrated at the inner circumference in this situation.
Also, inducing screening currents by a magnetic moment at $z_0 = \qty{500}{\nano\meter}$ (blue line) lead to a slightly higher coupling factor at $z = \qty{500}{\nano\meter}$.

Comparing Fig. \ref{Fig:khapaev_GL_big_loop} to Fig.~\ref{main-Fig_comparison_GL_London} of the main text, it is obvious that considering the correct current density distribution becomes more important for larger line widths.
Our evaluation of coupling factors at $z = 0$, calculated with a flux quantum in the loop gives a lower bound for loops with large line widths.
%

\begin{figure*}
\includegraphics[width = \linewidth]{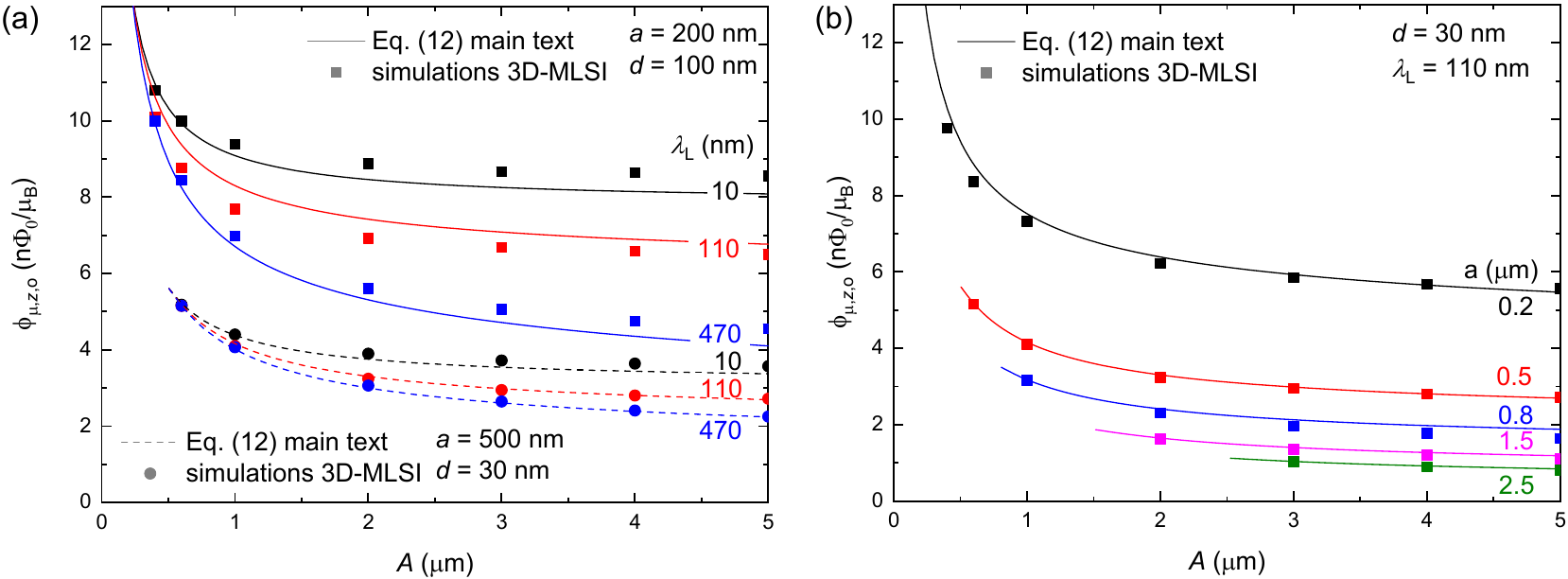}
\caption{Coupling factors vs outer loop radius for circular loops, simulated with 3D-MLSI (symbols) and calculated with Eq.~\eqref{main-Eq:washerCirc} (lines) of the main manuscript:
(a) for three different values of $\lambda_{\rm L}$ indicated by different colors, and two sets of loop geometry parameters $a = \qty{200}{\nano\meter}$, $d = \qty{100}{\nano\meter}$ (solid lines and squares) and $a = \qty{500}{\nano\meter}$, $d = \qty{30}{\nano\meter}$ (dashed lines and circles).
(b) for five different values of $a$ indicated by different colors and  $d = \qty{30}{\nano\meter}$ and $\lambda_{\rm L} = \qty{110}{\nano\meter}$.}
\label{Fig:circ_A}
\end{figure*}

Figure~\ref{Fig:circ_A} shows additional simulated coupling factors vs outer radius $A$ (at $z=0$) obtained by 3D-MLSI for circular loops with variable parameters $a$, $d$ and $\lambda_{\rm L}$, covering the investigated parameter range.
The simulation results are compared to Eq.~\eqref{main-Eq:washerCirc} given in the main manuscript.
It is evident that Eq.~\eqref{main-Eq:washerCirc} describes the simulation results well, over a wide range of parameters.
Largest deviations between simulation results and Eq.~\eqref{main-Eq:washerCirc} can be seen for the thicker film with smallest $\lambda_{\rm L}$ in Fig.~\ref{Fig:circ_A}(a).
Figure \ref{Fig:circ_A}(b) shows that for a typical parameter set $d = \qty{30}{\nano\meter}$ and $\lambda_{\rm L} = \qty{110}{\nano\meter}$ we find very good agreement between simulation results and Eq.~\eqref{main-Eq:washerCirc} for a wide range of inner loop sizes $a$ from 0.2 to 2.5\,$\mu$m.
Overall the fit formula Eq.~\eqref{main-Eq:washerCirc} from the main manuscript fits the simulation data with an accuracy of $\approx 10\%$.

\subsection{Simulation results of square shaped loops}
\label{subsec:far-field-square}

\begin{figure}[b]
\includegraphics[width = \linewidth]{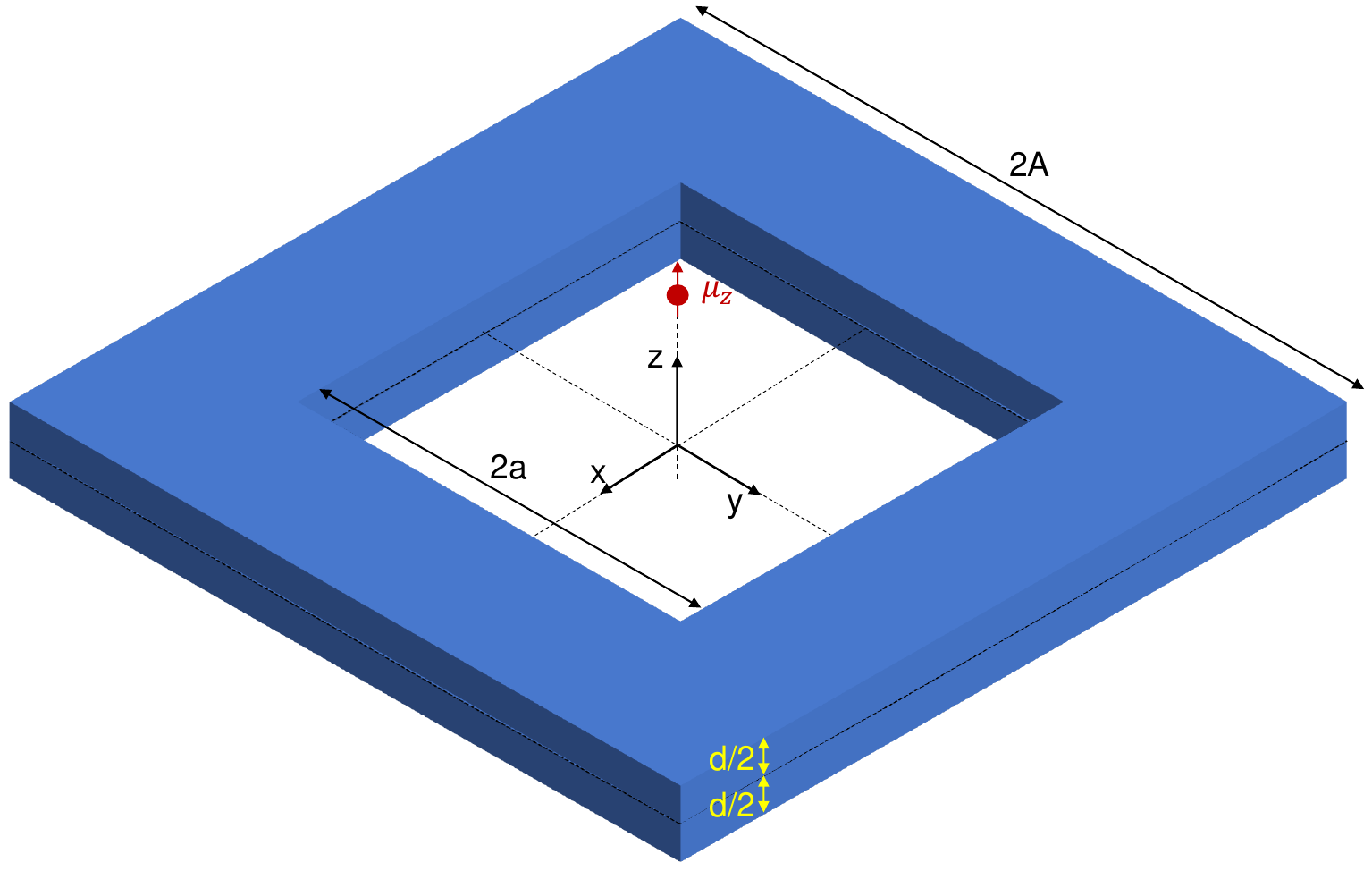}
\caption{Far-field regime: square SQUID loop (inner length $2a$, outer length $2A$, film thickness $d$) with the magnetic moment $\bm\mu$ of a MNP on the $z$-axis above the origin pointing along $\hat{\bm e}_z$.
The origin is centered in the loop.}
\label{Fig:sketch_far_field_square}
\end{figure}

Here we consider the coupling factor for square shaped loops (with inner length $2a$, outer length $2A$, film thickness $d$) in the far field regime.
A sketch of the geometry is depicted in Fig.~\ref{Fig:sketch_far_field_square}.

\begin{figure}[b]
\includegraphics[width=\linewidth]{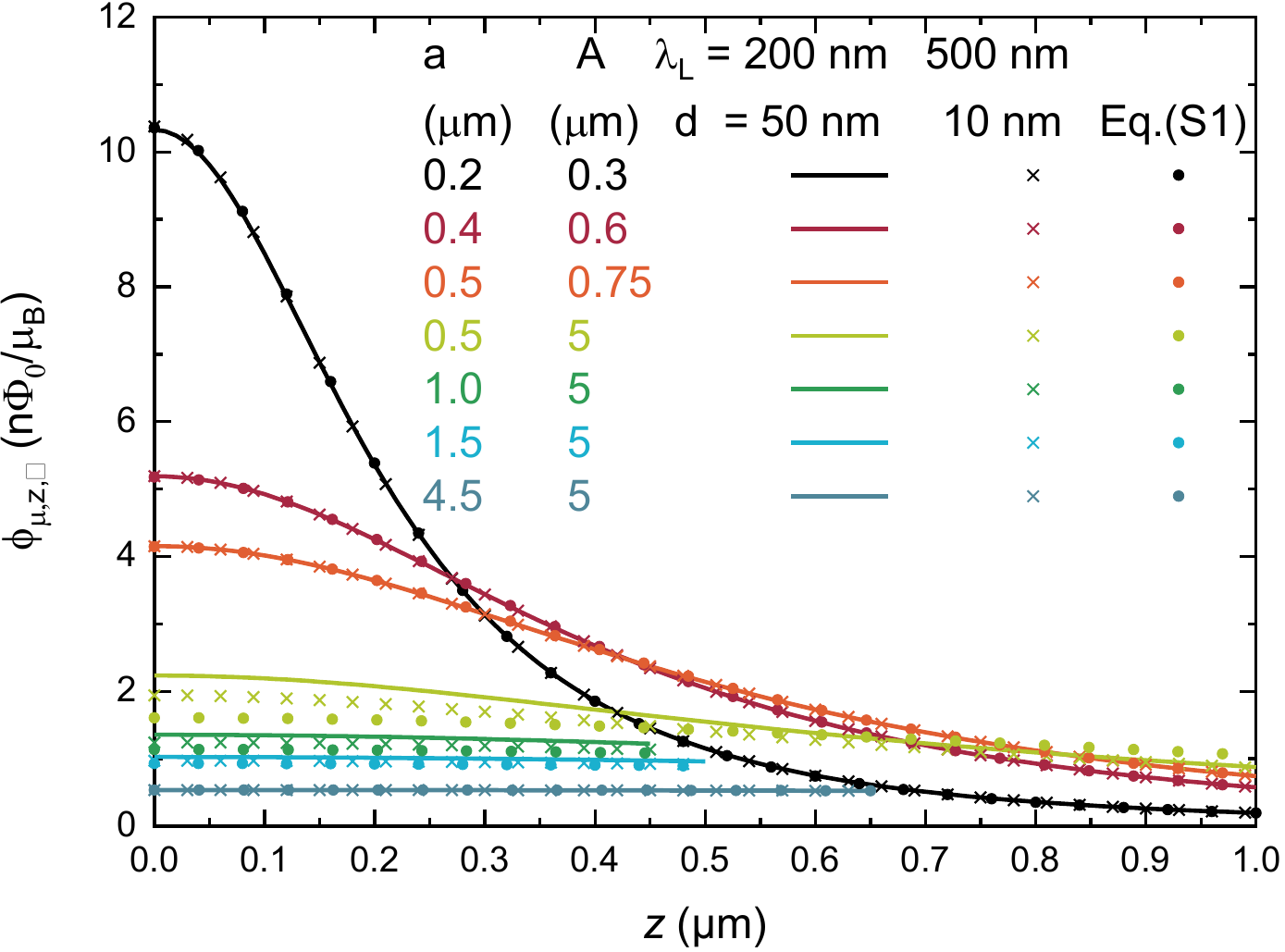}
\caption{Simulated coupling factor $\phi_{\mu,z,\square}$ vs $z$ position of a MNP placed on the $z$ symmetry axis of a square shaped SQUID loop, for different values of the inner length $2a$ and outer length $2A$.
Simulations are shown for two sets of $\lambda_{\rm L}$ and $d$ and are compared to Eq.~\eqref{Eq:Square_Loop}, where $a$ is replaced by an effective length $a_{\rm eff}=\sqrt{a\cdot A}$.}
\label{fig:square_washer_z_dependence}
\end{figure}

\begin{figure*}[t]
\includegraphics[width = \linewidth]{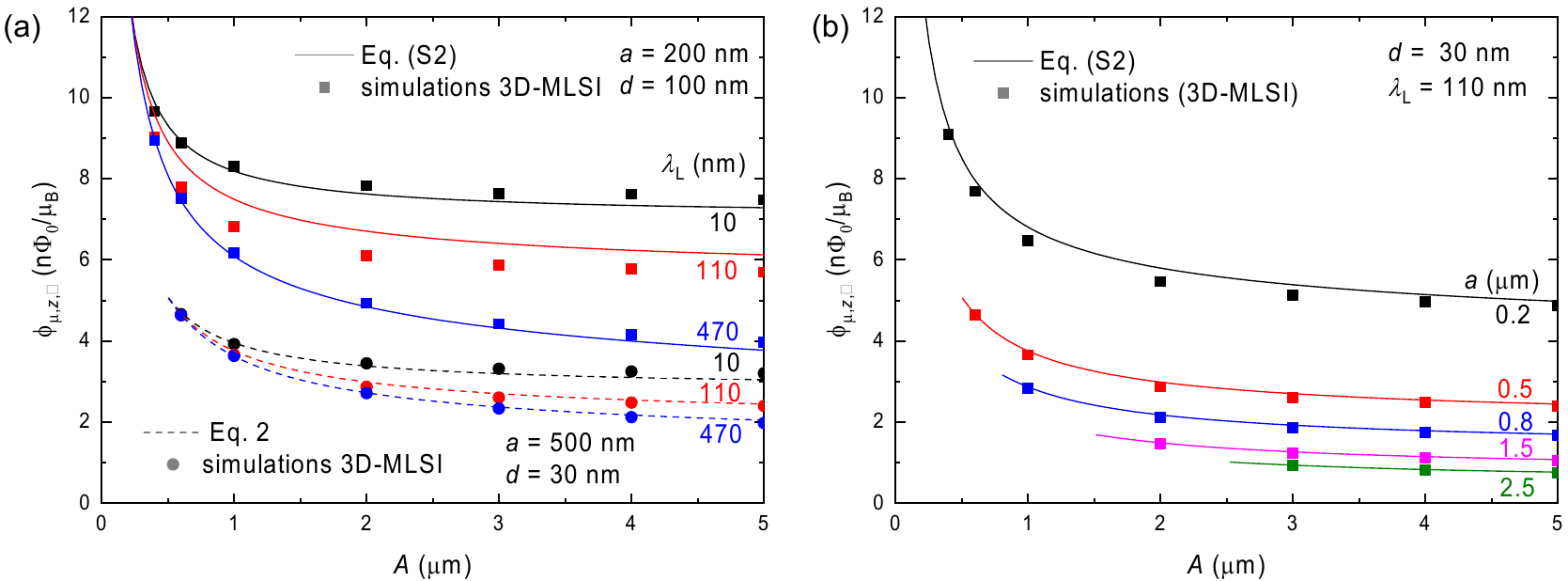}
\caption{Coupling factors vs outer loop size for square loops, simulated with 3D-MLSI (symbols) and calculated with Eq.~\eqref{Eq:washerSquare} (lines).
(a) for three different values of $\lambda_{\rm L}$ indicated by different colors, and two sets of loop geometry parameters $a = \qty{200}{\nano\meter}$, $d = \qty{100}{\nano\meter}$ (solid lines and squares) and $a = \qty{500}{\nano\meter}$, $d = \qty{30}{\nano\meter}$ (dashed lines and circles).
(b) for five different values of $a$ indicated by different colors and  $d = \qty{30}{\nano\meter}$ and $\lambda_{\rm L} = \qty{110}{\nano\meter}$.}
\label{Fig:square_A}
\end{figure*}

As stated in the main manuscript Eq.(\ref{main-Eq:Square_Loop}), the $z$-dependence of an infinitesimally thin and narrrow square loop with loop size $2a$ can be calculated as
%
\begin{equation}
\phi_{\mu, \square}(a,z) = \frac{r_0}{\Delta\cdot a}\frac{1}{(1+\frac{z^2}{a^2})^{3/2}}\sqrt{\frac{2(z^2+a^2)}{z^2+2a^2}}\,\cpl\;.
\label{Eq:Square_Loop}
\end{equation}
%
Figure \ref{fig:square_washer_z_dependence} compares simulations of the coupling factor vs $z$ position for square loops with inner dimension $2a$ and outer dimension $2A$ to Eq.~\eqref{Eq:Square_Loop}, where the length $a$ has been been replaced by an effective length $a_{\rm eff}=\sqrt{a\cdot A}$. 
The data yield a similar behavior as shown in Fig.~\ref{main-Fig:washer-z-dep} in the main manuscript for circular loops, but with slightly reduced coupling factors.

For $z = 0$ the coupling is strongest. To describe the simulation data at $z = 0$ for square loops, we use a fit formula which is analogous to the fit formula for circular loops (Eq~\ref{main-Eq:washerCirc} in the main manuscript) 
%
\begin{eqnarray}
\label{Eq:washerSquare}
\frac{\phi_{\mu, z, \square} (A', a', d', \lambda'_\mathrm{L})}{\cpl} &=& \frac{1.57}{a' \cdot \Delta} + \frac{1.247}{A' \cdot \Delta}\\
&-& \frac{([A'-a']\cdot \Delta)^{1.5}}{(A' \cdot a' \cdot \Delta^2)^{1.35}}\cdot\frac{0.43}{1 + \frac{0.29 d'}{\Lambda'_\mathrm{L}}}\;,\nonumber
\end{eqnarray}
%
where $\Delta = \pi/2\sqrt{2}$ and variables with primes are given in \unit{\micro\meter}.
Equation~\eqref{Eq:washerSquare} matches the 3D-MLSI results with an error $\lesssim 10\,\%$ for the parameter range
\begin{align*}
a &= 0.2 \ldots 5\,\unit{\mu m} \\
A &= (a + 0.1\unit{\mu m}) \ldots 5\,\unit{\mu m} \\
d &= 10 \ldots 100\,\unit{nm} \\
\lambda_\mathrm{L} &= 10 \ldots 500\,\unit{nm}\;.
\end{align*}
By replacing $A'$ by $ga'$ (with $g = A/a$), one finds the same expression for $\phi_{\mu, z, \square}(g,a,\Lambda)/\phi_{\mu, K, \square}(a)$ as in Eq.~\eqref{main-Eq:washerCirc-norm} of the main manuscript, i.e.
%
\begin{equation}
	\label{Eq:washerSquare-norm}
	\frac{\phi_{\mu, z, \square} (g, a, \Lambda)}{\phi_{\mu, {\rm K}, \square}(a)}  = 0.557 + \frac{0.443}{g} - f_1(g)\cdot f^*_2(a)\cdot f_3(\Lambda)\;;
\end{equation}
%
however, with the modified function
%
\begin{equation}
	\label{Eq:washerSquare-norm2}
	f^*_2(a) = \left(\frac{r_0}{\Delta\cdot a}\right)^{0.2}\;.
\end{equation}

Figure \ref{Fig:square_A} shows simulated coupling factors vs outer loop size $A$ (at  $z=0$) obtained by 3D-MLSI (symbols) for square loops with variable parameters $a$, $d$ and $\lambda_{\rm L}$, covering the investigated parameter range.
The simulation results are compared to Eq.~\eqref{Eq:washerSquare} (lines).
For the wide parameter range shown, the maximum deviation is below 10\%.

\begin{figure}[h]
\includegraphics[width = \linewidth]{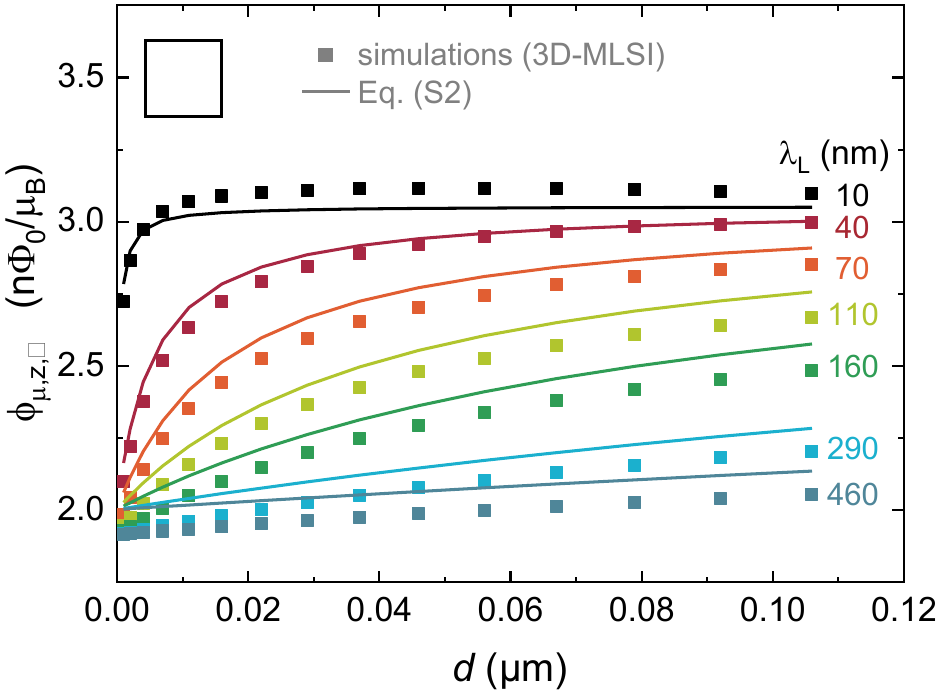}
\caption{Simulated (squares) and calculated (lines) coupling factors $\phi_{\mu, z}$ vs film thickness $d$ for a square loops, with $A = \qty{5}{\micro\meter}$ and $a = \qty{500}{\nano\meter}$ for different $\lambda_\mathrm{L}$.}
\label{Fig:Square_a0_5_A5}
\end{figure}

In Fig.~\ref{Fig:Square_a0_5_A5} simulated data of a square loop with $a = \SI{0.5}{\micro\meter}$ and $A = \SI{5}{\micro\meter}$ is compared to Eq.~\eqref{Eq:washerSquare} with variable $d$ for different values of $\lambda_\mathrm{L}$.
Overall the maximum deviation between simulation and Eq.~\eqref{Eq:washerSquare} is below $\sim 5\,\%$.
%

\clearpage
\section{Near field regime: coupling to SQUID arms with and without constrictions}
\label{sec:near-field}

In the main manuscript we have shown exemplary simulation results of the coupling factor where the magnetic moment oriented along $\hat{\bm e}_x$ is placed in close proximity at height $z$ above the center of an arm of the SQUID which is oriented along the $y$-axis (see Fig.~\ref{main-Fig:constriction} in the main manuscript).
Instead of simulating the full loop we showed that it is sufficient to only consider a single strip representing the SQUID arm, were the geometry of the single strip is described by the width $w$, thickness $d$ and length $l$.
In the main manuscript the simulation results for for some representative sets of geometric parameters and $\lambda_{\rm L}$ have been compared to Eq.~\eqref{main-Eq:coupling_constriction}, which was calculated for a homogeneous current density.
The simulations were done with current sheets oriented along $(x,y)$ and $(y,z)$ to account for $\lambda_{\rm L}<w$ and $\lambda_{\rm L}<d$ respectively.

\begin{figure}[b]
\includegraphics[width = \linewidth]{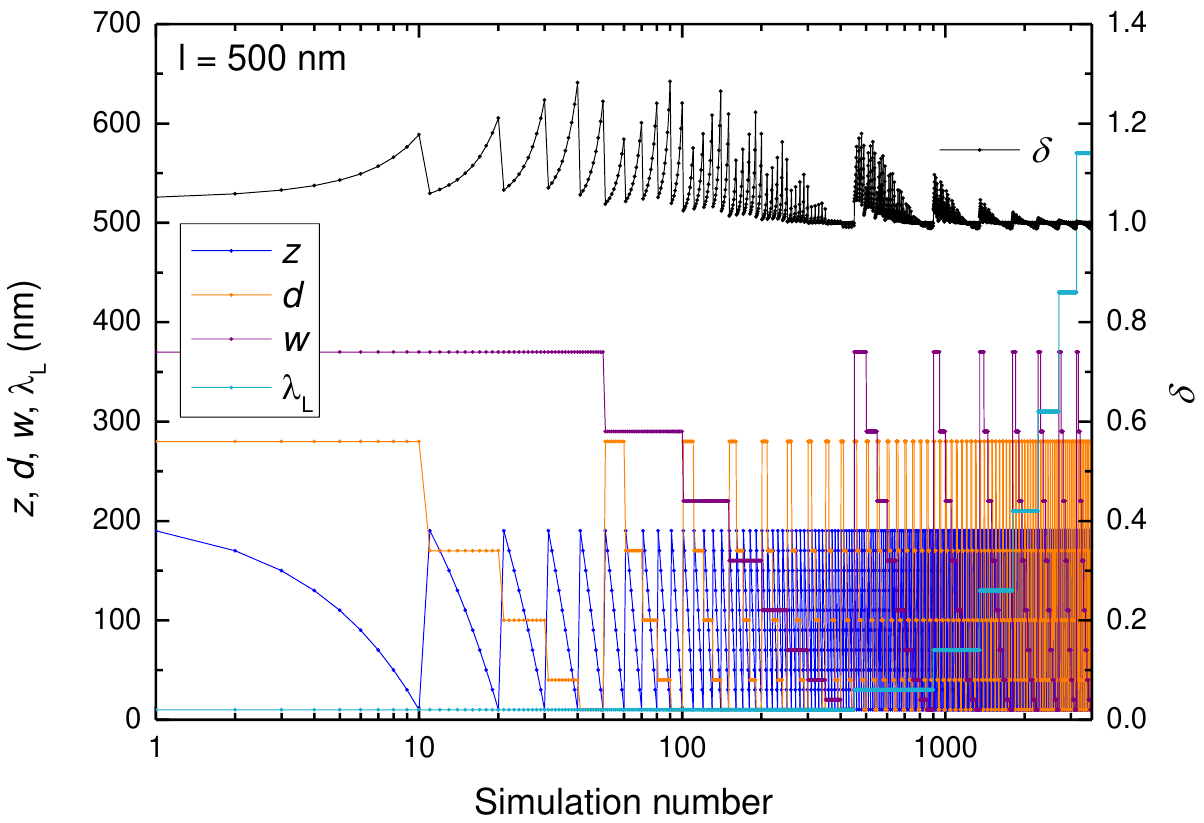}
\caption{
Ratio $\delta = \phi_{\mu, x,\text{calc}}/\phi_{\mu, x,\text{sim}}$ (black) between coupling factors calculated from Eq.~\eqref{main-Eq:coupling_constriction} and simulated with 3D-MLSI for a magnetic moment coupled to a single strip with variable $\lambda_{\rm L}$ and geometry ($z$, $d$, $w$; with fixed $l = \qty{500}{\nano\meter}$), with current sheets oriented along the $(x,y)$ plane.
In total 3600 values for $\delta$ are shown, for values of the corresponding quadruples  ($z$, $d$, $w$, $\lambda_{\rm L}$) shown by the coloured lines.
Each set of parameters is labeled by a simulation number (horizontal axis) from 1 to 3600.}
\label{Fig:error_near_field_xy}
\end{figure}

Here we analyze for most simulation results, obtained for strips with fixed length $l = \qty{500}{\nano\meter}$, and variable $w$, $d$, $z$ and $\lambda_{\rm L}$, the ratio $\delta = \phi_{\mu, x,\text{calc}}/\phi_{\mu, x,\text{sim}}$ between coupling factors $\phi_{\mu, x,\text{calc}}$ calculated with Eq.~\eqref{main-Eq:coupling_constriction} and $\phi_{\mu, x,\text{sim}}$ simulated with 3D-MLSI.
Figure \ref{Fig:error_near_field_xy} shows this for current sheets oriented along the $(x,y)$ plane, where current density variations along $w$ are captured in the simulations.
In this case, calculations with Eq.~\eqref{main-Eq:coupling_constriction} have the tendency to overestimate the coupling factor, which is indicated by $\delta>1$ (black).
With increasing $\lambda_{\rm L}$ (cyan) the deviation between calculation and simulation decreases; for $\lambda_{\rm L}>\qty{30}{\nano\meter}$ the deviation is below 10\,\%, independent of the geometry of the single strip and the distance $z$ (blue).
For the smallest value of $\lambda_{\rm L} = \qty{10}{\nano\meter}$ the inhomogeneity of the simulated current density distribution is most pronounced; i.e., the deviation between calculation and simulation reaches a maximum of $\approx 30\%$ in a few cases.
Moreover, it is obvious from Fig.~\ref{Fig:error_near_field_xy}, that the deviations always peak at minimum $z$.
Also, with decreasing $d$ (orange) the deviations increase.
With decreasing $w$ (purple) the deviations between simulation and calculation decrease.
This can best be seen best for the data sets with  $\lambda_{\rm L}>\qty{10}{\nano\meter}$.

\begin{figure}
\includegraphics[width = \linewidth]{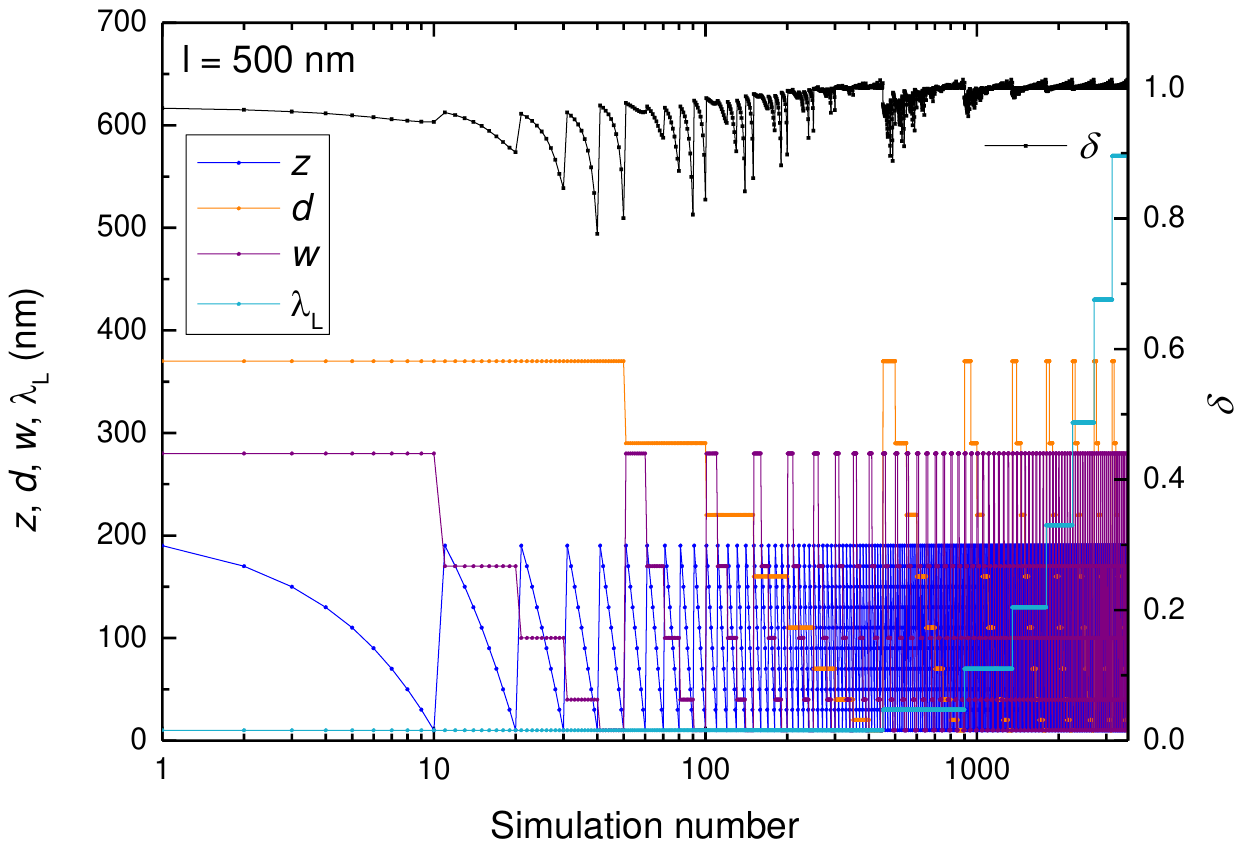}
\caption{Ratio $\delta = \phi_{\mu, x,\text{calc}}/\phi_{\mu, x,\text{sim}}$ (black) between coupling factors calculated from Eq.~\eqref{main-Eq:coupling_constriction} and simulated with 3D-MLSI for a magnetic moment coupled to a single strip with variable $\lambda_{\rm L}$ and geometry ($z$, $d$, $w$; with fixed $l = \qty{500}{\nano\meter}$), with current sheets oriented along the $(y,z)$ plane.
In total 3600 values for $\delta$ are shown, for values of the corresponding quadruples  ($z$, $d$, $w$, $\lambda_{\rm L}$) shown by the coloured lines.
Each set of parameters is labeled by a simulation number (horizontal axis) from 1 to 3600.}
\label{Fig:error_near_field_yz}
\end{figure}

Figure \ref{Fig:error_near_field_yz} shows accordingly the comparison between calculations with Eq.~\eqref{main-Eq:coupling_constriction} and 3D-MLSI simulations, with current sheets along $(y,z)$, where current density variations along $d$ are captured in the simulations.
For this current sheet orientation, Eq.~\eqref{main-Eq:coupling_constriction} has the tendency to underestimate the coupling factor, which is indicated by $\delta<1$ (black).
The behavior of deviations is the same like in Fig.~\ref{Fig:error_near_field_xy}, however with exchanged roles of $d$ and $w$.
Again for $\lambda_{\rm L}>\qty{30}{\nano\meter}$ deviations are below 10\,\% independent of the geometry of the simulated strip.

\section{Simulated current densities}
\label{sec:current-density}

In this section sheet current density distributions $j_{2D}(x,y)$ obtained from 3D-MLSI and GL simulations for various geometries are shown.
For better comparability we normalize the current densities to their maximum value $j_\mathrm{max}$.
All current densities shown, are extracted in the middle of the film thickness. 
In case of simulations of loops with 3D-MLSI, the current density distribution corresponds to $1\Phi_0$ stored in the loop.
Current densities obtained from GL simulations use two different ways of inducing the circulating current.
To compare the results of the simulation tools also current density distributions corresponding to $1\Phi_0$ stored in the loop are used.
Further, the current density can be induced by placing a magnetic moment ($\bm{\mu} \approx 35.5\cdot10^6\mu_{\bm B}$) at a height $z_0$ on the symmetry axis on the loop.

\subsection{Simulated current densities for the far field regime: circular and square shaped loops}

\begin{figure}[b]
\includegraphics[width=\linewidth]{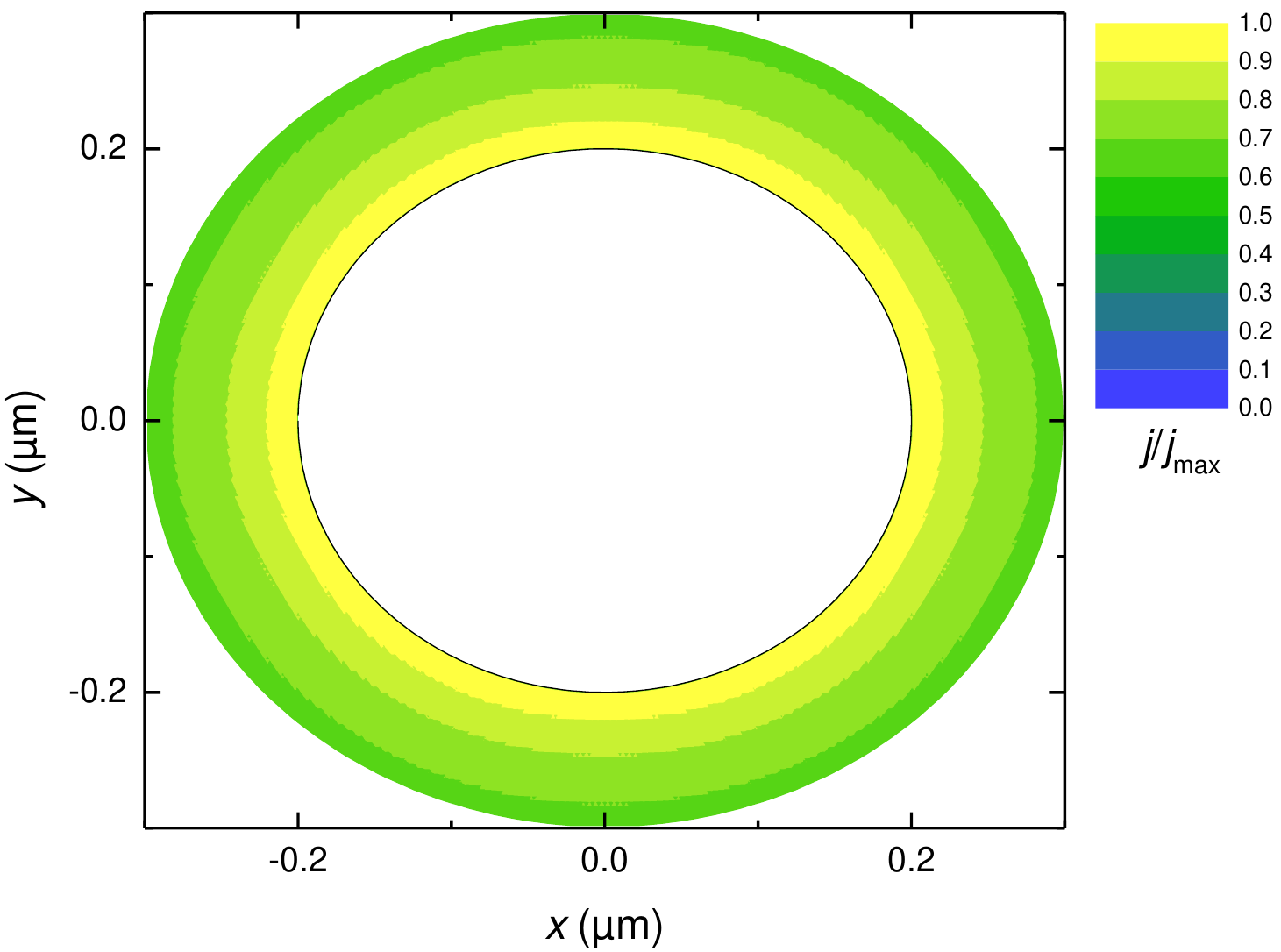}
\caption{Normalized current density for a circular loop with inner radius $a = \qty{200}{\nano\meter}$, outer radius $A = \qty{300}{\nano\meter}$, $\lambda_{\rm L} = \qty{200}{\nano\meter}$ and film thickness $d = \qty{50}{\nano\meter}$, simulated with 3D-MLSI.
}
\label{Fig:j_khapaev_lambda_200_d_50_200_300_round}
\end{figure}

In Fig.~\ref{Fig:j_khapaev_lambda_200_d_50_200_300_round} the normalized current density of a circular loop with inner radius $a = \qty{200}{\nano\meter}$, outer radius $A = \qty{300}{\nano\meter}$, $\lambda_{\rm L} = \qty{200}{\nano\meter}$ and film thickness $d = \qty{50}{\nano\meter}$, simulated with 3D-MLSI is shown.
%
%
A decrease down to roughly 70\,\% of the normalized current density from the inner circumference to the outer circumference is visible.
This current density distribution was used, to calculate the coupling factors of this geometry in Fig.~\ref{main-Fig_comparison_GL_London} and Fig.~\ref{main-Fig:washer-z-dep} of the main manuscript.

\begin{figure}
\includegraphics[width=\linewidth]{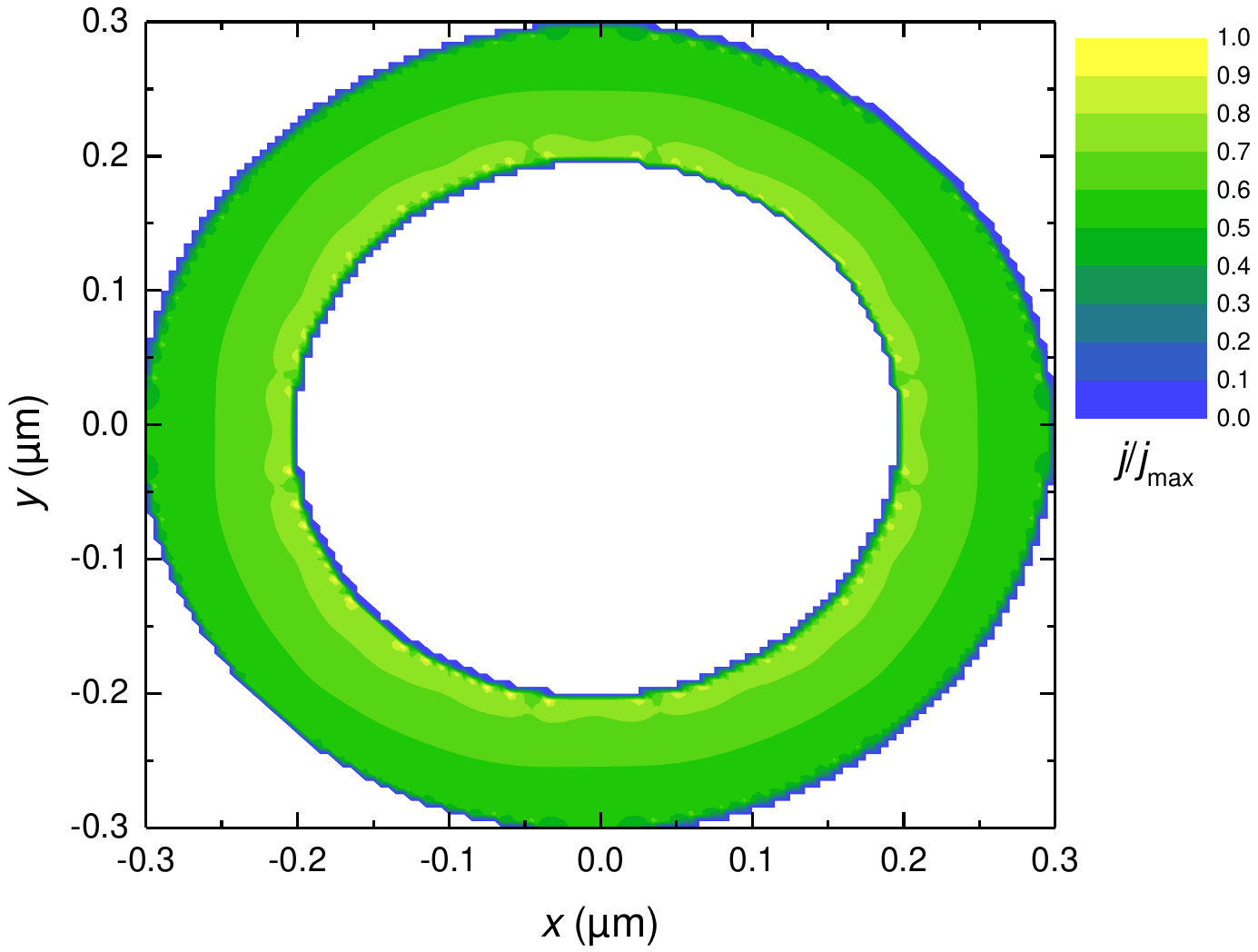}
\caption{Normalized current density for a circular loop with inner radius $a = \qty{200}{\nano\meter}$, outer radius $A = \qty{300}{\nano\meter}$, $\lambda_{\rm L} = \qty{200}{\nano\meter}$, GL coherence length $\xi_{\rm GL} = \qty{20}{\nano\meter}$ and film thickness $d = \qty{50}{\nano\meter}$, calculated with GL simulations. 
The current density is induced by field cooling and subsequently turning off the field to trap $1\,\Phi_0$.}
\label{Fig:j_GL_q_xi_20_lambda_200_d_50_200_300_round}
\end{figure}

A similar decrease is shown in Fig.~\ref{Fig:j_GL_q_xi_20_lambda_200_d_50_200_300_round}, where the current density for the same geometry was simulated with GL simulations using a GL coherence length $\xi_{\rm GL} = \qty{20}{\nano\meter}$.
%
%
In this simulation the current density was created by field cooling the circular loop to the superconducting state and turning off the field subsequently to achieve one trapped flux quantum in the loop.
In this figure the current density decreases from $\approx0.7$ to $\approx0.5$ moving from the inner to the outer radius, which also gives roughly a decrease of 70\%.
The lower values of normalized current density result from simulation artifacts, namely the current density peaking at several points located at the inner circumference due to the square lattice used in GL simulations.

\begin{figure}[b]
\includegraphics[width=\linewidth]{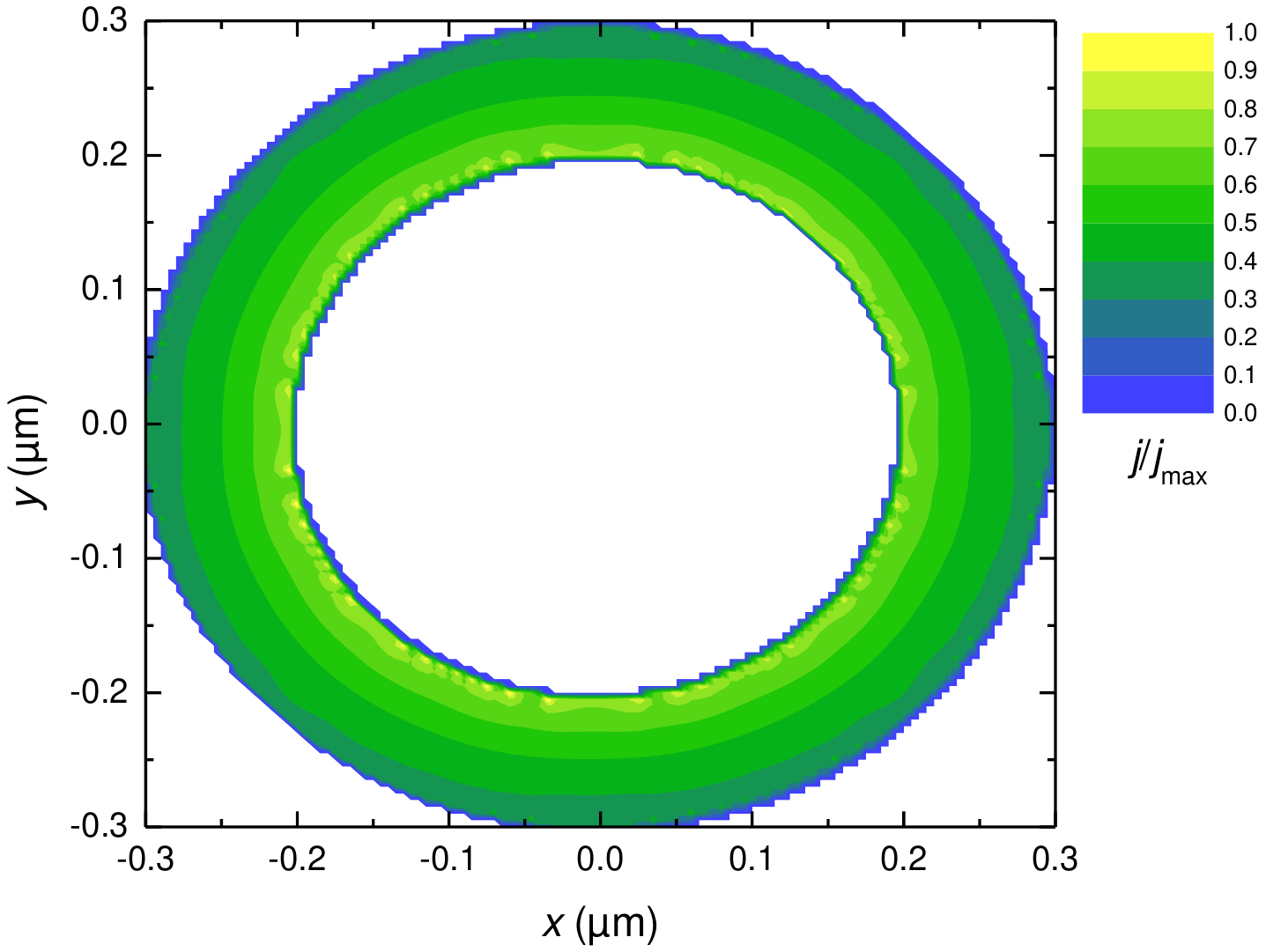}
\caption{Normalized current density for a circular loop with inner radius $a = \qty{200}{\nano\meter}$, outer radius $A = \qty{300}{\nano\meter}$, $\lambda_{\rm L} = \qty{200}{\nano\meter}$, GL coherence length $\xi_{\rm GL} = \qty{20}{\nano\meter}$ and film thickness $d = \qty{50}{\nano\meter}$, calculated with GL simulations.
The current density is induced by placing a dipole on the symmetry axis of this geometry at a position $z_0 = 0$.}
\label{Fig:j_GL_0_xi_20_lambda_200_d_50_200_300_round}
\end{figure}

\begin{figure}
\includegraphics[width=\linewidth]{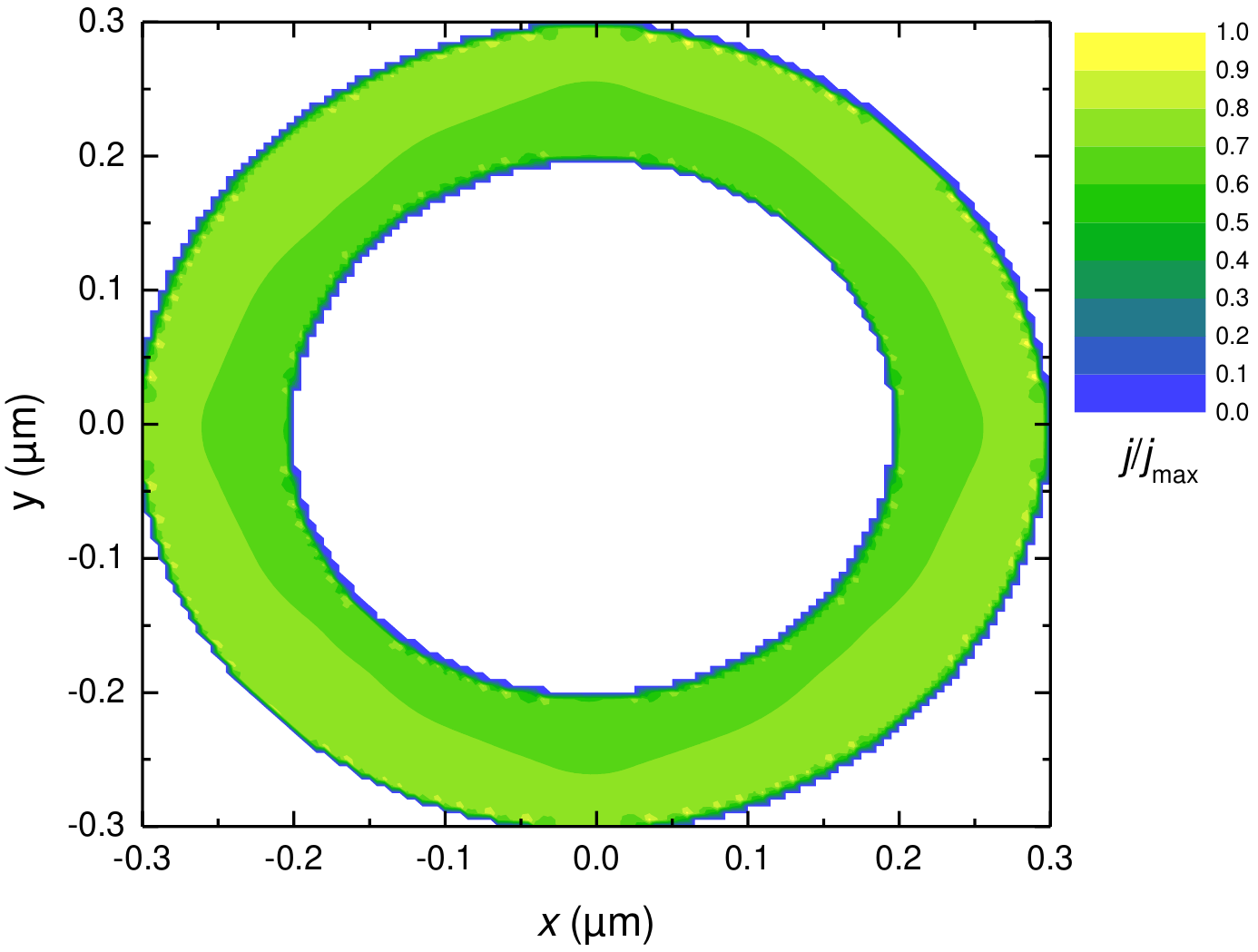}
\caption{Normalized current density for a circular loop with inner radius $a = \qty{200}{\nano\meter}$, outer radius $A = \qty{300}{\nano\meter}$, $\lambda_{\rm L} = \qty{200}{\nano\meter}$, GL coherence length $\xi_{\rm GL} = \qty{20}{\nano\meter}$ and film thickness $d = \qty{50}{\nano\meter}$, calculated with GL simulations. 
The current density is induced by placing a dipole on the symmetry axis of this geometry at a position $z_0 = \qty{500}{\nano\meter}$.}
\label{Fig:j_GL_500_xi_20_lambda_200_d_50_200_300_round}
\end{figure}

Again for the same set of parameters, in Fig.~\ref{Fig:j_GL_0_xi_20_lambda_200_d_50_200_300_round} and Fig.~\ref{Fig:j_GL_500_xi_20_lambda_200_d_50_200_300_round} the current density distribution is shown, resulting from GL simulations where a dipole is placed at the symmetry axis of the already superconducting loop at $z_0 = 0$ and $\qty{500}{\nano\meter}$ respectively.
%
%
The decrease of current density from the inner to the outer circumference for the simulation with the dipole at $z_0 = 0$ (from $~0.7$ to $~0.35$, i.e. roughly 50\%) is more pronounced compared to the simulation with the trapped flux quantum.
In contrast, the simulation with the dipole far from the structure ($z_0 = \qty{500}{\nano\meter}$) shows a quite homogeneous current density, which exhibits a slight increase in current density when moving from the inner to the outer circumference.

\begin{figure}[b]
	\includegraphics[width=\linewidth]{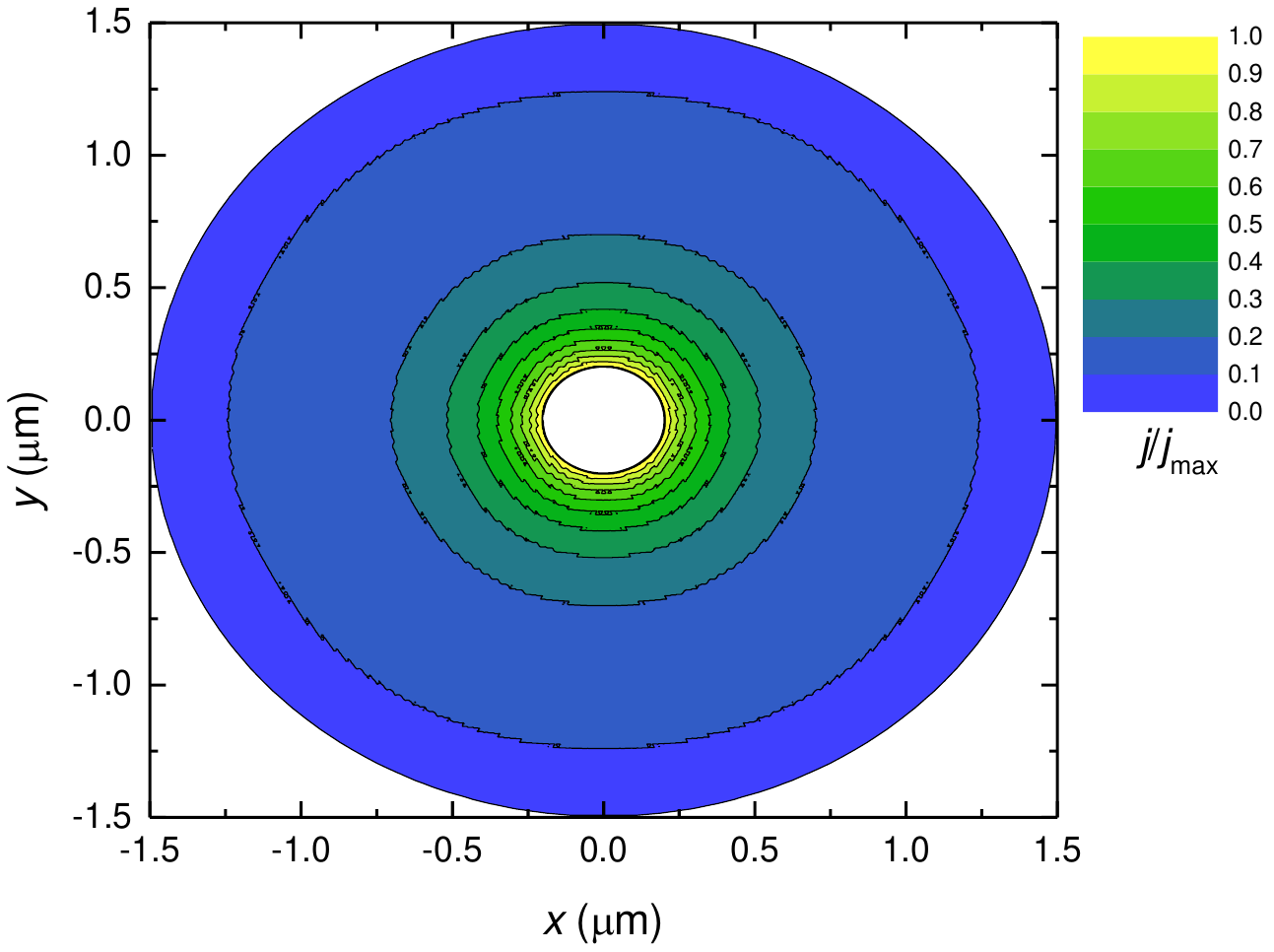}
	\caption{Normalized current density for a circular loop with inner radius $a = \qty{200}{\nano\meter}$, outer radius $A = \qty{1.5}{\micro\meter}$, $\lambda_{\rm L} = \qty{200}{\nano\meter}$ and film thickness $d = \qty{50}{\nano\meter}$, simulated with 3D-MLSI.}
	\label{Fig:j_khapaev_lambda_200_d_50_200_1500_round}
\end{figure}

In Fig.~\ref{Fig:j_khapaev_lambda_200_d_50_200_1500_round} the normalized current density for a loop with $a = \qty{200}{\nano\meter}$, outer radius $A = \qty{1.5}{\micro\meter}$, $\lambda_{\rm L} = \qty{200}{\nano\meter}$ and film thickness $d = \qty{50}{\nano\meter}$, simulated with 3D-MLSI is shown.
Due to the larger line width, the decrease of the current density towards the outer circumference is more pronounced as compared to Fig.~\ref{Fig:j_khapaev_lambda_200_d_50_200_300_round}.

\begin{figure}
	\includegraphics[width=\linewidth]{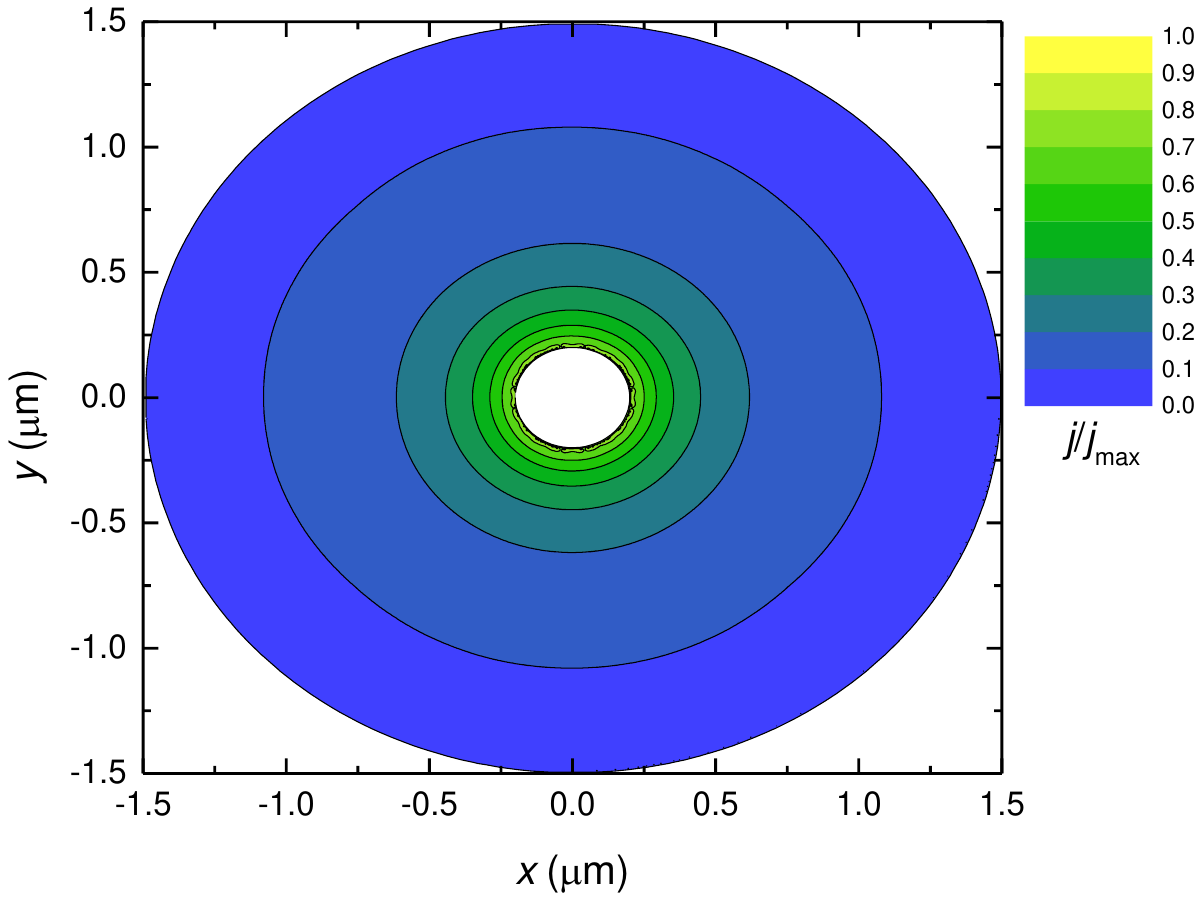}
	\caption{Normalized current density for a circular loop with inner radius $a = \qty{200}{\nano\meter}$, outer radius $A = \qty{1500}{\nano\meter}$, $\lambda_{\rm L} = \qty{200}{\nano\meter}$, GL coherence length $\xi_{\rm GL} = \qty{20}{\nano\meter}$ and film thickness $d = \qty{50}{\nano\meter}$, calculated with GL simulations.
		The current density is induced by field cooling and subsequently turning off the field to trap $1\,\Phi_0$.}
	\label{Fig:j_GL_q_xi_20_lambda_200_d_50_200_1500_round}
\end{figure}

Figure \ref{Fig:j_GL_q_xi_20_lambda_200_d_50_200_1500_round} shows the same situation simulated with GL-simulation.
The overall current density distribution looks similar to Fig.~\ref{Fig:j_khapaev_lambda_200_d_50_200_1500_round}, however with slight deviations due to the square lattice in GL simulations leading to current peaks at the inner circumference and therefore a different $j_{\rm max}$.

\begin{figure}[b]
\includegraphics[width=\linewidth]{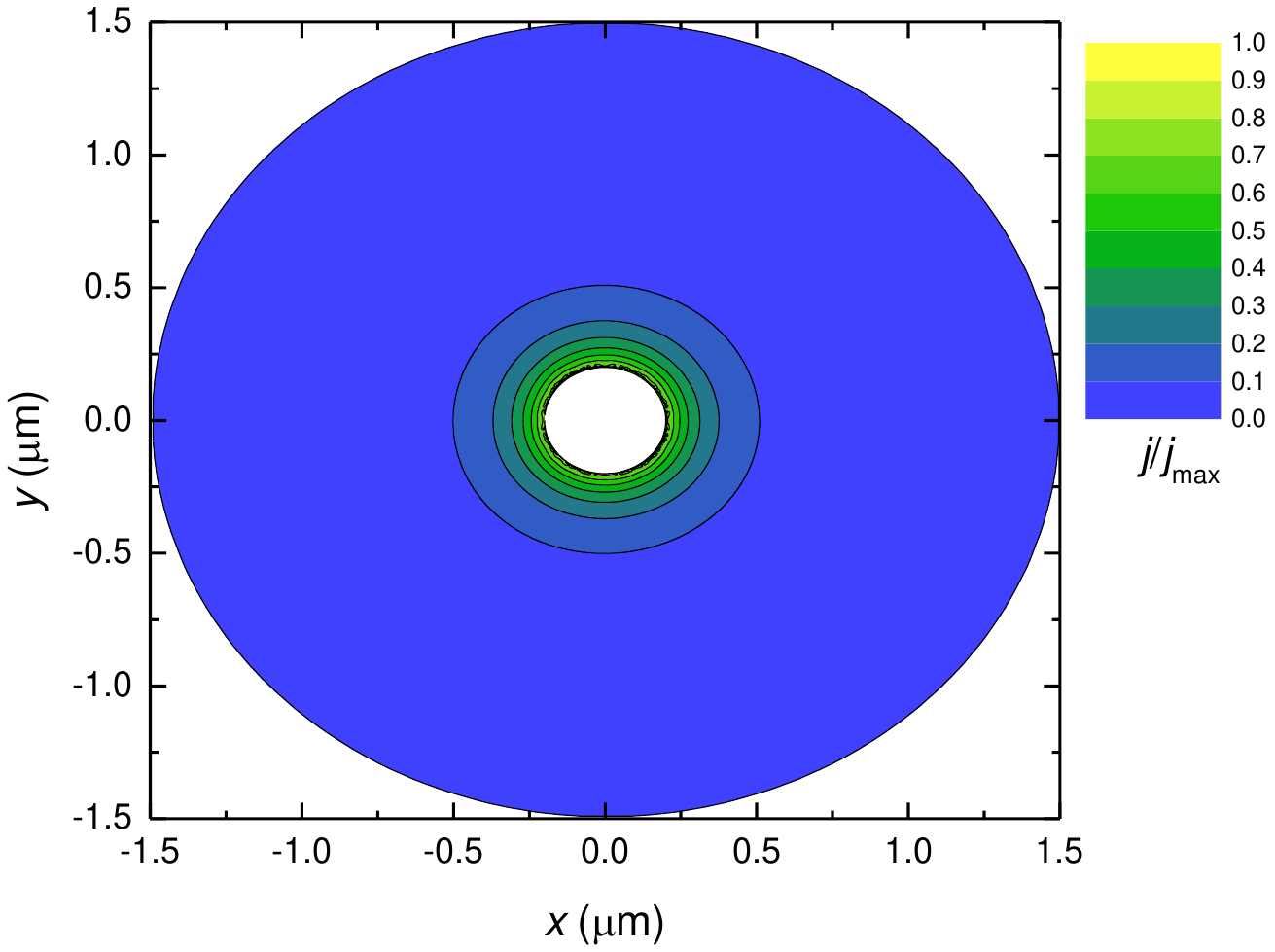}
\caption{Normalized current density for a circular loop with inner radius $a = \qty{200}{\nano\meter}$, outer radius $A = \qty{1500}{\nano\meter}$, $\lambda_{\rm L} = \qty{200}{\nano\meter}$, GL coherence length $\xi_{\rm GL} = \qty{20}{\nano\meter}$ and film thickness $d = \qty{50}{\nano\meter}$, calculated with GL simulations.
The current density is induced by placing a dipole on the symmetry axis of this geometry at a position $z_0 = \qty{0}{\nano\meter}$.}
\label{Fig:j_GL_0_xi_20_lambda_200_d_50_200_1500_round}
\end{figure}

Figure \ref{Fig:j_GL_0_xi_20_lambda_200_d_50_200_1500_round} shows the same geometry calculated with GL simulations ($\xi_{\rm GL} = \qty{20}{\nano\meter}$), where the current is induced by a magnetic moment placed at $z_0 = \qty{0}{\nano\meter}$.
This again results in more current being concentrated at the inner circumference, compared to the current density distribution belonging to a stored flux quantum.
Calculated coupling factors from the current density shown in Fig.~\ref{Fig:j_GL_0_xi_20_lambda_200_d_50_200_1500_round} are larger for small $z$ compared to calculations with the current density belonging to a stored flux quantum (compare Fig.~\ref{Fig:khapaev_GL_big_loop}).

\begin{figure}
	\includegraphics[width=\linewidth]{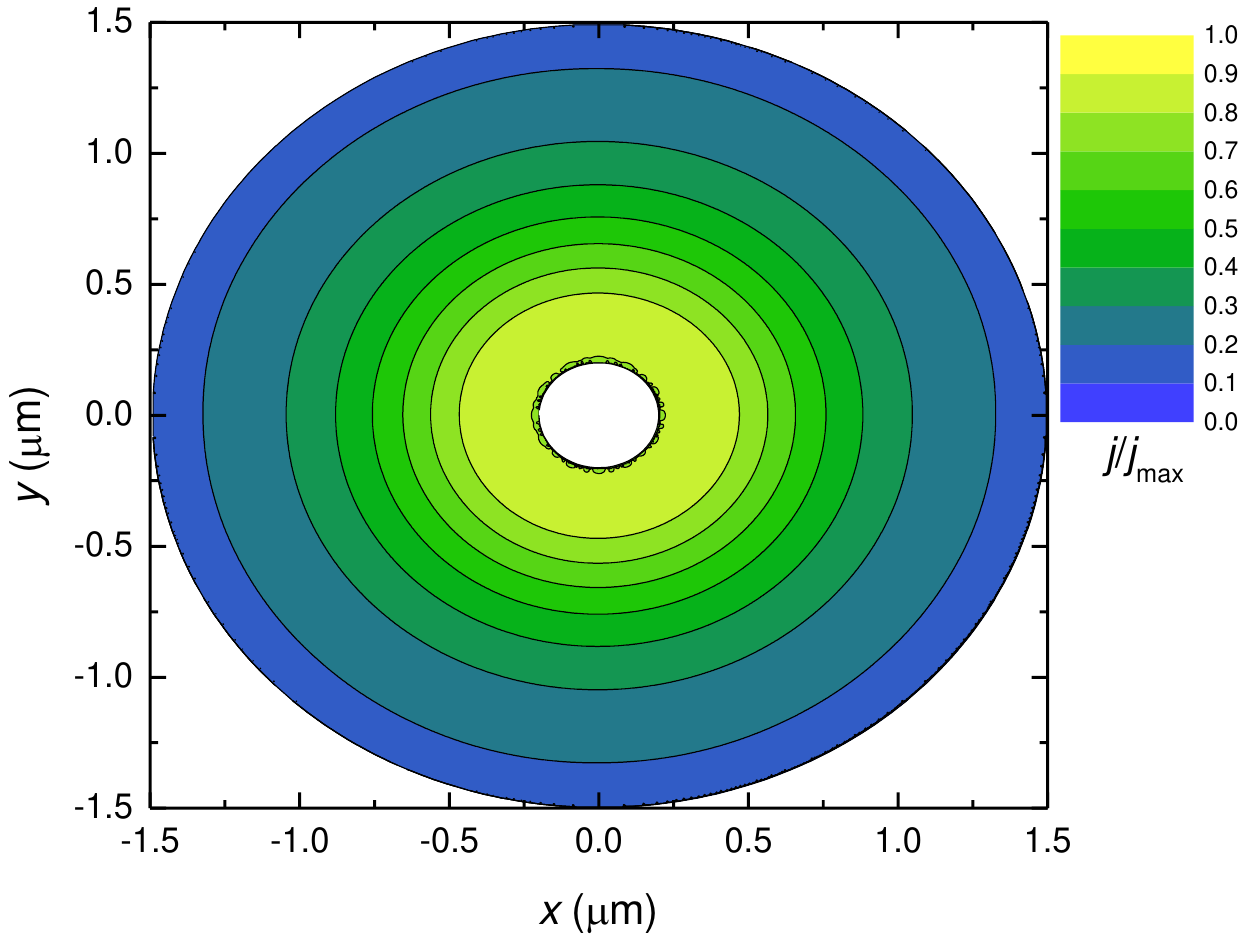}
	\caption{Normalized current density for a circular loop with inner radius $a = \qty{200}{\nano\meter}$, outer radius $A = \qty{1500}{\nano\meter}$, $\lambda_{\rm L} = \qty{200}{\nano\meter}$, GL coherence length $\xi_{\rm GL} = \qty{20}{\nano\meter}$ and film thickness $d = \qty{50}{\nano\meter}$, calculated with GL simulations.
		The current density is induced by placing a dipole on the symmetry axis of this geometry at a position $z_0 = \qty{500}{\nano\meter}$.}
	\label{Fig:j_GL_500_xi_20_lambda_200_d_50_200_1500_round}
\end{figure}

Figure \ref{Fig:j_GL_500_xi_20_lambda_200_d_50_200_1500_round} shows again the same geometry calculated with GL simulations ($\xi_{\rm GL} = \qty{20}{\nano\meter}$), where the current is induced by a magnetic moment placed at $z_0 = \qty{500}{\nano\meter}$.
This current density distribution shows a slower decay towards the outer circumference compared to Figs.~\ref{Fig:j_khapaev_lambda_200_d_50_200_1500_round}~$-$~\ref{Fig:j_GL_0_xi_20_lambda_200_d_50_200_1500_round}.

\begin{figure}
	\includegraphics[width=\linewidth]{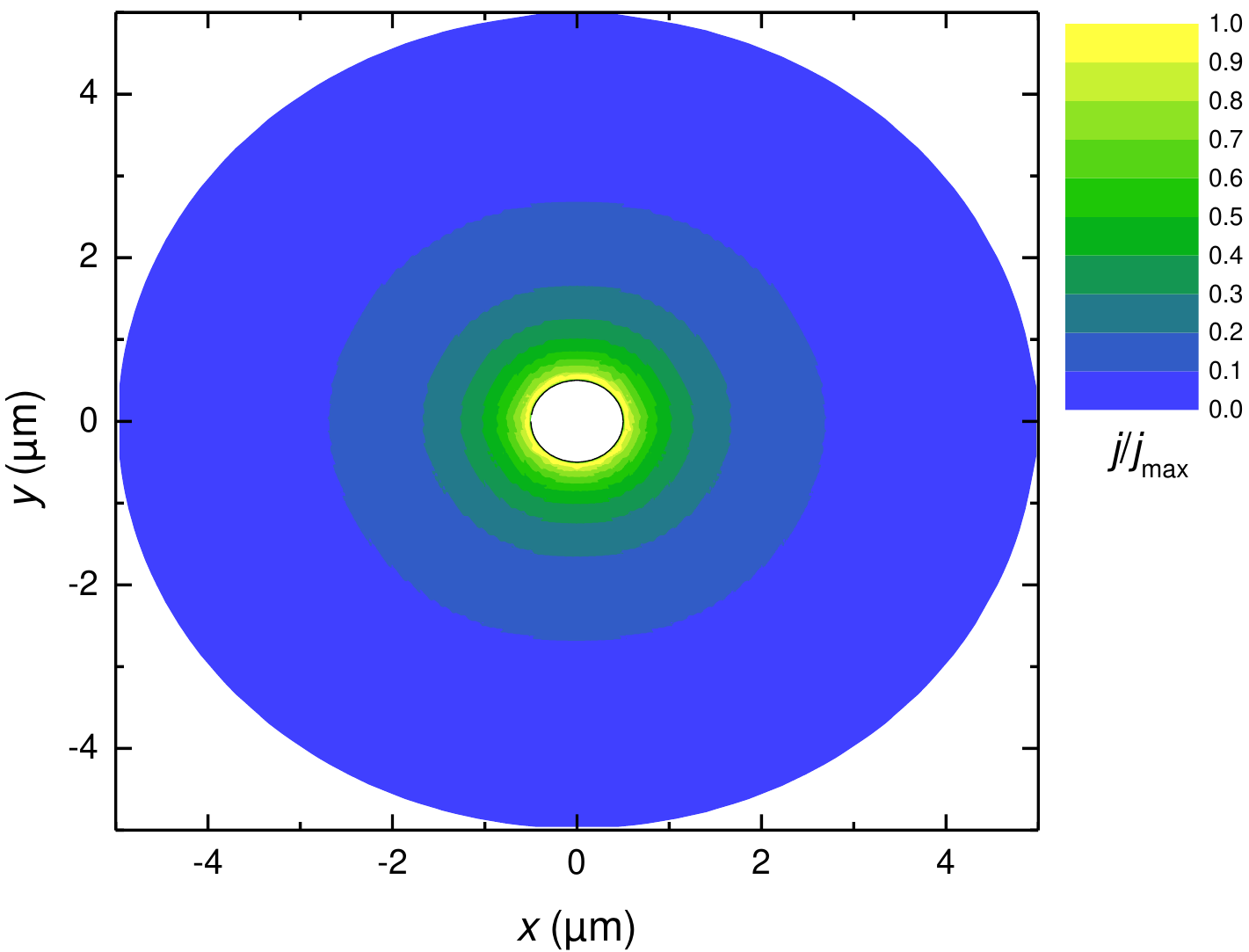}
	\caption{Normalized current density for a circular loop with inner radius $a = \qty{500}{\nano\meter}$, outer radius $A = \qty{5}{\micro\meter}$, $\lambda_{\rm L} = \qty{200}{\nano\meter}$ and film thickness $d = \qty{50}{\nano\meter}$, simulated with 3D-MLSI.}
	\label{Fig:j_khapaev_lambda_200_d_50_500_5000_round}
\end{figure}

Figure \ref{Fig:j_khapaev_lambda_200_d_50_500_5000_round} shows the current density distribution for a even more massive circular loop with $a = \qty{500}{\nano\meter}$, $A = \qty{5}{\micro\meter}$, $\lambda_{\rm L} = \qty{200}{\nano\meter}$ and  $d = \qty{50}{\nano\meter}$, simulated with 3D-MLSI.
The larger line width ($A-a$) of this geometry leads to a much more inhomogeneous current density, compared to the previously shown, where only a small amount of current flows at the outer circumference.

\begin{figure}
	\includegraphics[width=\linewidth]{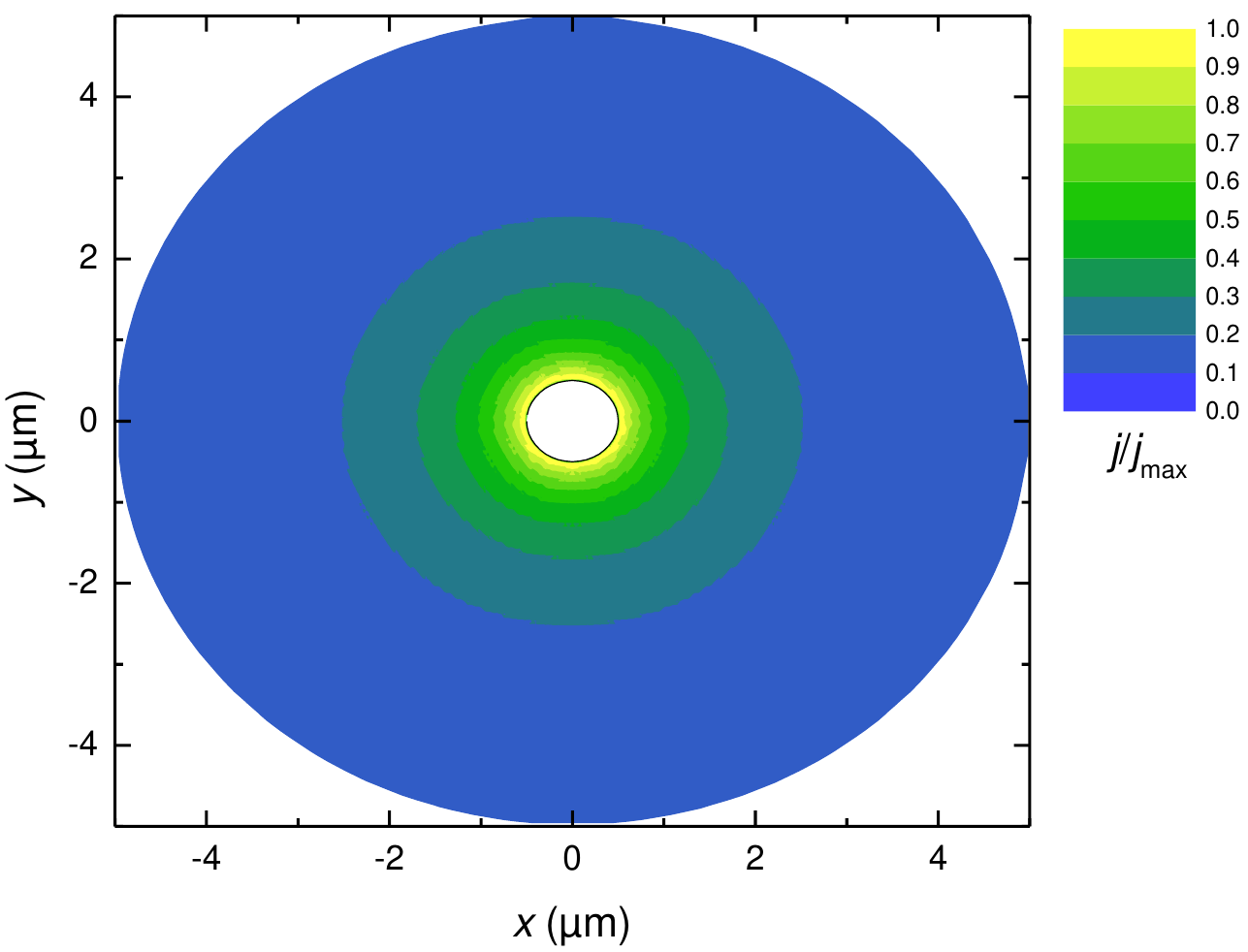}
	\caption{Normalized current density for a circular loop with inner radius $a = \qty{500}{\nano\meter}$, outer radius $A = \qty{5}{\micro\meter}$, $\lambda_{\rm L} = \qty{500}{\nano\meter}$ and film thickness $d = \qty{10}{\nano\meter}$, simulated with 3D-MLSI.}
	\label{Fig:j_khapaev_lambda_500_d_10_500_5000_round}
\end{figure}

Figure \ref{Fig:j_khapaev_lambda_500_d_10_500_5000_round} shows again the same geometry, with $\lambda_{\rm L} = \qty{500}{\nano\meter}$ and $d = \qty{10}{\nano\meter}$.
%
%
In this case the current concentration at the inner circumference is slightly less pronounced compared to the simulation with smaller $\lambda_{\rm L}$ and bigger $d$, which results in the difference in coupling factors seen in Fig.~\ref{main-Fig:washer-z-dep} in the main manuscript.

\newpage

\begin{figure}
	\includegraphics[width=\linewidth]{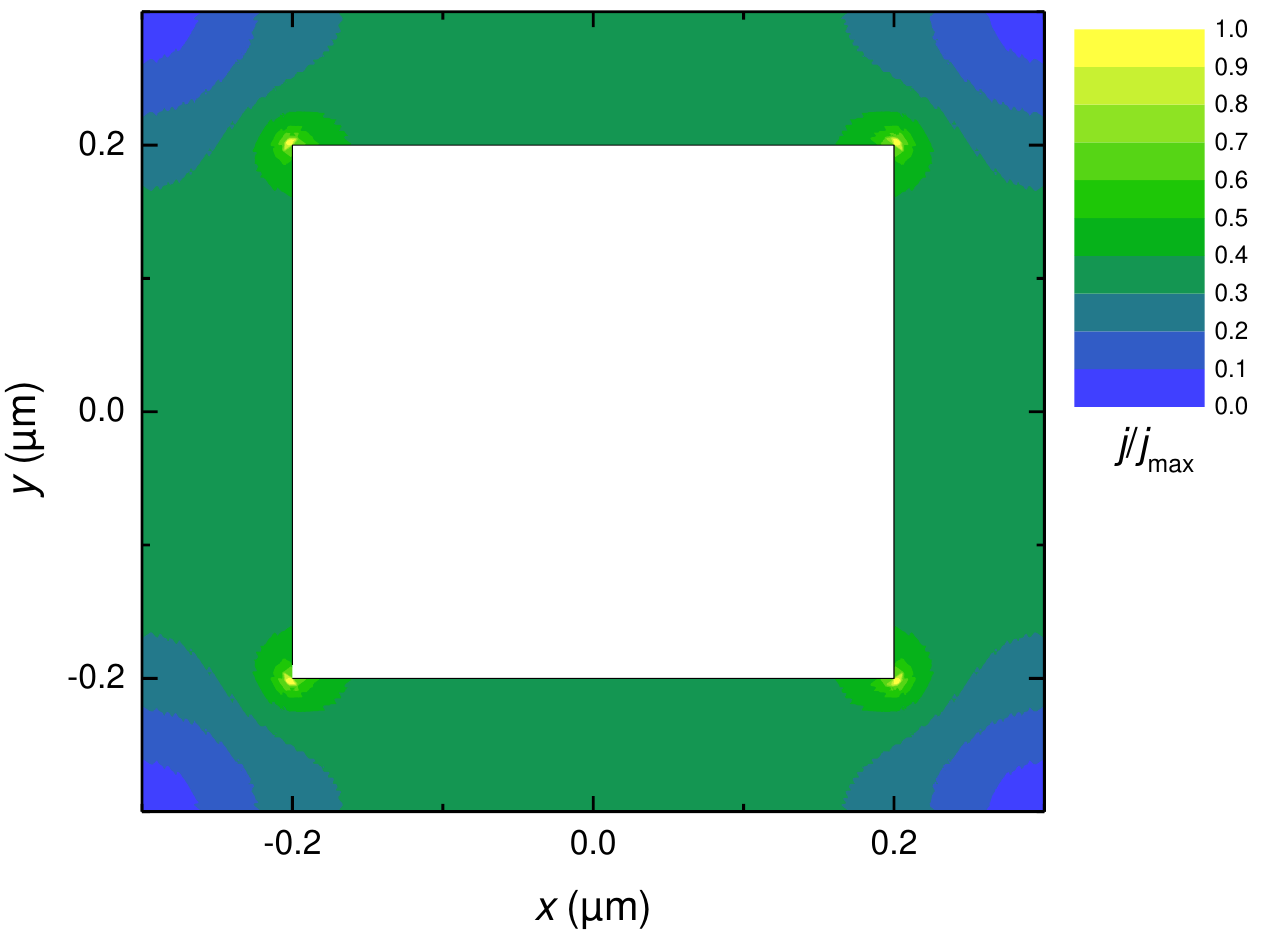}
	\caption{Normalized current density for a square shaped loop with $a = \qty{200}{\nano\meter}$,  $A = \qty{300}{\nano\meter}$, $\lambda_{\rm L} = \qty{200}{\nano\meter}$ and film thickness $d = \qty{50}{\nano\meter}$, simulated with 3D-MLSI.}
	\label{Fig:j_khapaev_lambda_200_d_50_200_300}
\end{figure}

\begin{figure}
	\includegraphics[width=\linewidth]{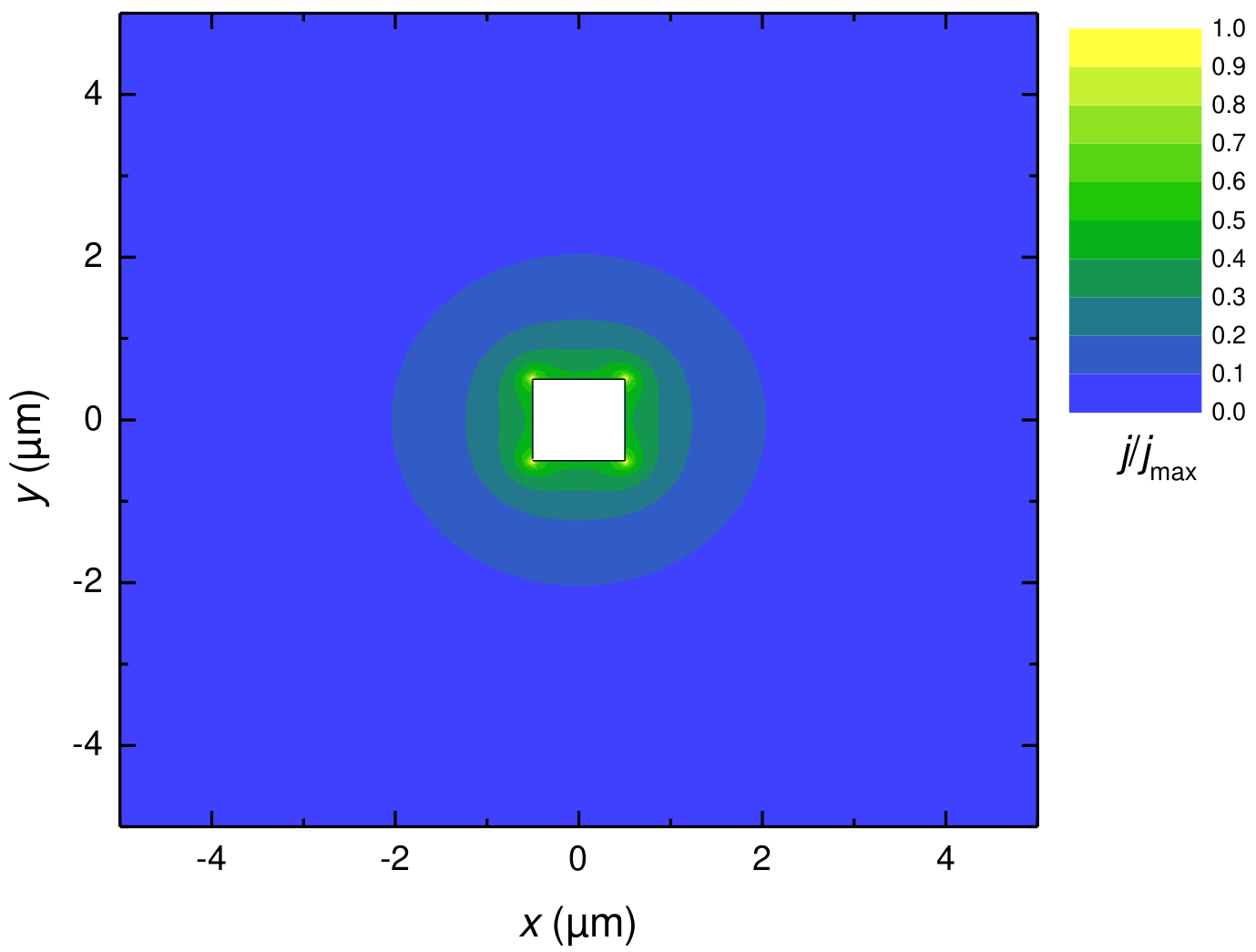}
	\caption{Normalized current density for a square shaped loop with $a = \qty{500}{\nano\meter}$, $A =\qty{5}{\micro\meter}$, $\lambda_{\rm L} = \qty{200}{\nano\meter}$ and film thickness $d = \qty{50}{\nano\meter}$, simulated with 3D-MLSI.}
	\label{Fig:j_khapaev_lambda_200_d_50_500_5000}
\end{figure}

The following figures show the simulated normalized current densities with 3D-MLSI of square shaped loops with inner dimension $2a$ and outer dimension $2A$ for $\lambda_{\rm L} = \qty{200}{\nano\meter}$ and $d = \qty{50}{\nano\meter}$.
The simulation for a square shaped loop with small line width ($a = \qty{200}{\nano\meter}$, $A = \qty{300}{\nano\meter}$) in Fig.~\ref{Fig:j_khapaev_lambda_200_d_50_200_300} shows, like for circular loops, a more homogeneous current density compared to the simulation with larger line width ($a = \qty{500}{\nano\meter}$, $A = \qty{5}{\micro\meter}$) in Fig.~\ref{Fig:j_khapaev_lambda_200_d_50_500_5000}.
The current density shows a current concentration at the inner corners of the square for both simulations.
%
%

\newpage

\subsection{Simulated current densities in the near field regime: square loops and single strips}

\begin{figure}
	\includegraphics[width=\linewidth]{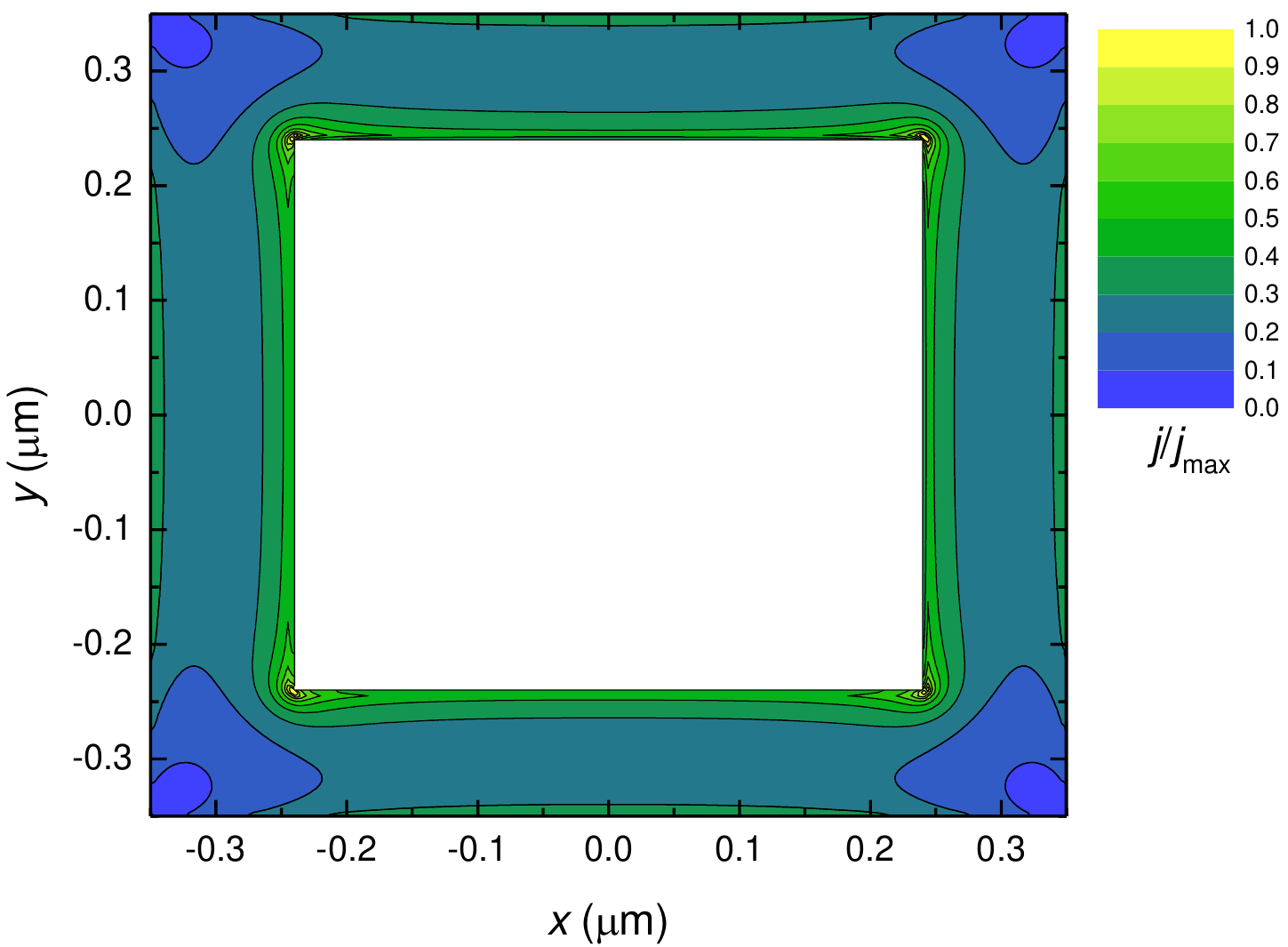}
	\caption{Normalized current density for a square shaped loop with $a = \qty{240}{\nano\meter}$,  $A = \qty{350}{\nano\meter}$, $\lambda_{\rm L} = \qty{10}{\nano\meter}$ and film thickness $d = \qty{10}{\nano\meter}$, simulated with 3D-MLSI.}
  \label{Fig:j_khapaev_full_loop_lambda_10}
\end{figure}

\begin{figure}
\includegraphics[width=\linewidth]{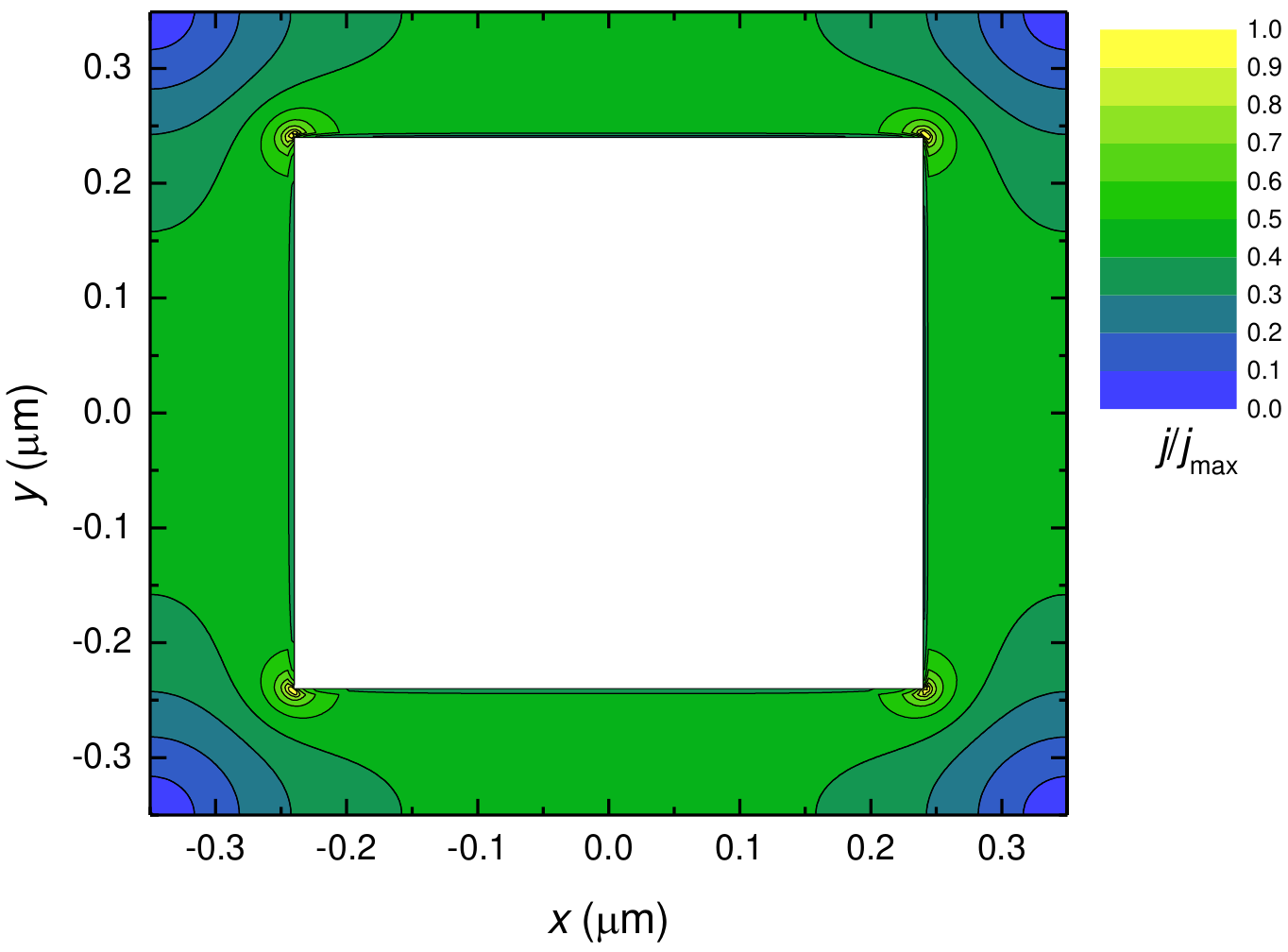}
\caption{Normalized current density for a square shaped loop with $a = \qty{240}{\nano\meter}$,  $A = \qty{350}{\nano\meter}$, $\lambda_{\rm L} = \qty{570}{\nano\meter}$ and film thickness $d = \qty{10}{\nano\meter}$, simulated with 3D-MLSI.}
  \label{Fig:j_khapaev_full_loop_lambda_570}
\end{figure}

\begin{figure}
	\includegraphics[width=\linewidth]{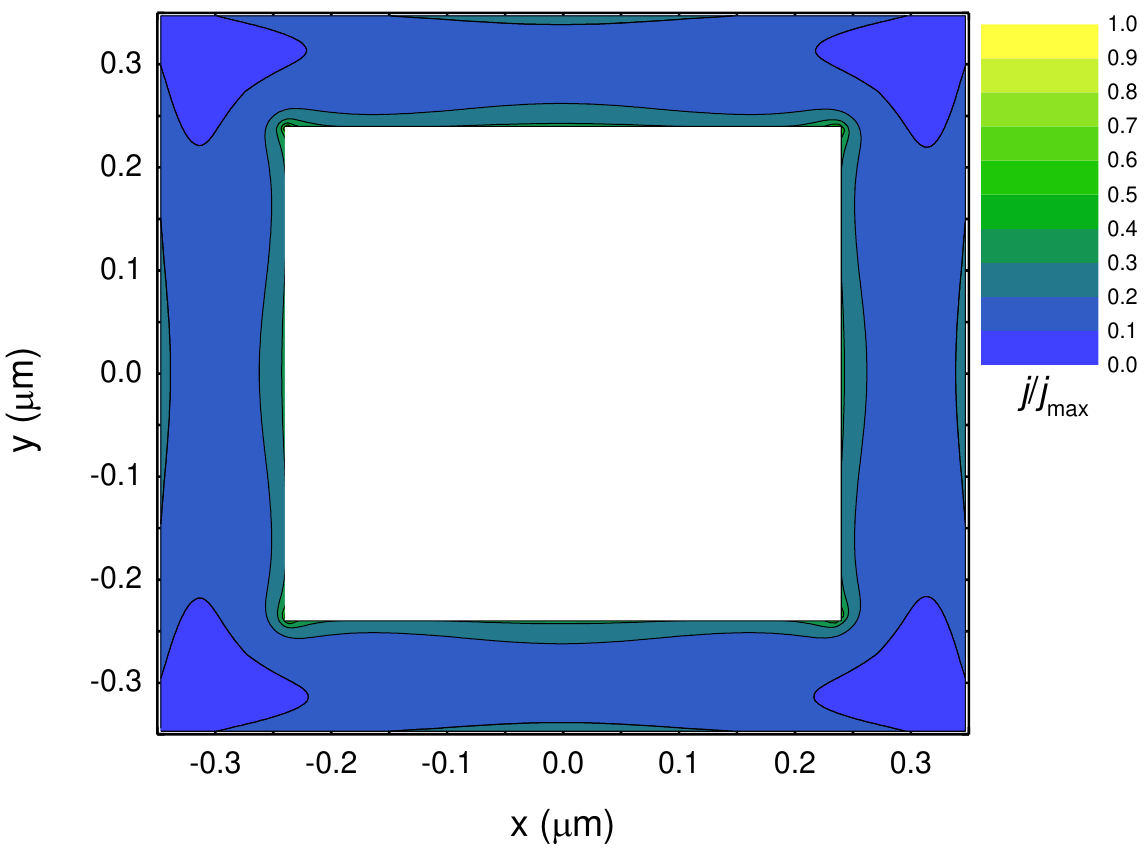}
	\caption{Normalized current density for a square shaped loop with $a = \qty{240}{\nano\meter}$,  $A = \qty{350}{\nano\meter}$, $\lambda_{\rm L} = \qty{10}{\nano\meter}$ and film thickness $d = \qty{10}{\nano\meter}$, calculated with GL simulations via 1 $\Phi_0$ trapped in the loop.}
	\label{Fig_j_GL_full_loop_lambda_10}
\end{figure}

\begin{figure}
	\includegraphics[width=\linewidth]{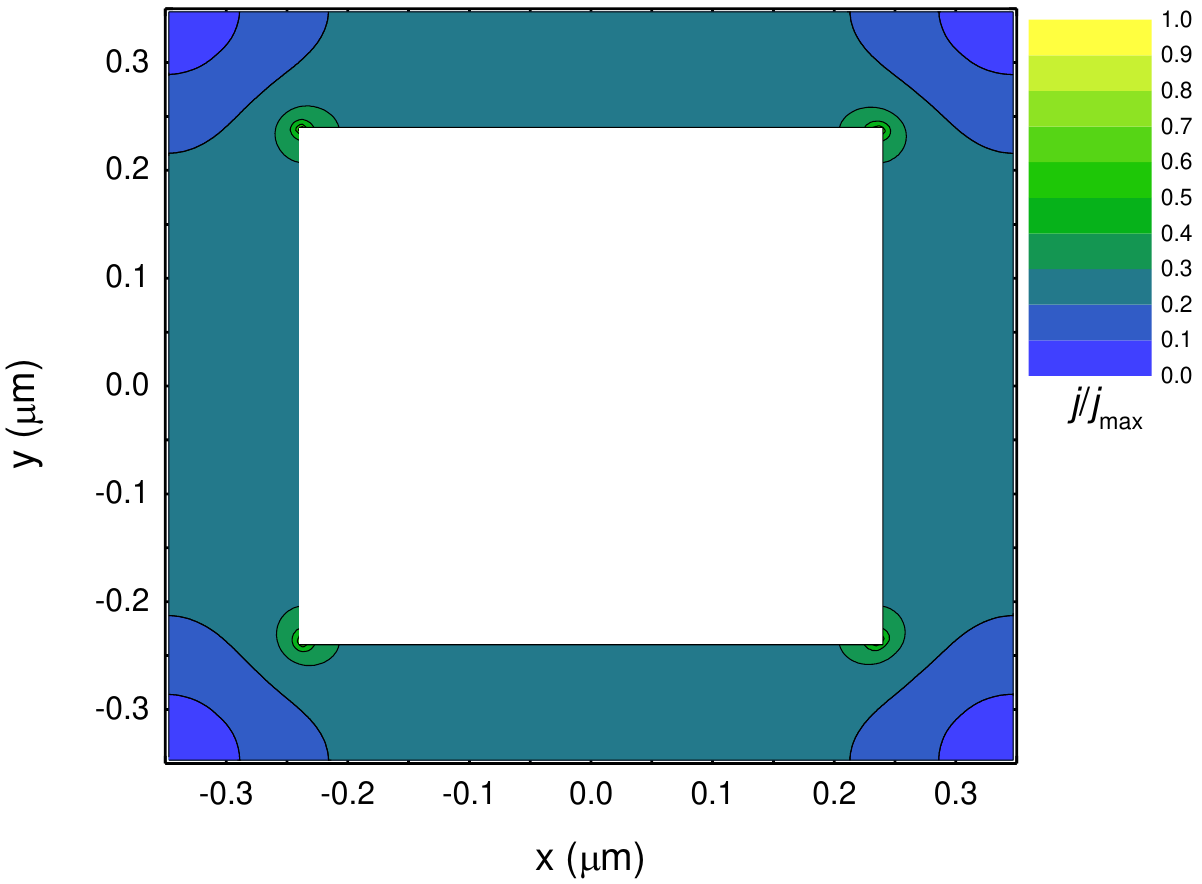}
	\caption{Normalized current density for a square shaped loop with $a = \qty{240}{\nano\meter}$,  $A = \qty{350}{\nano\meter}$, $\lambda_{\rm L} = \qty{570}{\nano\meter}$ and film thickness $d = \qty{10}{\nano\meter}$, calculated with GL simulations via 1 $\Phi_0$ trapped in the loop.}
	\label{Fig_j_GL_full_loop_lambda_570}
\end{figure}

\begin{figure}
	\includegraphics[width=\linewidth]{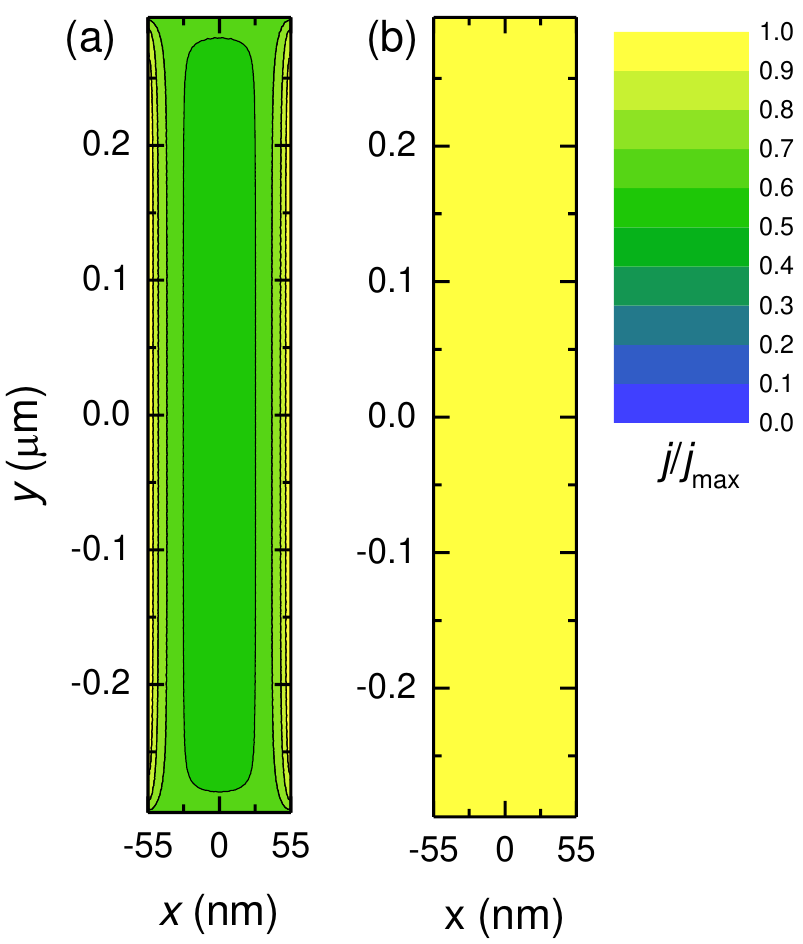}
	\caption{Normalized current density of a single strip with $l = \qty{590}{\nano\meter}$, $w = \qty{110}{\nano\meter}$ and $d = \qty{10}{\nano\meter}$ simulated with 3D-MLSI.
  (a): Simulation with $\lambda_{\rm L} = \qty{10}{\nano\meter}$.
  (b): Simulation with $\lambda_{\rm L} = \qty{570}{\nano\meter}$.}
  \label{Fig:j_khapaev_conductor}
\end{figure}

In the main manuscript a comparison of coupling factors obtained above one arm of a full SQUID loop with coupling factors obtained above a single strip is shown (Fig.~\ref{main-Fig:loop_vs_constriction} in the main manuscript).
Figures~\ref{Fig:j_khapaev_full_loop_lambda_10}~-~\ref{Fig:j_khapaev_conductor} are showing the current densities utilized for calculating these coupling factors.
In Fig.~\ref{Fig:j_khapaev_full_loop_lambda_10} the normalized current density for a square shaped loop with $a = \qty{240}{\nano\meter}$,  $A = \qty{350}{\nano\meter}$, $\lambda_{\rm L} = \qty{10}{\nano\meter}$ and film thickness $d = \qty{10}{\nano\meter}$ simulated with 3D-MLSI is shown.
Due to the small value of $\lambda_{\rm L}$ the overall current density is inhomogeneous.
The current density is highest at the inner circumference, especially at the inner corners of the structure.
Moving towards the outer circumference of the loop, it can be seen that first the current decreases.
At the outer circumference again a slight increase is observed.
The current is lowest at the outer corners of the structure. 
%
%
Figure \ref{Fig:j_khapaev_full_loop_lambda_570} shows the normalized current density of the same loop with $\lambda_{\rm L} = \qty{570}{\nano\meter}$, which is bigger than the line width of the loop.
Compared to Fig.~\ref{Fig:j_khapaev_full_loop_lambda_10} the resulting current density is homogeneous over wide parts of the structure.
Again the current is highest at the inner and lowest at the outer corners.
%

Figures \ref{Fig_j_GL_full_loop_lambda_10} and \ref{Fig_j_GL_full_loop_lambda_570} show current density distributions of the corresponding geometry obtained from GL simulations with $\lambda_{\rm L} = \qty{10}{\nano\meter}$ and \qty{570}{\nano\meter} respectively, where $1 \Phi_0$ is trapped in the loop.
The resulting current density distributions look similar, but with slight differences which are partially explained by different values of the current peaks at the inner corners and therefore different $j_{\rm max}$.

Figure \ref{Fig:j_khapaev_conductor} shows the normalized current density of a single strip with $l = \qty{590}{\nano\meter}$, $w = \qty{110}{\nano\meter}$ and $d = \qty{10}{\nano\meter}$ for $\lambda_{\rm L} = \qty{10}{\nano\meter}$ and $\lambda_{\rm L} = \qty{570}{\nano\meter}$ in (a) and (b) respectively.
In these simulations the current is induced via terminals at the edges $y = \pm \qty{290}{\nano\meter}$, acting as current source and drain.
At the terminals, boundary conditions induce a homogeneous current flow.
Both current densities are symmetric around the center of the strip.
For $\lambda_{\rm L} = \qty{10}{\nano\meter}$ the resulting normalized current density in Fig.~\ref{Fig:j_khapaev_conductor}(a) drops from 1 to roughly 0.6 from the edges to the middle of the strip.
For $\lambda_{\rm L} = \qty{570}{\nano\meter}$ the current density is homogeneous with changes far below 1\%.
%

\begin{figure}
	\includegraphics[width=\linewidth]{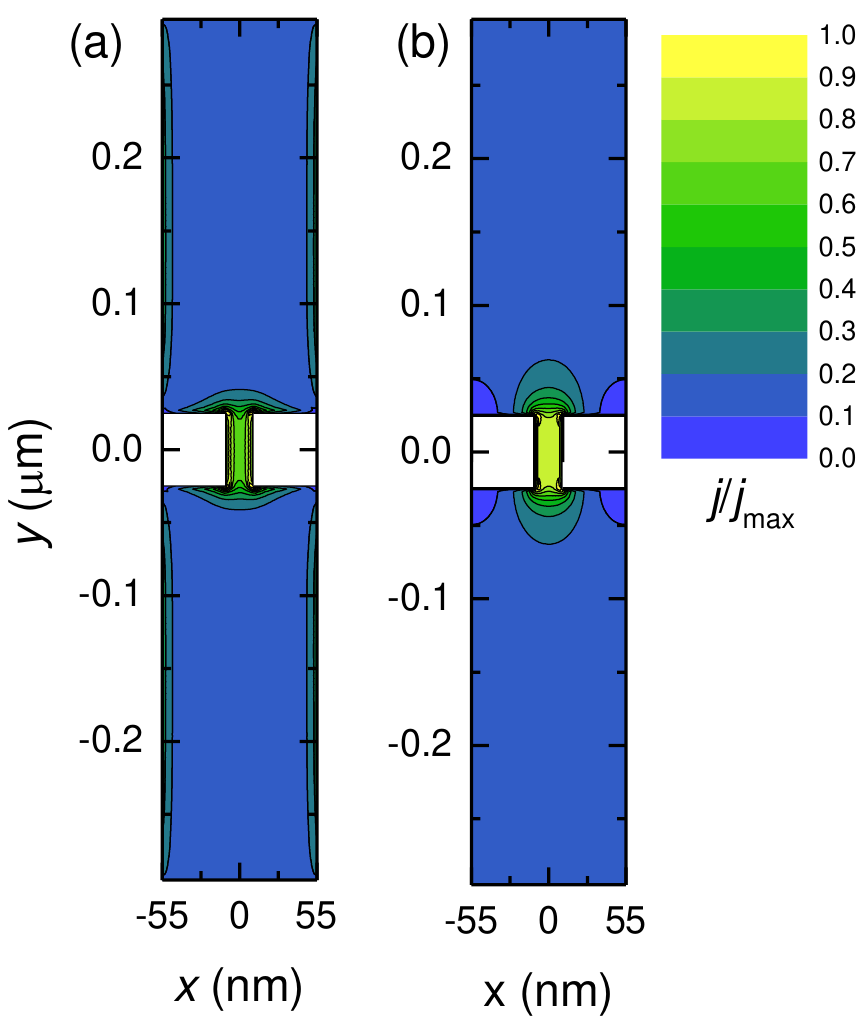}
	\caption{Normalized current density of a single strip containing a constriction simulated with 3D-MLSI. 
  Dimensions of the strip are $l = \qty{590}{\nano\meter}$, $w = \qty{110}{\nano\meter}$ and $d = \qty{50}{\nano\meter}$.
  Parameters of the constriction are $l_{\rm c} = \qty{50}{\nano\meter}$, $w_{\rm c} = \qty{20}{\nano\meter}$ and $d_{\rm c} = d$.
  (a): Simulation with $\lambda_{\rm L} = \qty{10}{\nano\meter}$.
  (b): Simulation with $\lambda_{\rm L} = \qty{570}{\nano\meter}$.}
  \label{Fig:j_khapaev_conductor_with_constriction}
\end{figure}

Figure \ref{Fig:j_khapaev_conductor_with_constriction} shows two normalized current density distributions of a strip containing a constriction simulated with 3D-MLSI, used to calculate the coupling factors shown in Fig.~\ref{main-Fig:strip_with_constriction} of the main manuscript.
The dimensions of the strip are $l = \qty{590}{\nano\meter}$, $w = \qty{110}{\nano\meter}$ and $d = \qty{50}{\nano\meter}$, with a constriction with $l_{\rm c} = \qty{50}{\nano\meter}$, $w_{\rm c} = \qty{20}{\nano\meter}$ and $d_{\rm c} = d$.
The current density in Fig.~\ref{Fig:j_khapaev_conductor_with_constriction}~(a) and (b) are obtained for $\lambda_{\rm L} = \qty{10}{\nano\meter}$ and $\lambda_{\rm L} = \qty{570}{\nano\meter}$ respectively.
Both show a current concentration in the region of the constriction according to the reduced cross section.
At the transition between constriction and wider strip, a region can be seen where the current decreases to the value observed further away from the constriction.
For $\lambda_{\rm L} = \qty{10}{\nano\meter}$ this region is smaller (Fig.~\ref{Fig:j_khapaev_conductor_with_constriction}~(a)) compared to $\lambda_{\rm L} = \qty{570}{\nano\meter}$ (Fig.~\ref{Fig:j_khapaev_conductor_with_constriction}~(b)).
In Fig.~\ref{main-Fig:strip_with_constriction} in the main manuscript coupling factors obtained from these current density distributions is compared to calculations by Eq.~\eqref{main-Eq:coupling_constriction} in the main manuscript.
In the calculations, the total coupling factor is composed of the contribution of the constriction and the contribution of the surrounding arms, with the assumption of homogeneous current density in each of the regions. 
Due to the smaller transition region for smaller $\lambda_{\rm L}$ simulation and calculation are in better agreement for this case.
%

\makeatletter\@input{xx.tex}\makeatother